\documentclass[usenatbib]{mn2e}
\usepackage{graphicx, amsmath, aas_macros, deluxetable}

\def \Msun {\, \rm M_\odot}

\title[Microlensing Towards the LMC]{The Origin of the Microlensing Events Observed Towards the LMC and the 
Stellar Counterpart of the Magellanic Stream} 
\author[Besla et al.]{
{\parbox{\textwidth}{Gurtina Besla$^{1}$\thanks{$\!\!$Hubble Fellow   e-mail:gbesla@astro.columbia.edu}, 
Lars Hernquist$^2$ \& Abraham Loeb$^2$} \vspace{0.4cm}}\\
$\!\!$$^{1}$Department of Astronomy \& Astrophysics, Columbia University, 550 West 120th Street, New York, NY 10027, USA\\
$\!\!$$^{2}$ Harvard-Smithsonian Center for Astrophysics, 60 Garden Street, Cambridge, MA 02138 
}
\begin{document}

\pagerange{\pageref{firstpage}--\pageref{lastpage}}
\pubyear{2012}

\maketitle

\label{firstpage}

\begin{abstract}
 
We introduce a novel theoretical model to explain the long-standing puzzle of the nature of the microlensing events 
reported towards the Large Magellanic Cloud (LMC) by the MACHO and OGLE collaborations.  We propose 
that a population of tidally stripped stars from the Small Magellanic Cloud (SMC)
 located $\sim$4-10 kpc behind a lensing population of LMC disk stars 
can naturally explain the observed event durations (17-71 days), 
event frequencies and spatial distribution of the reported events.
 Differences in the event frequencies reported by the OGLE ($\sim$0.33 yr$^{-1}$) 
and MACHO ($\sim$1.75 yr$^{-1}$) surveys appear to be naturally 
accounted for by their different detection efficiencies and sensitivity to faint sources. 
 The presented models of the Magellanic System were constructed without 
prior consideration of the microlensing implications.  
 These results favor a scenario for the interaction history of the Magellanic Clouds, 
 wherein the Clouds are on their first infall towards the Milky Way and the 
SMC has recently collided with the LMC 100-300 Myr ago, leading to a large number of faint sources 
distributed non-uniformly behind the LMC disk.  
In contrast to self-lensing models, microlensing events are also expected to occur in fields off
 the LMC's stellar bar since the stellar debris is not expected to be concentrated in the bar region. 
This scenario leads to a number of observational tests:  the sources are low-metallicity
 SMC stars, they exhibit high velocities relative to LMC 
disk stars that may be detectable via proper motion studies, and, most notably, 
there should exist a stellar counterpart to the gaseous 
Magellanic Stream and Bridge with a V-band surface brightness of $>$ 34 mag/arcsec$^2$.
In particular, the stellar Bridge should contain enough RR Lyrae stars to be
detected by the ongoing OGLE survey of this region. 

\end{abstract}


\section{Introduction}
\label{sec:intro}

\citet{paczynski1986} suggested that massive compact halo objects (MACHOs) could be
 found by monitoring the brightness of several million stars in the Magellanic Clouds (MCs) in
 search for microlensing by unseen foreground lenses.  
Despite the seemingly daunting scale of such a project, 
there have been a number of surveys conducted towards the MCs in search of microlensing events, 
such as the MACHO survey \citep{alcock2000},  the Experience pour la Recherche D'Objets Sombres survey
 \citep[EROS;][]{tisserand2007} and  
the Optical Gravitational Lensing Experiment \citep[OGLE;][]{udalski1997}.   
The goal of these surveys was to test the hypothesis that MACHOs could be a major component of the dark matter halo of the 
Milky Way (MW). 
After 20 years of work, a number of microlensing events have been detected towards the Large Magellanic Cloud (LMC),
 but a clear explanation for their origin remains elusive.

The results from the MACHO, OGLE and EROS surveys (summarized in $\S$~\ref{sec:Surveys})
 have demonstrated that objects of astrophysical mass 
($10^{-7} - 10 \Msun$) do not comprise the bulk of the dark matter in the halo \citep{Alcock2009, Moniez2010}. 

The idea that the MW's dark matter halo is {\it completely} populated by MACHOs has been ruled out based on 
elemental abundances and constraints on the total baryon fraction from baryon acoustic oscillations observed in the 
cosmic microwave background \citep{komatsu2011}.  
If the observed events result from lensing by MACHOs, the implied MACHO mass fraction 
(assuming these objects are $\sim$0.5 $\Msun$ in mass) 
of the MW dark matter halo is 
$\sim$16 \% \citep{bennett2005}.  The nature of such MACHOs is unknown. Low mass hydrogen burning stars
have been ruled out as a possibility by red star counts from the Hubble Space Telescope \citep{flynn1996}. 
 White dwarfs have also been proposed as 
potential candidates, supported by claims that $>$ 2 \% of the halo is possibly made up of old
 white dwarfs \citep{oppenheimer2001}. 
However, it appears likely that the observed white dwarfs are part of the thick disk of the
 MW \citep[etc.]{torres2002, flynn2003, 
spagna2004}.  \citet{brook2003} also rule out the possibility of a white dwarf halo fraction in excess of 5\% based on 
the inability of an initial mass function biased heavily toward white dwarf precursors to reproduce the observed 
C/N/O abundances in old MW stars
\citep[see also][]{gibson1997}.

Alternatively, the observed microlensing signal might be a result of lensing by normal stellar populations. 
Self-lensing by stars in the LMC's disk may explain the events detected by the OGLE team 
\citep{calchi2009, calchi2011, Wyrzy2011, sahu1994}.
Also binary star lenses were found to be associated with the LMC/SMC
 \citep{bennett1996, afonso2000, udalski1994, dominik1996, mao1995, alard1995}. 
 \citet{sahu2003} raise a number of additional arguments disfavoring a halo origin for the lensing events.  In particular, if the lenses
are MACHOs in the halo, the durations of the events in the LMC should be similar to those in the SMC.  However, 
the SMC event durations (75-125 days) are observed to be much longer than those towards the LMC (17-71 days). 
This discrepancy can be reconciled in models where the lenses originate within the Clouds themselves
 since the SMC has a much larger line-of-sight depth than the LMC \citep{haschke2012, sub2012, crowl2001}.
 
However, attempts to explain the MACHO team's larger number of reported events using a 
variety of disk and spheroid models for the LMC and MW have failed to explain their reported
 microlensing optical depth at the 99.9\% level 
\citep[and references therein]{gyuk2000, bennett2005}.  \citet{gould1995a} 
presents simple dynamical arguments against self-lensing models 
by LMC stars \citep[such as those proposed by][]{sahu1994, wu1994}, finding that the required
 velocity dispersions would be much larger than 
observed for known LMC populations in order to reproduce the observed lensing optical depth. 

Non-standard models for the LMC's structure have been explored to try and get around such constraints.
\citet{zhao2000} suggest that the off-center bar in the LMC is an unvirialized structure, misaligned with and offset from the plane 
of the LMC disk by 25 degrees.  They argue that this offset allows the bar structure to lens stars in the LMC's disk. 
As discussed in \citet[][hereafter, B12]{besla2012}, our simulations agree with the theory of \citet{zhao2000} that the bar of the LMC has become 
warped by its recent interactions with the SMC - but the simulated warp of $\sim$10 degrees is less severe than what they require.  
Moreover, microlensing events have been identified off the bar of the LMC, meaning that an offset bar 
cannot be the main explanation.

\citet{mancini2004} refine the \citet{gould1995a} argument and account for our updated
 understanding of the geometry of the LMC's disk \citep{vanderMarel2001,vanderMarel}. 
 Like \citet{zhao2000}, they argue that a non-coplanar bar strengthens the self-lensing argument. But
ultimately they find that self-lensing cannot account for all the events reported by the MACHO team
\citep{calchi2011}.

 \citet{weinberg2000} and \citet{evans2000}
claim that the LMC may have a non-virialized stellar halo, differing from a traditional stellar halo 
in that it has a low velocity dispersion
(for that reason it is referred to as a ``shroud" rather than a halo).
A stellar halo is reported to have been detected around the LMC, extending as much as 20 degrees
(20 kpc) from the LMC center
\citep[][and Nidever et al. 2012, in prep.]{majewski2009, munoz2006}\footnote{Although it is 
unclear whether the very extended stellar population detected by \citet{munoz2006} includes 
main sequence stars (Olszewski et al. 2012 in prep), shedding doubt on their membership as part 
of an extended LMC stellar halo.}.
\citet{pejcha2009} used RR Lyrae stars from the OGLE III catalogue to claim that such an 
LMC stellar halo is structured as a triaxial ellipsoid elongated along the line of sight to the observer, 
which would aid in explaining the microlensing events.
But the total mass of such a halo is likely no more than 5\% of the mass of the LMC \citep{kinman1991}, 
and is thus unlikely to explain the observed microlensing optical depth: 
\citet{alves2000} place a limit on the fractional contribution to the optical depth from 
such a low mass LMC stellar halo component to 20\%.  
\citet{gyuk2000} argue that in order to explain the total optical depth, the mass of the stellar halo 
must be comparable to that of the LMC disk + bar, a scenario that is ruled out by observations \citep[see also the discussion in][]{alcock2001}.

\citet{zaritsky1997} claim to have detected a foreground intervening stellar population 
with a stellar surface density of $\Sigma \sim 16 \Msun $ pc$^{-2}$, which could serve as the 
lensing stars. 
However, \citet{gould1998} argues that such a population would need an anomalously
 high mass-to-light ratio in order to explain a large
fraction of the observed events and remain unseen.  \citet{bennett1998} also point out 
that a foreground stellar population that could explain the LMC optical depth would 
require more red clump stars than inferred by \citet{zaritsky1997}. Furthermore, both
 \citet{gallart1998} and \citet{beaulieu1998} argue that the feature 
 seen by \citet{zaritsky1997} in the color magnitude diagram is a natural feature of the
  red giant branch, rather than an indication of a foreground population.

Thus, 10 years after the MACHO team's findings, we are still left with unsatisfactory solutions to the
nature of the observed microlensing events.

In this study, we consider the possibility that LMC stars are not the sources of the microlensing 
events, but rather the lenses of background stellar debris stripped from the SMC.  
This theoretical model is based on our recently published
simulations of the interaction history of the MCs and the formation of the
 Magellanic Stream, a stream of HI gas that trails behind the 
MCs 150 degrees across the sky \citep{nidever2010, Putman2003, Mathewson}, in a scenario where the MCs are on 
their first infall towards our MW \citep{besla2010, besla2012}. 

 \citet{zhao1998} (hereafter Z98) suggested that the microlensing events towards the LMC
  could be explained by tidal debris enshrouding
the MCs. This debris is assumed to be associated 
with the formation of the Magellanic Stream and could be either in front or behind the LMC, i.e. acting as 
sources or lenses.  In particular, Z98 was able to predict a lower concentration of events towards the
 LMC center than the self-lensing 
models, which rely strongly on the bar.  Our model differs in that the debris (source) stars are a combination of 
 SMC stars captured by the LMC over previous close encounters between the MCs ($<$1 Gyr ago)
  and a population of younger stars 
 pulled out from the SMC (or forming in the Bridge) during the recent collision with the LMC 100-300 Myr ago
  \citep[Model 2 from][]{besla2012}.  
 Z98, on the other hand, hypothesize that the debris is a significantly older stellar population
  associated with the Magellanic Stream, spanning a great circle about the MW.  The resulting kinematics and spatial distribution
 of the sources in our model is consequently quite different than in Z98. Moreover, the great circle picture of stellar debris 
 is inconsistent with a first infall scenario, as the MCs have not yet completed an orbit about the MW. 

In the B12 model, the Magellanic Stream owes its origin
 to the action of LMC tides stripping material from the SMC before the MCs have been accreted by the MW. 
 This scenario is in sharp contrast to prevailing models for the origin of the Stream, which all rely 
 to some degree on the action of MW tides at previous pericentric passages to explain its full extent 
 and the existence of material leading to the MCs, referred to as the Leading Arm Feature \citep{Heller, Lin95, GN, Bekki, 
 Connors, Mastro, ruzicka2010, diaz2011, diaz2012}. Such a scenario is at odds with a first infall, which 
 is the expected orbital solution for the MCs based on updated cosmological models for the dark matter halo of the MW 
 \citep{besla2007, BoylanBesla2011, busha2010} and new HST proper motion measurements
 of the MCs \citep[][ Kallivayalil et al., 2012 in prep]{Nitya1, Nitya2}.

The B12 LMC-SMC tidal model should remove stars as well as gas; however, 
to date no old stars have been detected in the Stream or the 
Magellanic Bridge that connects the two galaxies. Indeed the apparent
 absence of stars has been the main argument in favor of 
hydrodynamic models, where the Stream forms from ram pressure stripping by the MW's diffuse
 hot gaseous halo or from stellar outflows. All tidal models for the Magellanic Stream predict a stellar 
counterpart to some degree \citep[e.g.,][]{diaz2012, GN}, whereas the hydrodynamic models naturally 
explain their absence \citep[e.g.,][]{nidever2008, Mastro, Moore}.   
In this work we demonstrate that the non-detection of a stellar 
counterpart to the Magellanic Stream is reconcilable with the B12 model, as tides remove material 
from the outskirts of the original SMC's extended gaseous disk, where the stellar density is low.  The corresponding 
surface brightness of the predicted stellar counterpart to the Magellanic Stream and Bridge is below the sensitivity 
of existing surveys, but should be observable by future and ongoing photometric surveys.  In fact, such a debris field 
may already have been detected:  recently, \citet{olsen2011} discovered a population of metal
 poor RGB stars in the LMC field that have different kinematics from those of local stars in the LMC disk \citep[see also][]{graff2000}.
 In B12 we showed that such kinematically distinct debris is expected from tidal LMC-SMC interactions and 
 represents $\sim$1.5\% of the LMC's disk mass. 

The overall goal of the work presented here
is to use the simulated stellar debris predicted to exist behind the LMC by the B12 models to determine the corresponding
 microlensing event properties (durations, frequency, distribution) towards the LMC owing to the lensing of this debris by 
 LMC disk stars. These values will be compared directly to the MACHO and OGLE survey results.

 We begin our analysis with a description of the results of the three main microlensing
 surveys ($\S$~\ref{sec:Methods})
  and our methodology and simulations ($\S$~\ref{sec:Sims}).  We follow with a detailed
   characterization of the surface densities and distribution of the sources and lenses 
   predicted by the B12 tidal models for the past interaction history of the MCs ($\S$~\ref{sec:Stellar}). 
 We will henceforth refer to SMC debris stars as sources and LMC disk stars as lenses. 
 We then estimate the expected effective distance between the simulated lens and sources ($\S$~\ref{sec:Dist})
 and the relative velocities
transverse to our line of sight ($\S$~\ref{sec:Vperp}). Using these quantities we derive 
the event durations ($\S$~\ref{sec:Dur}) and event frequency ($\S$~\ref{sec:Freq}) from 
lensed SMC debris and compare 
directly to the MACHO and OGLE results described in $\S$~\ref{sec:Surveys}.  Finally we outline 
a number of observationally testable consequences of the presented scenario ($\S$~\ref{sec:Discuss}).

\section{Microlensing Observations} 
\label{sec:Methods}

First we provide an overview of the relevant equations ($\S$~\ref{sec:Definitions}) and 
summarize the results of the MACHO, OGLE and EROS surveys ($\S$~\ref{sec:Surveys}). 
In this study we aim to compute the duration and number of expected microlensing events 
that would be observable over 
the duration of the MACHO and OGLE surveys from two
different models of the interaction history of the LMC-SMC binary system, as introduced in the B12 study.

\subsection{Microlensing Equations} 
\label{sec:Definitions}

The term microlensing refers to the action of compact low mass objects along the line of sight to 
distant sources, where ``compact" 
refers to objects smaller than the size of their own Einstein Radii, and ``low mass" refers 
to objects between 1 M$_{\oplus}$ and 
$10^{3} \Msun$.
  The resulting splitting of the image of the distant source is typically unresolvable, appearing 
instead as an overall increase in the apparent brightness of the source.

The microlensing optical depth, $\tau$, is defined to be the instantaneous probability that a random 
star is magnified by a lens by more than a factor of 1.34 (corresponding to an impact parameter
equal to the Einstein radius). 
The optical depth depends only on the density profile of lenses and the effective
 separation between the lens and sources ($D$)  \citep{paczynski1986}:   

\begin{equation}
D = \frac{D_{L} D_{SL}}{D_S}
\label{eq:Deff}
\end{equation}

{\noindent}where $D_{L}$ is the distance between the observer and the lens, 
$D_{S}$ is the distance between the observer and the source and $D_{SL} = D_S - D_L$. 
Throughout this study we compute quantities in gridded regions spanning the face of the 
LMC disk, as defined in $\S$~\ref{sec:Grid}.
We take $D_L$ as the mean distance to LMC disk stars in each grid cell  
but compute $D_S$ explicitly for each SMC star particle individually. 
We identify source debris stars as those SMC stars with $D_{SL} > 0$ to ensure that we only 
choose source stars that are behind the LMC disk on average. 

The microlensing optical depth is formally defined as:

\begin{equation}
\tau = \frac{4 \pi \rm{G}}{\rm{c}^2} D \hspace{0.01in}  \Sigma_{\rm lens}
\end{equation}

\begin{equation}
\tau = 6.0 \times 10^{-10} \left ( \frac{D}{\rm{kpc}} \right ) \left ( \frac{\Sigma_{\rm lens}}{\Msun/{\rm pc}^2} \right).
\label{ch8:eqTAU}
\end{equation}

{\noindent}The optical depth reflects the fraction of source 
stars that are magnified by a factor A $>$ 1.3416 at any given time. 
Observationally, $\tau$ is estimated as:  
\begin{equation}
\tau_{\rm obs}  = \frac{1}{E} \frac{\pi}{2} \sum_i \frac{t_i}{\xi(t_i)}
\end{equation}

{\noindent}where $E= 6.12 \times 10^7$ object years is the total 
exposure time ($10.7 \times 10^6$ objects 
times the survey duration of 2087 days), $t_i$ is the Einstein ring diameter
crossing time of the $i$th event,  and $\xi(t_i)$ is its detection efficiency, which 
peaks at 0.5 for durations of 200 days for the MACHO survey. The value of $E$ is dependent on the assumed source population.  
Thus, $\tau$ can be approximated as:

\begin{equation} 
\label{ch8:eqTAU2}
\tau \sim \frac{\pi \Gamma t_e}{ 2 N_{\rm source}}
\end{equation}

{\noindent}where $\Gamma$ is the event frequency and $t_e$ is the event duration.
We aim to derive these two quantities from the simulations. 

From equation (\ref{ch8:eqTAU2}), it is clear that the value of $\tau$ is strongly dependent on the 
adopted definition of the source population.
In the model explored here, we assume the source population consists of stars stripped from the SMC. This differs markedly from the choices made 
by the OGLE, MACHO and EROS teams, who assumed that the source population consists of LMC stars. As such, our value of $N_{\rm source}$
is significantly smaller than that adopted by these surveys. Consequently, our computed lensing probability, $\tau$, cannot be directly 
compared to the expectations from these surveys. Instead we must compare our simulations to the quantities that were observed directly,
 i.e. the event durations and event frequency.

The event duration ($t_e$) is defined as the Einstein radius crossing time: 
\begin{equation}
t_e = \frac{ R_e}{ V_{\perp}} \,
\end{equation}
After substituting for the Einstein radius $R_e$, we get : 
\begin{equation}
t_e = 4944.7 \left ( \frac{D}{\rm kpc} \right )^{0.5} \left ( \frac{V_\perp}{\rm{km/s}}\right )^{-1} \left (\frac{M_{\rm lens}}{\Msun} \right)^{0.5}  \hspace{0.1in} {\rm days}.
\label{eq:Te}
\end{equation}

{\noindent} Following \citet{alcock2000}, we do not include event durations longer than 300 days.
Note that the MACHO team adopt a non-standard definition for the event duration, which is twice the value we adopt here.
We compute $t_e$ using the average value of $\langle M_{\rm lens}^{0.5} \rangle$ from an adopted initial mass function, 
as described later in the text.

To compute $t_e$ we need to substitute both $D$ (as defined in equation~\ref{eq:Deff}) and $V_\perp$, the relative velocity 
of the source and lens perpendicular to our line-of-sight and projected in the plane of the lens. 
Following \citet{han1996},

\begin{equation}
V_{\perp} = \left(V_L - V_O \right)  - \left( V_S - V_O \right) \frac{D_L}{D_S}, 
\label{eq:V}
\end{equation}

{\noindent}where $V_S$, $V_L$ and $V_O$ are the velocities of the source, lens and observer, respectively, perpendicular to our line-of-sight. 
 These are defined by taking the cross product of the Galactocentric velocity of each particle with the normalized line-of-sight position vector of that particle. 
 $V_O$ is defined by taking the cross product of the observer velocity, $V_O = V_{\rm LSR} + V_{\rm pec}$, 
 with the average normalized line of sight position vector of the LMC particles. 

Recently, the values of both $V_{\rm LSR}$ and $V_{\rm pec}$ have come under debate. 
 The standard IAU value \citep{kerr1986} for the circular velocity of the Local Standard of 
Rest (LSR) is $V_{\rm LSR}$ = 220 km/s.  However, models based on the proper motion of Sgr A$^\ast$ 
\citep{reid2004} and masers in high-mass star-formation regions \citep{reid2009} have 
suggested that the circular velocity may be higher. \citet{McMillan2011} has presented 
an analysis that includes all relevant observational constraints, from which he 
derived $V_{\rm LSR} = 239 \pm 5$ km/s . In the following, we adopt this value for $V_{\rm LSR}$
and his distance for the Sun from the MW center of R$_\odot$ = 8.29 $\pm$ 0.16 kpc. 
For the peculiar velocity of the Sun with respect to the LSR we take the recent estimate 
from \citet{Schonrich2010}: $V_{\rm pec}$ = (U$_{\rm pec}$ , V$_{\rm pec}$, W$_{\rm pec}$ ) = (11.1, 12.24, 7.25),
 with uncertainties of (1.23, 2.05, 0.62) km/s. 
 
This was not the approach adopted in B12.  The revised $V_{\rm LSR}$ and $V_{\rm pec}$ imply a solar velocity in the 
Galactocentric-Y direction that is 26 km/s larger than that 
used by \citet{Nitya1} and \citet{Nitya2}. This directly impacts 
the determination of the 3D velocity vector of the LMC \citep[see also,][]{shattow2009}.  
We discuss the consequences of these revised values for the simulations analyzed in this study in $\S$~\ref{sec:Sims}. 

Because stellar particles representing source/lens stars will exhibit a wide variation in $V_\perp$ and $D$, we also expect to find a range of plausible 
$t_e$ in a given field of view. This range of values can be compared directly to the MACHO and OGLE results. 
We compute the average value of $t_e$ by integrating over the probability distribution of $P(D^{0.5}/V_\perp)$:

\begin{align}
\begin{split}
\langle t_e \rangle &= 4944.7   \langle \left (\frac{M_{\rm lens}}{\Msun} \right)^{0.5} \rangle \hspace{0.1in} {\rm days} \\
	  & \int  \left ( \frac{D}{\rm kpc} \right) ^{0.5} \left ( \frac{V_\perp}{\rm{km/s}}\right )^{-1}   P(D^{0.5}/V_\perp) {\rm d}(D^{0.5}/V_\perp)     .
\label{eq:TeAvg}
\end{split}
\end{align}

To compute $\langle \left (\frac{M_{\rm lens}}{\Msun} \right)^{0.5} \rangle$, we define a mass spectrum 
following \citet{Bastian2010} and \citet[][supplementary information]{sumi2011} for the low mass end:

\begin{equation}
\label{eq:IMF}
\frac{dN}{dM_{\ast}} \propto M_{\ast}^{-\alpha}
\end{equation}

{\noindent}where 
\begin{eqnarray}
\alpha &=& 0.3  \left (0.01 \Msun < M_{\ast} < 0.075 \Msun \right) \\
\alpha &=& 1.3    \left( 0.075 \Msun < M_{\ast} < 0.5 \Msun \right) \\
\alpha &=& 2.35  \left ( 0.5 \Msun < M_{\ast} < 10 \Msun  \right). 
\end{eqnarray}

This IMF follows a Salpeter index ($ \alpha -1 = 1.35$) at the high mass end \citep{Salpeter1955}
and a shallower power law ($ \alpha - 1 = 0$) for lower mass stars down to the hydrogen burning limit of 0.075$\Msun$. 

We adopt a mass range of 0.01-3 $\Msun$ for lensing stars. Stars more massive than 3 $\Msun$ are 
bright enough to have been easily observable. The lower limit accounts for the possibility that 
brown dwarfs could act as lenses; brown dwarfs are as common as main sequence stars in the Milky Way, 
meaning that a large fraction of Galactic bulge microlensing events owe to brown dwarfs \citep{sumi2011}.
 Reliable statistics for the mass function of brown dwarfs exist down to 
about 0.01 $\Msun$ \citep{sumi2011}.  
With these given mass limits, the average square root of the lensing mass is:
\begin{equation}
 \langle M_{\rm lens}^{0.5} \rangle =  \frac{ \int m^{0.5} \frac{dN}{dm} dm}{\int \frac{dN}{dm} dm}  =   0.61 \Msun^{0.5}
\label{eq:MTe}
\end{equation}


Finally, the event frequency, $\Gamma$, is computed from equations (\ref{ch8:eqTAU}), (\ref{ch8:eqTAU2})
 and (\ref{eq:Te}) as : 
 
\begin{equation}
\Gamma = 2.8 \times 10^{-11}  N_{\rm source} \left( \frac{\Sigma_{\rm lens} }{\Msun/{\rm pc}^2} \right) V_\perp D^{0.5} \left( \frac{M_{\rm lens}}{\Msun} \right)^{-0.5} \, {\rm yr}^{-1} 
\label{eq:Freq}
\end{equation}

{\noindent}where $\Sigma_{\rm lens}$ is the average surface mass density of lenses in each grid. 
 To determine $N_{\rm source}$, the number of source stars per grid cell, we normalize the IMF given in eqn~\ref{eq:IMF} 
using the total mass of stellar debris from the SMC per field of view.
We then compute the number stars within a subset mass range of 0.075-3 $\Msun$, which we adopt for source stars.  
The lower limit corresponds to the hydrogen 
burning limit and the upper limit to $\sim$ the mass of a main sequence A star. 

 To compute the average number of events expected per year within a given field of view, we must account for the probability distribution of 
  $P(V_\perp D^{0.5})$ for the stripped SMC stellar particles (sources) within that region:
  
 \begin{align}
 \begin{split}
\langle \Gamma \rangle &=   2.8 \times 10^{-11}  N_{\rm source} \left( \frac{\Sigma_{\rm lens}}{\Msun/{\rm pc}^2} \right)  \langle \left (\frac{M_{\rm lens}}{\Msun} \right )^{-0.5} \rangle  \\
	&  \int V_\perp D^{0.5} P(V_\perp D^{0.5}) \rm{d}(V_\perp D^{0.5})  \, {\rm yr}^{-1}  . 
\label{eq:FreqAvg}
\end{split}
 \end{align}
   
{\noindent}Here, $P(V_\perp D^{0.5})$ is the normalized probability distribution for the product of $V_\perp D^{0.5}$ 
 using equations (\ref{eq:Deff}) and (\ref{eq:V}) for all sources in a given field.   
 This method assumes that all sources are screened by the same lens population, meaning that $\Sigma_{\rm lens}$ is
 assumed to be uniform across the field of view (grid cell) considered. 

Again, using the IMF in eqn~\ref{eq:IMF}, the relevant average lensing mass (mass range 0.01-3 $\Msun$) in eqn~\ref{eq:FreqAvg} is: 
\begin{equation}
\langle M_{\rm lens}^{-0.5} \rangle =  \frac{ \int m^{-0.5} \frac{dN}{dm} dm}{\int \frac{dN}{dm} dm}  =   2.11 \Msun^{-0.5}
\label{eq:MFreq}
\end{equation}

\subsection{ Results of the MACHO, EROS and OGLE Surveys}
\label{sec:Surveys}

The MACHO survey examined the central $\sim$14 deg$^2$ region of the LMC over 5.7 years and 
 identified 17 candidate microlensing events \citep[set B of][]{alcock2000}, resulting 
in an estimate for the microlensing optical depth towards the LMC of $\tau = 1.2^{+0.4}_{-0.3} \times 10^{-7}$.
Since the original publication, a number of these candidates have been ruled out as transient events, 
such as SNe \citep[e.g. Event 22;][]{bennett2005},
 variable stars \citep[e.g. Event 23;][]{tisserand2005}
or as lensing events by stars in the thick disk of the MW \citep[e.g. Event 20 and Event 5;][]{kallivayalil2006,  nguyen2004, popowski2003, drake2004}.
OGLE III data recently shows that Event 7 exhibited a few additional brightening episodes, excluding it from being
a genuine microlensing event \citep{OGLEIII}.
Some authors have made different cuts; \citet{belokurov2004} claim that only 7 events are real.

After the listed corrections, 10 of the MACHO events survive as likely candidates and are listed in
 Table~\ref{ch8:MACHO} along with the duration of each event and source magnitudes. 
The event number is in reference to the event numbers assigned in \citet{alcock2000}. 
The corresponding event frequency is 10/5.7 $\sim$1.75 per year.  This is lower than would be expected if the dark halo
of the MW was made primarily of objects with masses of $\sim5 \Msun$ \citep{Alcock2009}. 

The EROS-1 survey covered a 27 deg$^2$ region containing the LMC bar and one field toward the SMC for 3 years.
Two events were reported towards the LMC \citep{aubourg1993}, 
but both have been rejected \citep{ansari1995, renault1997, laserre2000, tisserand2007}.
The followup EROS-2 survey covered a larger region of the LMC, spanning 84 deg$^2$, over 6.7 years found no 
 candidate microlensing event towards the LMC \citep{tisserand2007}; 
 combined with the EROS-1 results, this implied a microlensing optical depth 
 of less than $\tau < 0.36 \times 10^{-7}$
\citep{tisserand2007, Moniez2010}.  The discrepancy between the EROS-2 and MACHO 
results likely owes to the EROS team's choice to limit their 
candidate sources to only clearly identified {\it bright} stars in sparse fields throughout the LMC ($0.7 \times 10^7$ stars
over 84 deg$^2$), rather 
than including blended sources and faints stars in dense fields (MACHO: $1.2\times 10^7$ stars over 14 deg$^2$)
 \citep{Moniez2010}.   As also pointed out by \citet{Moniez2010}, only 2 of the 17 MACHO candidates in set B of \citet{alcock2000}
 were sufficiently bright to be compared to the EROS sample.  As such, we do not expect the EROS survey to have detected 
 the microlensing signal from our 
 predicted sources; i.e. a faint population of SMC debris behind the LMC should not be of similar magnitude as the brightest 
 LMC stars.  It should be noted that the EROS team was the first to point out that most lensing events were only 
 seen for faint stars \citep{tisserand2007}.

The OGLE II survey spanned a 4.7 deg$^2$ area centered on the bar region of the LMC. This is a somewhat smaller region than 
that covered by the MACHO survey. Following the MACHO strategy, blended sources were also considered. 
Over the 4 year span of the OGLE II survey (1996-2000), 2 events were reported \citep[LMC 1, 2,][ listed in 
Table ~\ref{ch8:OGLE}]{wyrzy2009, calchi2009}.
The follow-up OGLE III survey covered a larger 40 deg$^2$ area and reported 2 new events
 (LMC 3, 4; listed in Table~\ref{ch8:OGLE}), corresponding to an optical depth 
 of $\tau = (0.43 \pm 0.33) \times 10^{-7}$, 
and 2 less probable events over a total survey time of 8 years (2001-2009) \citep{OGLEIII}.  
The total event frequency is thus 4/12 $\sim$ 0.33 events/yr.  
This frequency is smaller than that detected by MACHO, but the spatial distribution
 of the events is in agreement with the MACHO team's findings: the 
events are not concentrated solely in the bar region.  The range of event durations is also similar.

 The OGLE III survey results imply a lower event frequency than observed by the MACHO team. 
 This discrepancy might be explained by differences
 in the sensitivity and detection efficiency between the two surveys. 
 The detection efficiency of the MACHO survey peaks at 50\% for event durations of 100 days, whereas that of 
 OGLE III peaks at 10-35\% for the same event duration, depending on the density of the field \citep{calchi2011}.
 More significantly, the detection efficiency of the OGLE survey is lower in more crowded fields \citep{OGLEIII, calchi2011};
  in contrast, the inner densest regions of the LMC disk is where the MACHO survey found the majority of their detections.
 Most of the LMC source stars are blue main sequence stars with little extinction, as a result the MACHO blue and red 
 bands have a much higher photon detection rate than the OGLE I-band. 
 Another factor is that the MACHO team spent a larger fraction of its observing time on the MCs, whereas the OGLE team 
 has higher sensitivity towards the Galactic bulge.   Furthermore,  the OGLE II and III survey threshold source magnitude
  was not as faint as that of the MACHO team \citep{calchi2011}.  Note that, since the OGLE survey
   is more sensitive that the EROS survey,  we opt to only directly compare the OGLE and MACHO results in this analysis.

Note that in Tables ~\ref{ch8:MACHO} and ~\ref{ch8:OGLE}, 
we have not listed the statistically corrected values of the blended\footnote{The term 
blending indicates that more than one star is enclosed within 
the seeing disk of a given resolved object \citep{calchi2011}} durations
  \citep[Table 4,][]{alcock2000}, since the distribution of these durations is artificially narrowed, 
  as pointed out by \citet{bennett2005} \citep[see also,][]{green2002, rahvar2004}. 
  This is fine for the calculation of the optical depth, 
but not for a study of the true distribution of durations. Instead, the durations
listed in Table~\ref{ch8:MACHO} are taken from Table 5 of \citet{alcock2000} 
and Table 1 of \citet{bennettbecker2005}.
  The observed event durations reported by both teams range from (17-70) days, with 
  similar average durations of 43 days for MACHO and 37.2 days for OGLE.
No short duration events (1hr - 10 days) have been observed to date \citep{aubourg1995, renault1997, alcock1996, alcock2000}, 
ruling out light objects ($< 10^{-1} \Msun$) as the prime constituents of the MW halo.

\begin{table*}
\centering
\begin{minipage}{200mm}
\caption{MACHO Microlensing Events \label{ch8:MACHO}}
\begin{tabular}{@{}ccc@{}}
\hline
Event			& Event Duration\tablenotemark{a}		&   $V_{M}$\tablenotemark{b}\\
				&  (days)							& 						\\
\hline
\hline

1				&   17.3; 17.4 						&   		19.78					\\ 
4 				&    39.5  							& 		21.33				\\ 
6				&     46.0 							& 		20.0					\\   
8 				&   	33.2 							&		20.20					\\ 
13				& 	66.0 							&	      21.76						\\ 
14\tablenotemark{c} 	 &    53.3							&		19.48					\\  
15				&    22							& 		21.18				 \\ 
18				&   37.9							& 		19.55					\\ 
21				&   70.8							& 		19.37					\\
25				&    42.7							& 		19.04					 \\
\hline
\tablenotetext{a}{Event durations are taken from Table 1 of  \citet{bennettbecker2005} for events 4, 13 and 15,
while the rest were taken from Table 5 (fits with blending)  \\
 of \citet{alcock2000}, rather than the statistically corrected \\
  durations used for the optical depth calculation.  The average event duration \\
  is $t_e \sim$ 43 days and the event frequency is $\Gamma \sim$1.75 events/yr over the 5.7 year survey.} 
\tablenotetext{b}{ $V$-band source star brightness from \citet{bennettbecker2005, alcock2001, alcock2000}}
\tablenotetext{c}{Event 14 is likely to be an LMC self-lensing event \citep{alcock2001}}	
\end{tabular}
\end{minipage}
\end{table*}

\begin{table*}
\centering
\begin{minipage}{200mm}
\caption{OGLE Microlensing Events \label{ch8:OGLE}}
\begin{tabular}{@{}ccc@{}}
\hline

Event			& Event Duration\tablenotemark{a} 	&  $V_{M}$\tablenotemark{b}   \\
				&  (days)						& 		\\
\hline
\hline

O1		&  57.2 				& 	20.65 $\pm$0.013  		\\  
O2 		&  23.8				& 	20.68 $\pm$ 0.014 		\\
O3		&  35.0 				& 	19.52 $\pm 0.1$		\\  
O4		&  32.8  				&  	18.45 $\pm 0.1$		\\  
\hline
\tablenotetext{a}{Event durations from OGLE (five parameter fit):  OGLE II results, O1 and O2 from  \citet{wyrzy2009}  \\
 and OGLE III results, O3 and O4 from \citet{OGLEIII}.  
 The average event duration is $t_e \sim$ 37.2 days and  \\
 the event frequency is $\Gamma \sim0.33$ events/yr.} 	
 \tablenotetext{b}{ Baseline $V$-band source star brightness from \citet{wyrzy2009}
  for events O1 and O2 and \citet{OGLEIII} for O3 and O4}  \\ 	
\end{tabular}
\end{minipage}
\end{table*}

Assuming the source stars are LMC disk stars,
 \citet{bennett2005} recomputed the microlensing optical depth based on a refined list of candidates 
(Events 5 and 7 were also included in this reanalysis) and found  $\tau \sim1.0 \pm 0.3 \times 10^{-7}$.  

We aim here to explain the listed event duration, distribution and event frequencies determined by the MACHO and OGLE
teams (Tables~\ref{ch8:MACHO} and~\ref{ch8:OGLE}) in a model where tidally stripped SMC stars exist behind the LMC's 
disk along our line of sight. 
Note that a population of debris stars located further away from the LMC disk along our line of sight are not likely to
 have been been included in the EROS source selection of only the brightest (unblended) stars. 
The EROS null result is thus not a limiting factor to our analysis and we focus on the 
events detected by the MACHO and OGLE collaborations.

\section{The Simulations}
\label{sec:Sims}

We follow the methodology outlined in B12 to simulate the interaction history of the LMC-SMC-MW system using the 
smoothed-particle hydrodynamics code, Gadget-3 \citep{Gadget2}.   
We refer the reader to the B12 paper for details of the numerical
simulations, but outline some of the more salient features here. 

\subsection{Model Parameters} 

Both the LMC and SMC are initially modeled with exponential gaseous and stellar disks and 
massive dark matter halos.  The gaseous disk of the SMC is initially 3 times more extended than its stellar disk.
The LMC has a total mass of $1.8 \times 10^{11} \Msun$ and the SMC has a total mass of $2 \times 10^{10} \Msun$
(mass ratio of $\sim$1:10).  

The interaction history of the LMC and SMC, independent of the MW, is followed in isolation for $\sim5$ Gyr. 
The interacting pair is then captured by the MW and travels to its current Galactocentric location on an orbit defined by the HST 
proper motion analysis of \citet{Nitya1}.  The MW is modeled as a static NFW potential with a total mass of 
$1.5 \times 10^{12} \Msun$.  In such a potential, backward orbital integration schemes show that the MCs are 
necessarily on their first infall towards our system, having 
entered within R$_{200}$ = 220 kpc of the MW 1 Gyr ago.

As mentioned in $\S$~\ref{sec:Definitions}, in this study we adopt the revised values of $V_{\rm LSR}$ 
from \citet{McMillan2011} and of $V_{\rm pec}$ from \citet{Schonrich2010}, rather than the IAU standard values 
as adopted in \citet{Nitya1}.  The B12 models were designed to reproduce the 3D velocity vector and position of the LMC
as defined by \citet{Nitya1}, thus we need to apply a velocity correction to the simulations in this study. 
 In addition to the revisions to the Solar motion, our team has obtained a 3rd epoch 
of proper motion data using WFC3 on HST (Kallivayalil et al. 2012, in prep).  With the new data, and better models for the
internal kinematics of the LMC, the proper motions have changed  from 
those originally reported by \citet{Nitya1, Nitya2}.  Specifically, the
 total center of mass motion of the LMC is 56 km/s lower than that reported by \citet[][378 km/s]{Nitya1}.

To account for these updated values, we systematically subtract out the old \citet{Nitya1} 3D velocity vector for the center of mass 
motion of the LMC from {\it each} particle in the B12 simulations and then add back the new Galactocentric
3D vector from Kallivayalil et al. (2012, in prep).

 If the LMC's total mass is $1.8 \times 10^{11} \Msun$, as expected within the $\Lambda$CDM paradigm 
  \citep{BoylanBesla2011}, 
 even this new lower velocity in a $1.5 \times 10^{12} \Msun$ MW model still 
 represents a first infall scenario (Kallivayalil et al. 2012, in prep). 
  Thus, the resulting dynamical picture has not changed by making this velocity shift. 
 One can think of this velocity correction as changing the observer location, rather than the physics or dynamical 
 picture. Moreover, similar revisions must be applied to the SMC's 3D velocity vector of 
 \citet{Nitya2}; consequently we find that the {\it relative} motion between the MCs is unchanged, at 128 $\pm$ 32 km/s. 
In this study we are largely concerned with computing quantities that are dependent on the
 {\bf relative motion} of LMC stars (lenses) with respect to the background SMC 
stellar debris (sources).  
The exact value of the 3D velocity vector of the LMC is thus not a major source of error in this study. 

We caution that the B12 models do not reproduce every detail of the Magellanic System.  In particular, the exact position and 
velocity of the SMC as determined by \citet{Nitya2} is not reproduced.  The SMC's position, however, does not 
affect the analysis in this study since we are concerned with the field of view centered on the LMC disk. 
  Moreover, the goal of this study, and that of B12, is not to 
make detailed models of the Magellanic System, but rather to illustrate a physical mechanism (namely 
tidal interactions between the MCs) that can simultaneously 
explain the observed large scale morphology, internal structure and kinematics of the Magellanic System and, perhaps 
surprisingly, also the nature of the observed MACHO microlensing events towards the LMC.

\subsection{Definition of Model 1 and Model 2}

In this study we examine the final state of the simulated Magellanic System for the two favored models 
for the orbital interaction history of the MCs from B12 (Model 1 and Model 2). 
In B12 we characterized the gaseous debris associated with the tidal LMC-SMC interactions. 
Here we characterize the resulting {\it stellar} counterpart to this gaseous system. 

Model 1 and Model 2 differ in terms of the number of pericentric passages the SMC has made about the LMC
since they began interacting with each other $\sim$6 Gyr ago. 
 In Model 1, the SMC completes 3 passages about the LMC over the past 6 Gyr, but it never gets closer 
 than $\sim$20 kpc to the LMC. 
 In Model 2 the SMC completes an additional passage about the LMC. This additional passage
 results in a direct collision between the MCs and the formation of a pronounced gaseous bridge connecting the MCs. 
 A Bridge is reproduced in Model 1, but is of much lower density, having formed in the same tidal event that formed the 
 Magellanic Stream. 

In B12, we showed that both of these models are able to reproduce the global large scale 
gaseous structure of the Magellanic System (Bridge, Stream and Leading Arm).  
Overall, however, Model 2 provides significantly better agreement with the internal structure and kinematics of the LMC. 

In particular, in Model 2, the LMC's bar (i.e. the photometric center of the LMC) is offcenter from the geometric and 
kinematic center of the underlying disk and is warped by $\sim$10 degrees relative to the disk plane, as observed.
  This is a consequence of a recent direct collision with the SMC.  In contrast, in Model 1, the bar is centered and coplanar 
  with the disk-plane. 
Model 2 is thus likely to provide a fairer assessment of the role of internal distribution of the LMC disk stars as lenses for 
the predicted microlensing event properties.

\section{The Stellar Counterpart to the Magellanic Stream and Bridge} 
\label{sec:Stellar}

In \citet{besla2010} and B12 we proposed that tidal forces from the LMC acting on the SMC are sufficient to 
strip out gaseous material that ultimately forms the 
Magellanic Stream without relying on repeated encounters with the MW. 
However, the dominant mechanism that removes material
 from the MCs is still a subject of debate; stellar outflows \citep{nidever2008, olano2004} and ram pressure stripping 
 \citep{Mastro, Moore} have also 
 been proposed as possible explanations. Distinguishing between these formation scenarios is critical to 
 the development of an accurate model for the orbital and interaction histories of the MCs and to our understanding
 of how baryons are removed from low mass systems. 

In the B12 model, LMC tides are also able to remove stars from the outskirts of the SMC disk in addition to the gas.
  In contrast, stars are not expected to be removed in the ram pressure or stellar outflow models.  The detection of stellar debris in the 
  Magellanic Stream would prove conclusively that the Stream is in fact a tidal feature, ruling out models based on hydrodynamic processes. 
The theoretical prediction of a stellar counterpart to the Magellanic Stream, Bridge and Leading Arm is not novel; it is expected
 in all tidal models for the Magellanic System  \citep[e.g.,][]{GN, Connors2006, Lin95, ruzicka2010, diaz2012}.

Recently, Weinberg (2012, in prep) found that by using a kinetic theory approach to solve the evolution of gas,  
gas undergoing ram pressure stripping by some diffuse ambient material is heated and ablated away, rather than being 
physically pushed out of the disk. As such, one would not expect ram pressure stripped material to be comprised 
predominantly of neutral hydrogen. These preliminary results suggest that it is unlikely that ram pressure
 stripping by the ambient MW hot halo is responsible for the formation of the Magellanic Stream.  
 This further implies that the location of the Stream on the plane of the sky cannot be used to precisely constrain the past
 orbital trajectory of the MCs, since its orientation in a tidal model will deviate from the path followed by the MCs \citep{besla2010}.
Note that, in the Appendix of B12 we also argue against a stellar outflow origin
 based on the low metallicities observed along the Stream \citep{fox2010}.

Ultimately, testing the tidal model has proven challenging as previous searches for stars in sections of the Stream
 have yielded null results \citep{NoStars, brueck1983}. We argue that this is likely because
 these studies were insufficiently sensitive.  
 Based on their null detection, \citet{NoStars} constrain the stellar-to-gas ratio to $< $0.1 in the high gas density 
 clumps they observed in the region of the Stream known as MSIV \citep[material between Galactic Latitudes  
 of -70$^\circ$ and -80$^\circ$ and Galactic Longitudes of 90$^\circ$ and 45$^\circ$; ][]{Putman2003}. 
 Here we will compare the B12 model results with these constraints. 
 
\citet{munoz2006} claim to have detected LMC stars in fields near the Carina dSph galaxy, over 20 degrees away from the LMC. 
 The identified stars have mean metallicities 
$\sim$1 dex higher than that of Carina and have colors and magnitudes consistent with the red clump of the LMC.  
They also argue that red giant 
star candidates in fields between the LMC and 
Carina exhibit a smooth velocity gradient, providing further evidence of the kinematic association of these distant 
stars to the LMC.  Such a velocity gradient would also be expected of tidal debris 
stripped from the SMC and the metallicities do not rule out an SMC origin
(Ricardo Mu$\tilde{\rm n}$oz, private communication, 2011). In line with this idea, \citet{kunkel1997} argued that a polar ring of carbon
 stars from the SMC exists in orbit about the LMC.  Also, \citet{weinberg2001} inferred the possible presence
 of tidal debris about the LMC based on Two-Micron All Sky (2MASS) data and suggest that this may have important 
 implications for microlensing studies, depending on its spatial distribution.
Such observations provide evidence for extended stellar distributions associated with the MCs
 that could potentially be interpreted as tidal 
debris. In this section we will attempt to reconcile these
observations with the tidal B12 model. 
 Note that \citet{majewski2009} have recently argued for the existence of an extended stellar halo around the LMC
  (David Nidever, private communication), but we do not model this explicitly. 

In B12 we noted that there is a significant transfer of stellar material from the SMC to the LMC expected in our proposed model. 
We also showed that the recent detection by \citet{olsen2011} of a metal poor population of RGB stars in the LMC disk with
 distinct kinematics from the mean local velocity fields of the LMC is consistent with the properties of the 
 transferred stars in the B12 model.  A marginally kinematically distinct 
 population of carbon stars has also been reported by \citet{graff2000}, which they also suggest could be tidal debris in the 
 foreground or background of the LMC that could account for the microlensing events observed toward the LMC. 

If stars are transferred from the SMC to the LMC there
 must also be tidal debris in the bridge of gas that connects the MCs, known as the Magellanic Bridge \citep{kerr1954}.  
\citet{harris2007} examined the stellar population 
in the Magellanic Bridge and was unable to locate an older stellar population, concluding that the detected
young stars ($<$ 300 Myr old) were formed in-situ in the gaseous bridge rather than being tidal debris stripped from the SMC. 
More recently, \citet{monelli2011} reported the detection of a population of stars located across the Bridge that could 
be compatible with the old disk population of the LMC. The observations by \citet{harris2007}  
 focused on the leading ridgeline (location of the highest gas density) of the Magellanic Bridge,
  which would experience the maximal ram pressure. 
 It is possible that ram pressure from the motion 
 of the MCs through the ambient halo gas of the MW may have displaced the gas in the bridge from the tidal stellar population.
  In fact, \citet{casetti2012} recently found a population of young stars
  offset by 1-2 degrees from the HI ridgeline.
 \citet{harris2007} constrained the stellar density of a possible offset stellar population using 2MASS observations, 
 but the 2MASS sensitivity limit of 20 Ks mag/arcsec$^2$ 
 is likely too low to have observed the expected stellar bridge.

In the following we quantify the properties and observability of the predicted stellar counterparts to the Stream, 
Bridge and Leading Arm Feature for the B12 model.

\subsection{The Extended Stellar Distribution of the Magellanic System}
\label{sec:StellarStream}

\begin{figure*}
\mbox{{\includegraphics[width=5.44in]{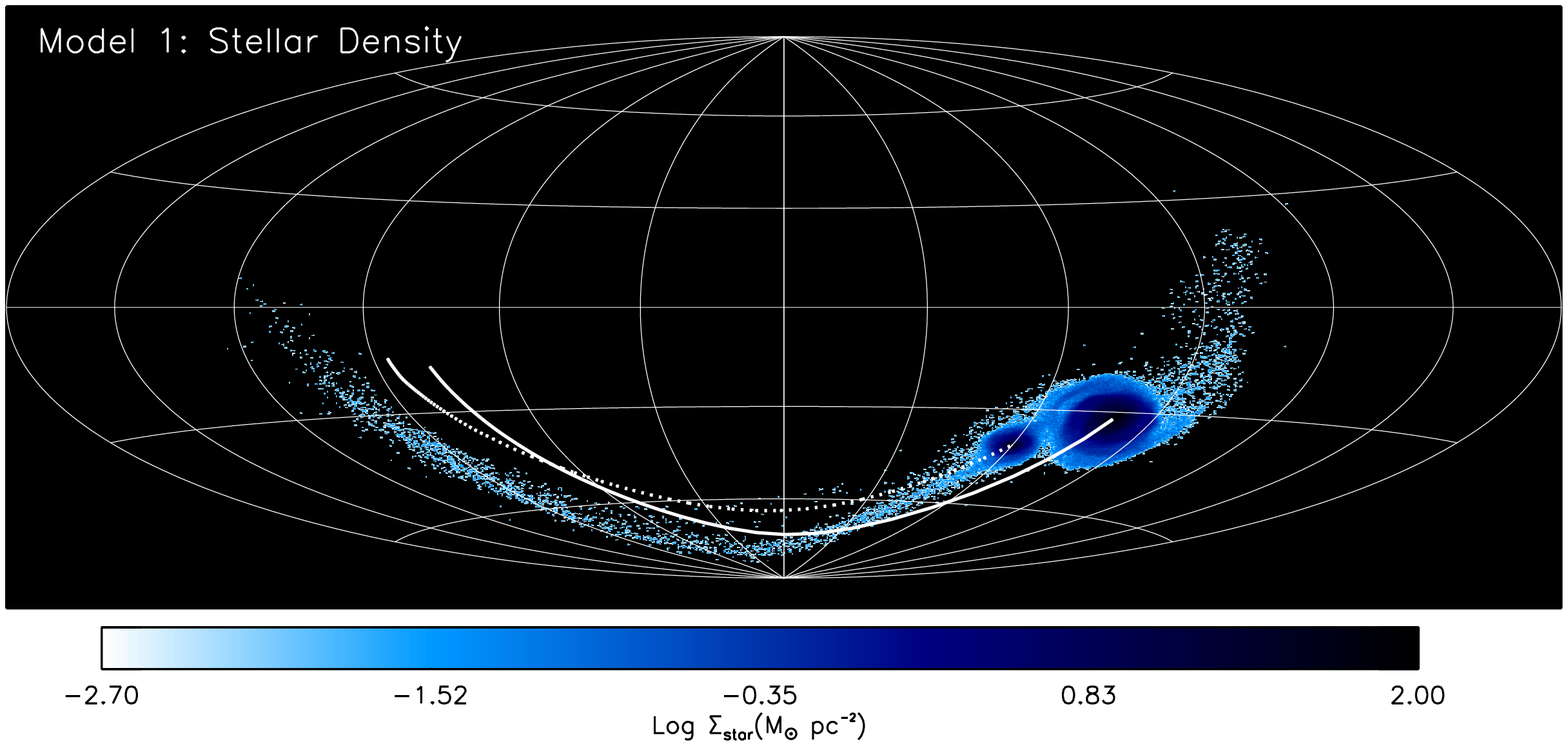}}}\\
   \mbox{\hspace{0.55in}{\includegraphics[width=6in]{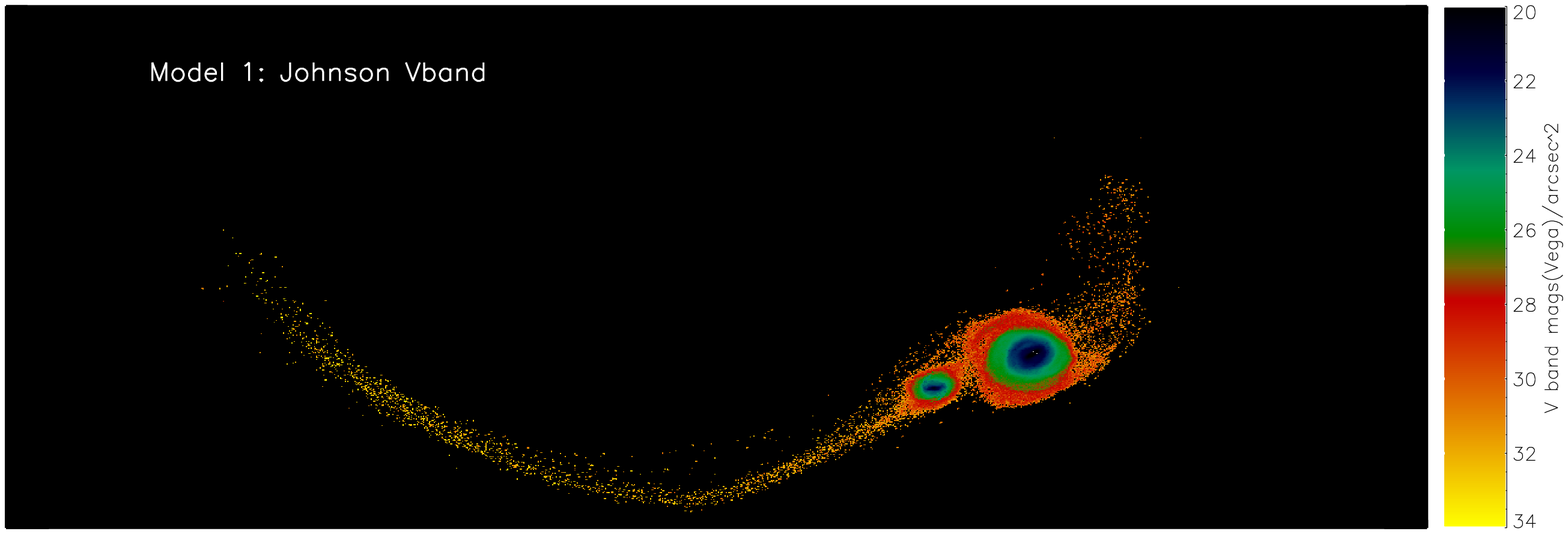}}}
 \caption{\label{ch7fig:AitoffModel1}  Hammer-Aitoff projection of the stellar density and
  Johnson V band surface brightness of the Magellanic System is plotted for Model 1. 
 The orbital trajectory of the LMC(SMC) is indicated by the solid(dashed) white lines in the stellar density plots. 
 The scale is the same in both images.   }
 \end{figure*}

\begin{figure*}
\mbox{{\includegraphics[width=5.44in]{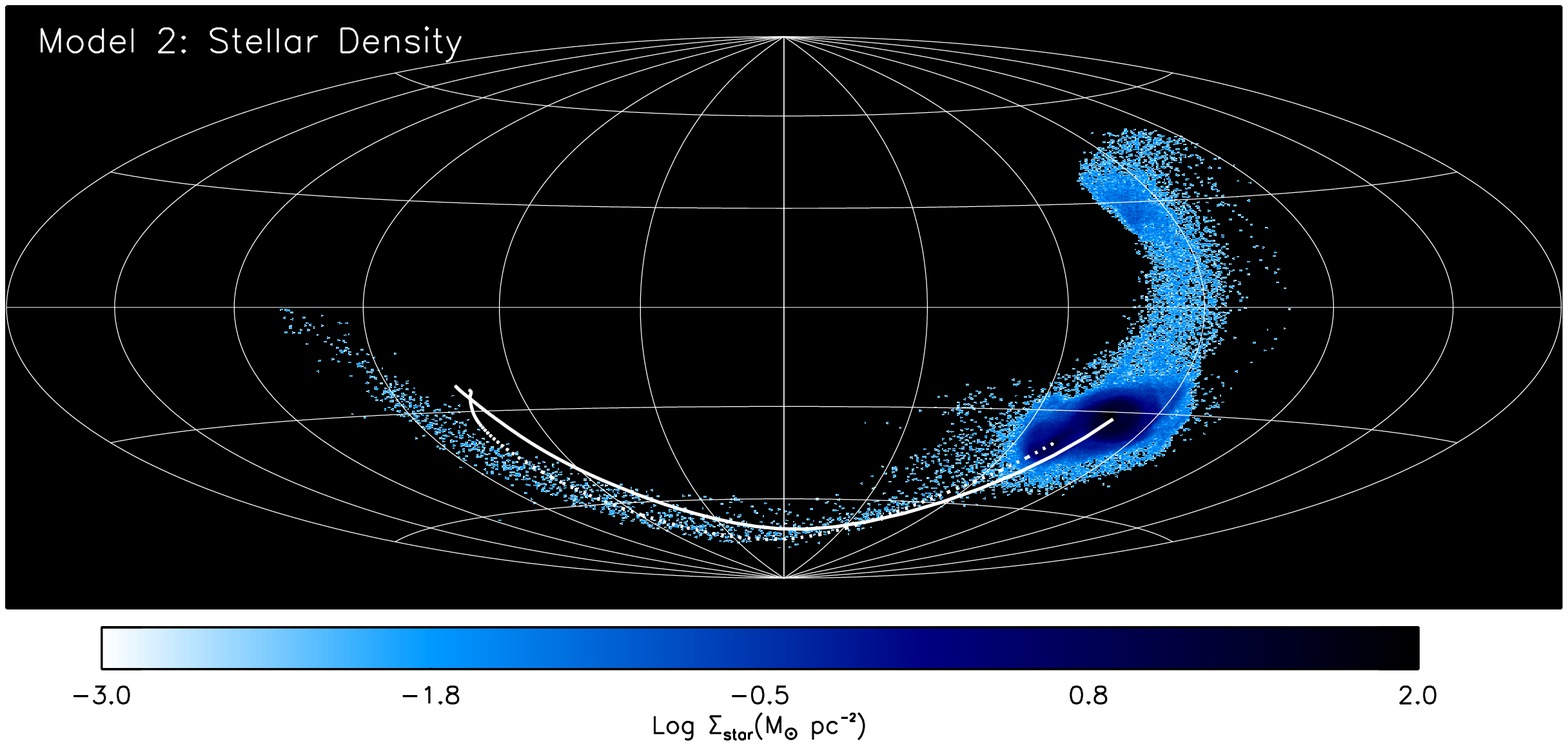}}}\\
   \mbox{\hspace{0.55in}{\includegraphics[width=6in]{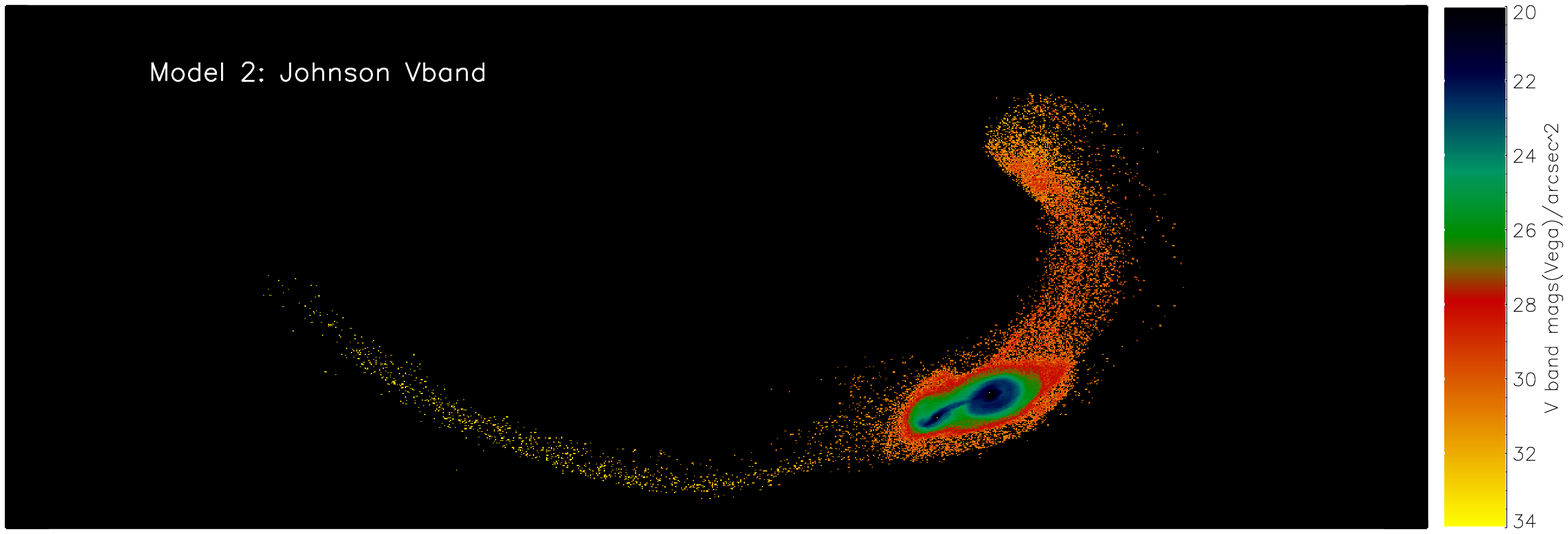}}}
 \caption{\label{ch7fig:AitoffModel2}  Same as Figure~\ref{ch7fig:AitoffModel1} but for Model 2.  
 The stellar leading arm feature differs markedly from that of Model 1. 
 The stellar density in the Bridge is also higher, however the majority of these stars are young, 
 having formed in-situ in the high gas-density Bridge as a result 
 of the recent SMC-LMC collision of Model 2.  }
 \end{figure*}

 Figures ~\ref{ch7fig:AitoffModel1} and ~\ref{ch7fig:AitoffModel2} show the Hammer-Aitoff projection of the stellar surface density 
and the corresponding Johnson V band surface brightness for Models 1 and 2, respectively. 
The expected Johnson V band emission is created from the simulations using the 
Monte Carlo dust radiative transfer code SUNRISE \citep{jonsson2010}. 
The bulk of the stellar material originates from the SMC. 
To model this stellar debris we adopt a Maraston stellar population model \citep{maraston2005}
 and the oldest stars are assumed to have an age of 8.7 Gyr.  SMC (LMC) stars are assigned metallicities of
  0.2 (0.5) Z$_{\odot}$, consistent with their observed values.

The average V band signal in the Stream is $>$ 30 mag/arcsec$^2$ for both Models, well below existing observational thresholds. 
This explains why the stellar counterpart has not been observed. The simulated stellar stream is
 more pronounced in Model 1 than in Model 2; this is 
because in Model 2, the Stream is older and the stars have more time to disperse. 

In particular, the stellar distribution in the Bridge region is well below 2MASS sensitivities (max of 20 mag/arcsec$^2$) in both models, 
making it unsurprising that \citet{harris2007} was unable to detect any offset 
stellar counterpart for the Bridge.   
The Bridge is clearly visible in the Vband in Model 2, but this is largely because of in-situ star formation (as the Bridge is known to be forming 
stars today). This is an important distinction between Models 1 and 2; in Model 1 stars are not forming in the Bridge, whereas in Model 2 the Bridge is 
higher gas density feature as a result of the recent direct collision between the LMC and SMC. In Model 2, the Bridge is formed via a combination of 
tides and ram pressure stripping, owing to the passage of two gaseous disks through one-another. This explains why the stellar density of {\it old} 
stars in the Model 2 Bridge is low. 
 Notice that the stellar distribution in the Bridge is fairly broad: only the youngest stars exist in the high density ``ridge" of the Bridge. 
The surface density of old stars in any given section of the bridge is low, making it plausible that 
\citet{harris2007} would have been unable to identify any older stars in the small fields they sampled along the Bridge ridgeline.

The two Models make very different predictions for the distribution of material leading the MCs.
  In Model 2, the leading stellar component is brighter than the Stream itself. 
This leading component is not dynamically old, having formed over the past 1 
Gyr and should therefore retain kinematic similarities to the motion of the overall Magellanic System. 

To compare against the upper limits on the star-to-gas ratio placed by \citet{NoStars} for the Magellanic Stream,  
we consider a high gas density section of the Model 1 simulated stream. The chosen region would 
correspond to the MSI region as defined by \citet{Putman2003} (between Magellanic Longitudes of -30 and -40).  
We choose to examine Model 1 because the gas and stellar densities in the modeled stream
 are higher than in Model 2: if Model 1 is consistent with the limits then Model 2 should be as well. 

\begin{figure*}
\begin{center}
\mbox{{\includegraphics[width=3in]{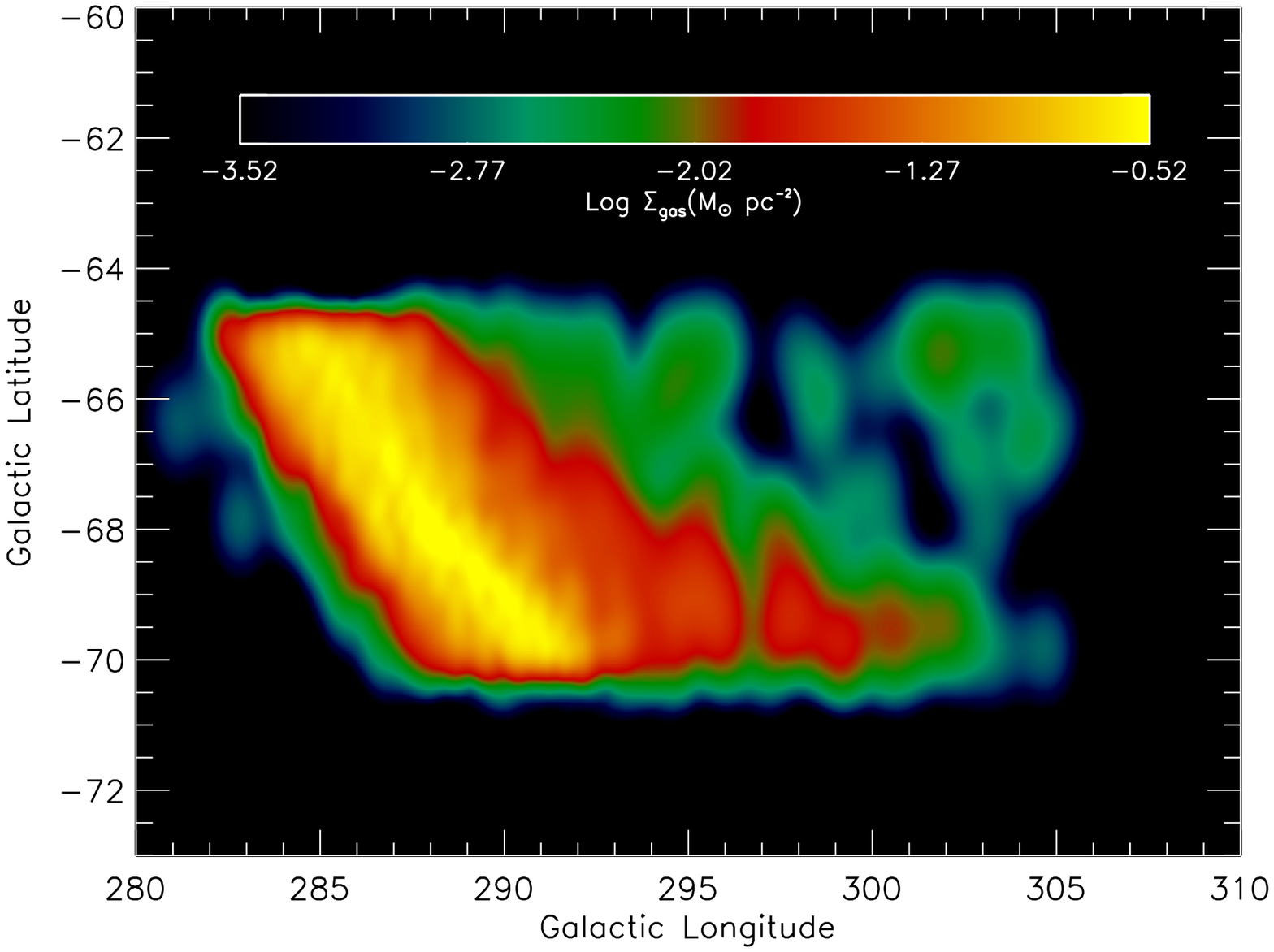}}
{\includegraphics[width=3in]{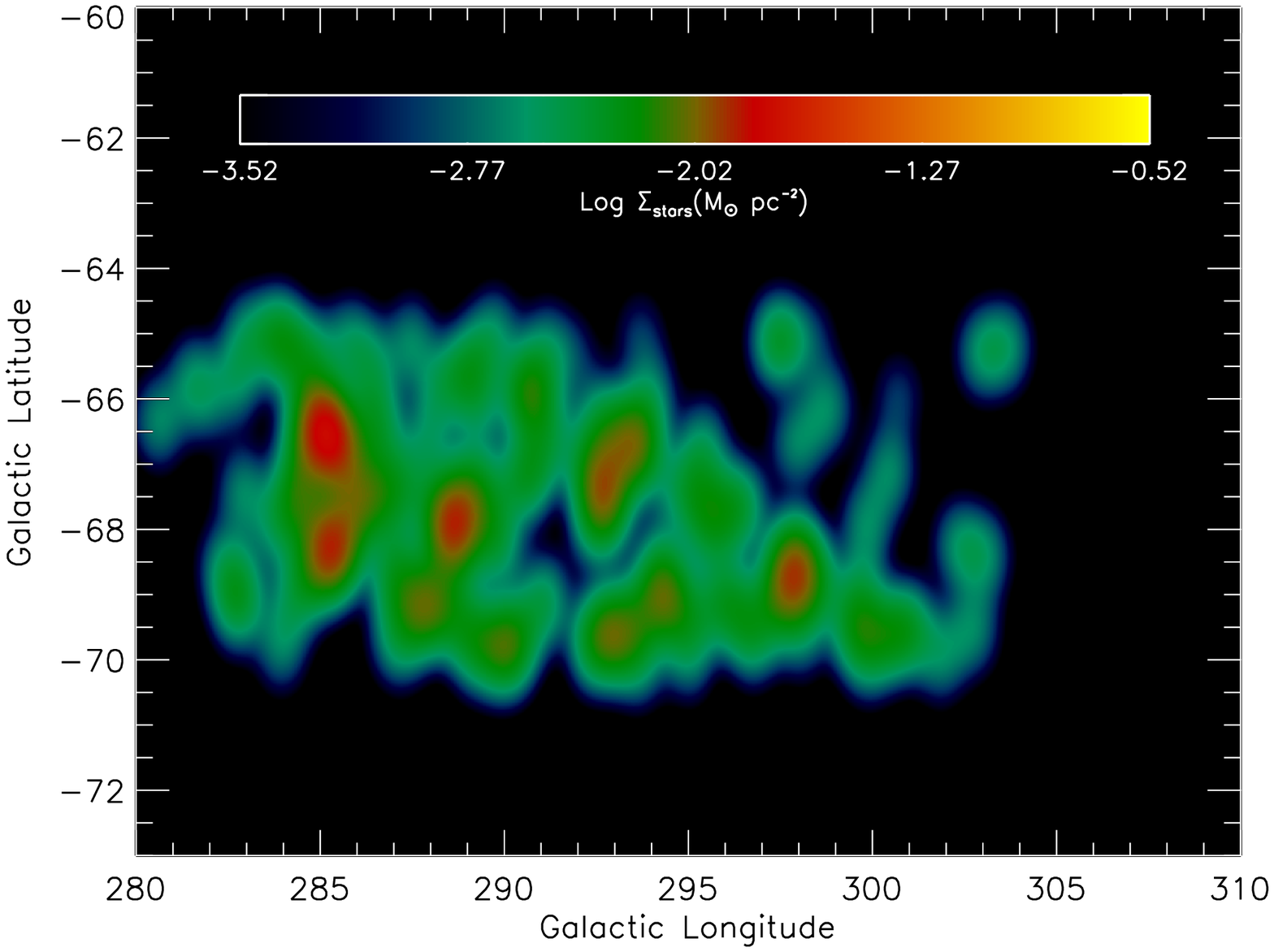}}}
 \end{center}
 \caption{\label{ch7fig:StarToGas}  The gaseous surface density (left) and stellar surface
  density (right) of a portion of the Magellanic Stream known as MSI \citep{Putman2003}
 is plotted in Magellanic Stream Coordinates \citep{nidever2008}.  
 The simulated stellar stream is coincident with the gaseous stream
  and the stellar (gas) densities peak at a value of $\sim$0.01 (0.3) $\Msun$/pc$^2$.  }
 \end{figure*}

Figure~\ref{ch7fig:StarToGas} shows the simulated gas and stellar surface densities in the MSI region for Model 1
in Magellanic Stream coordinates. 
This coordinate system is defined by \citet{nidever2008} such that 
the equator of this system is set 
 by finding the great circle best fitting the Magellanic Stream. The pole of this great circle is at 
 ($l$,$b$) = (188.5$^\circ$, -7.5$^\circ$) and the longitude scale is defined such that the LMC
 ($l$,$b$ = 280.47$^\circ$, -32.75$^\circ$)
 is centered at Magellanic Longitude of 0$^\circ$.  
From Figure~\ref{ch7fig:StarToGas}, we find that the peak gas surface density is $\sim$0.3 $\Msun$/pc$^2$
 and the stellar surface density peaks at $\sim$0.01$\Msun$/pc$^2$. 
The resulting stellar-to-gas ratio of 0.03 is well within the observational limits.

We thus conclude that the properties (density, surface brightness) of the predicted
 stellar counterpart of the Magellanic Stream and Bridge 
from both Models 1 and 2 of the B12 study are not at odds with 
the null-detections from existing surveys.

\subsection{The Source and Lens Populations}
\label{sec:Grid}

\begin{figure*}
\begin{center}
\mbox{\,{\includegraphics[width=6in, clip=true, trim= 0 4.4in 0 4in]{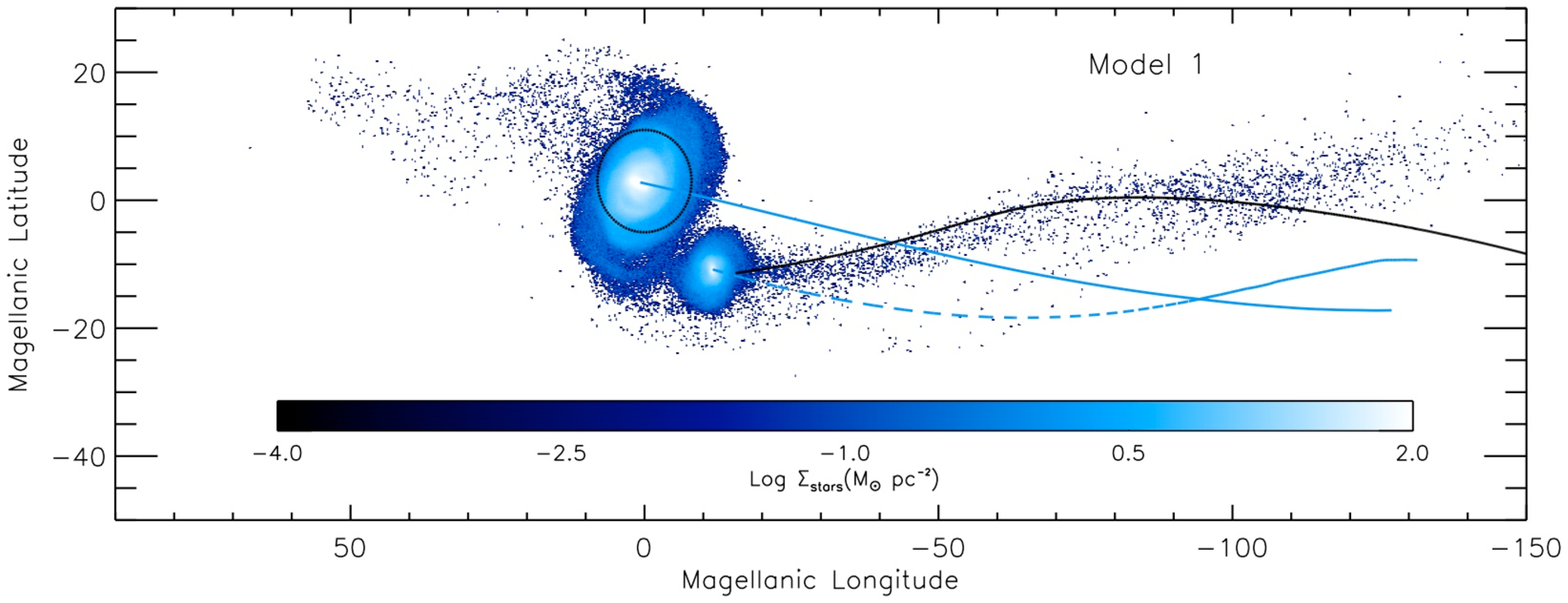}}}\\
   \mbox{ {\includegraphics[width=6in, clip=true, trim=0 4in 0 4in]{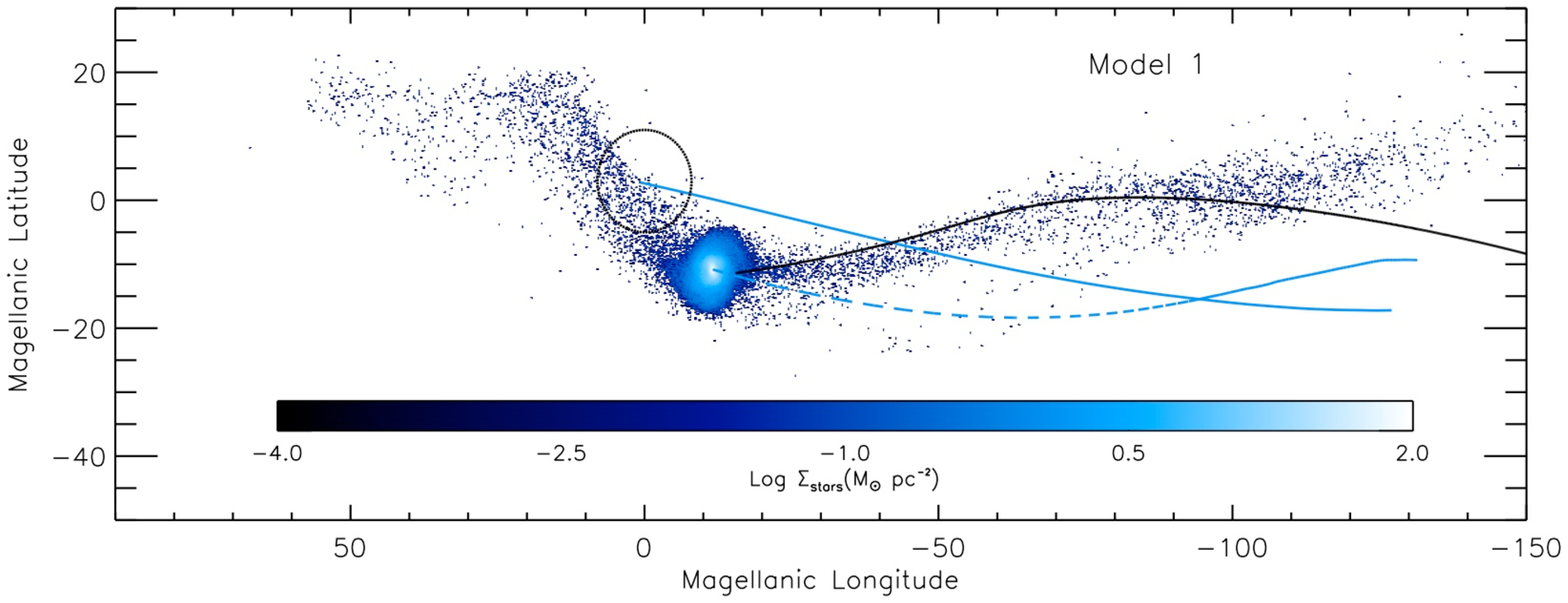}}}
 \end{center}
 \caption{\label{ch8fig:StarStreamModel1}  The stellar surface density of the Magellanic System 
 is plotted for Model 1 in Magellanic Stream Coordinates. Magellanic Longitude of 50 corresponds to a Galactic 
 Longitude of 0$\circ$, and the LMC ($l$,$b$ =  280.47$^\circ$, -32.75$^\circ$)
 is centered at Magellanic Longitude of 0$^\circ$ \citep{nidever2008, nidever2010}. Lines of 
 constant Galactic (l,b) are overplotted as dotted lines.  
 The orbital trajectory of the LMC(SMC) is indicated by the solid(dashed) blue lines. The thick black line traces the true location of the
 Magellanic Stream on the plane of the sky.  All particles are plotted in the top panel (this is the stellar counterpart of the top panel
 of Figure 7 in B12), whereas only particles originally part of the 
 SMC are plotted in the bottom panel. There are clearly SMC stellar particles in the same field as the LMC disk (indicated by the thick
 black circle).  }
 \end{figure*}

\begin{figure*}
\begin{center}
\mbox{\,{\includegraphics[width=6in, clip=true, trim= 0 0.5in 0 0]{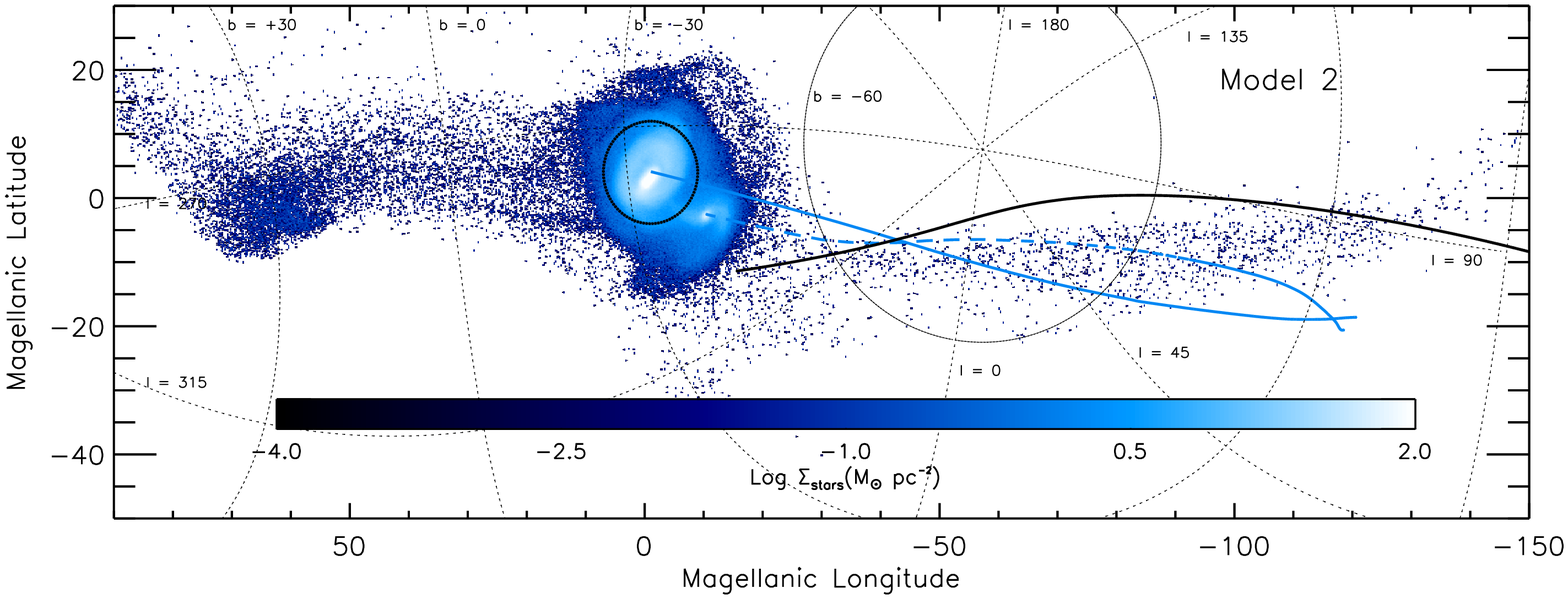}}}\\
   \mbox{ {\includegraphics[width=6in]{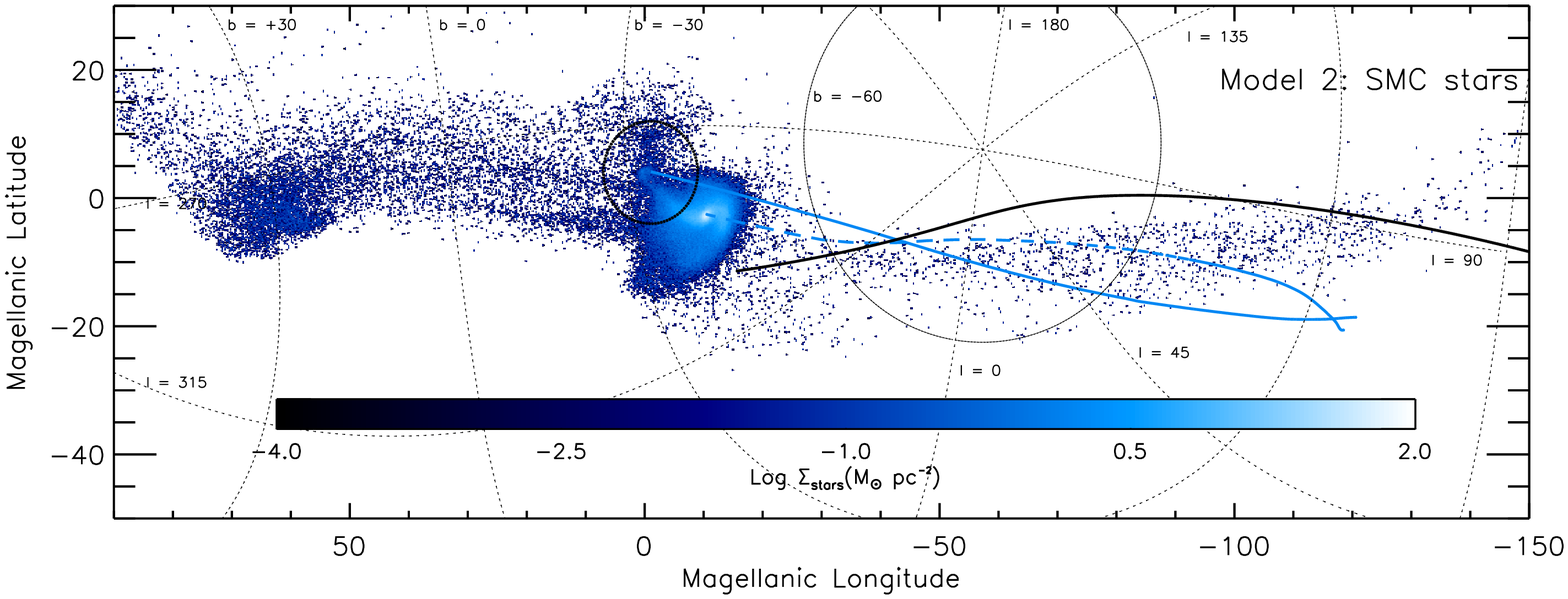}}}
 \end{center}
 \caption{\label{ch8fig:StarStream}  Same as Figure~\ref{ch8fig:StarStreamModel1} except for Model 2. The top panel is 
 the stellar counterpart of the bottom panel of Figure 7 in B12.  During the collision between the LMC and SMC stars are 
 stripped from the SMC from deeper within its potential than in Model 1. The corresponding debris has higher energy and 
 appears more dispersed spatially than in Model 1.    }
 \end{figure*}

In Figures~\ref{ch8fig:StarStreamModel1} and~\ref{ch8fig:StarStream} we illustrate the stellar distribution of the 
Magellanic System in Magellanic Coordinates, a variation of the galactic coordinate system where 
the Stream is straight \citep{nidever2008},   
 for both Models 1 and 2. The top panel shows the star particles from 
both of the LMC and SMC whereas the bottom panel depicts only stellar particles from the SMC. 
The location of the LMC is indicated by the white 
dotted circle.  There are clearly SMC stars in the same location as the LMC disk when projected on the plane of the sky. 
The distribution of stripped stars in Model 2 differs markedly from that in Model 1; in Model 2 the SMC collides directly with 
the LMC, the smaller impact parameter allows for the removal of stars from deeper within the SMC's potential.  
Consequently, the stellar debris in the Magellanic Bridge is more dispersed than in Model 1. 

\begin{figure*}
\begin{center}
\mbox{{\includegraphics[width=4.5in, clip=true, trim=0 0.5in 0 0]{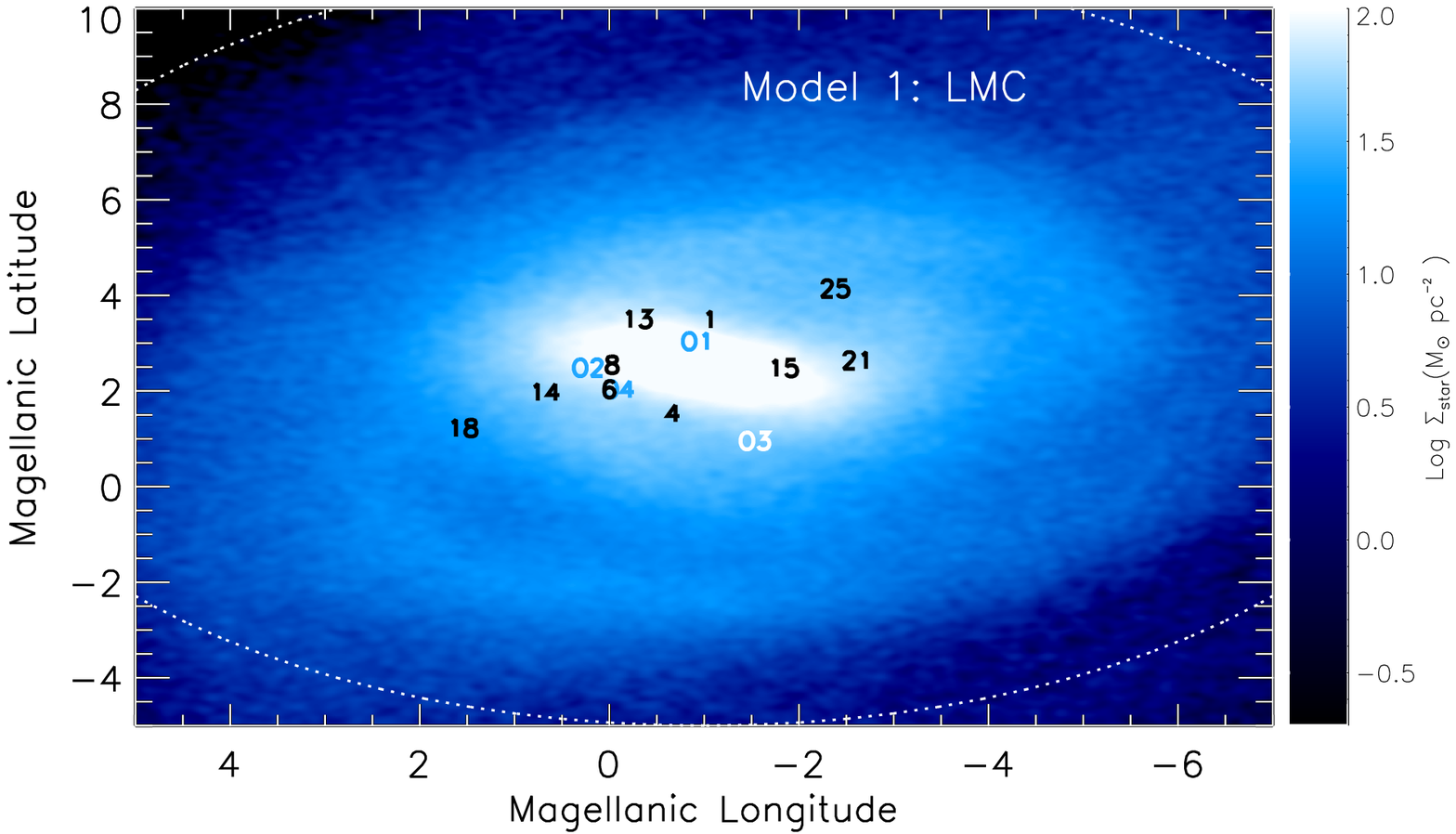}}}\\
\mbox{{\includegraphics[width=4.507in]{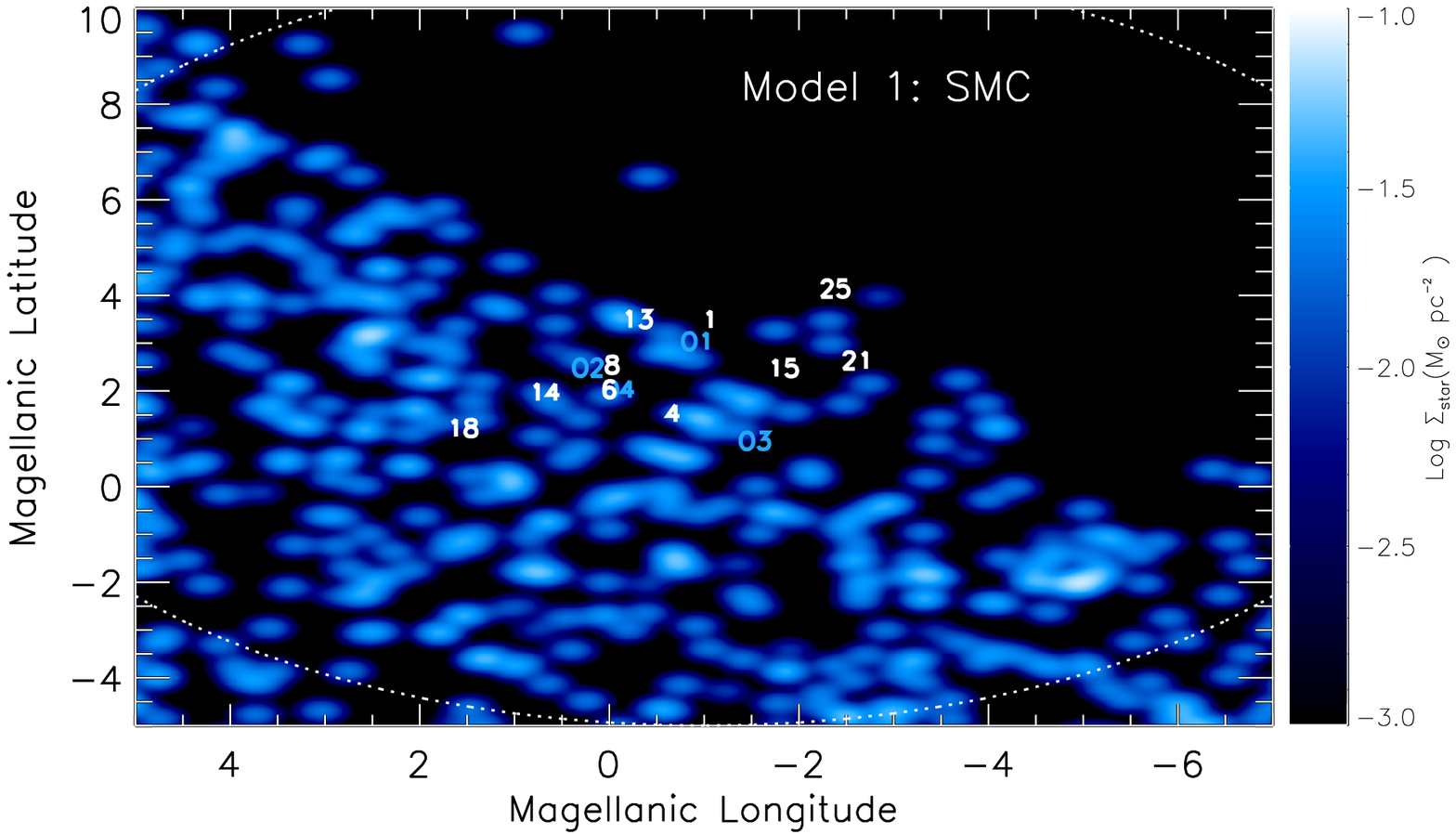}}}
 \end{center}
 \caption{\label{ch8fig:ZoomModel1}  The stellar surface density of lenses (LMC stars) for Model 1 is plotted in the top panel for a 10 degree region centered
 on the LMC. The middle panel shows the surface density of sources (stripped SMC stars) located in the same field. The location of the 
 MACHO and OGLE candidate microlensing events are marked by event number.  }
 \end{figure*}

\begin{figure*}
\begin{center}
\mbox{{\includegraphics[width=4.5in, clip=true, trim=0 0.5in 0 0]{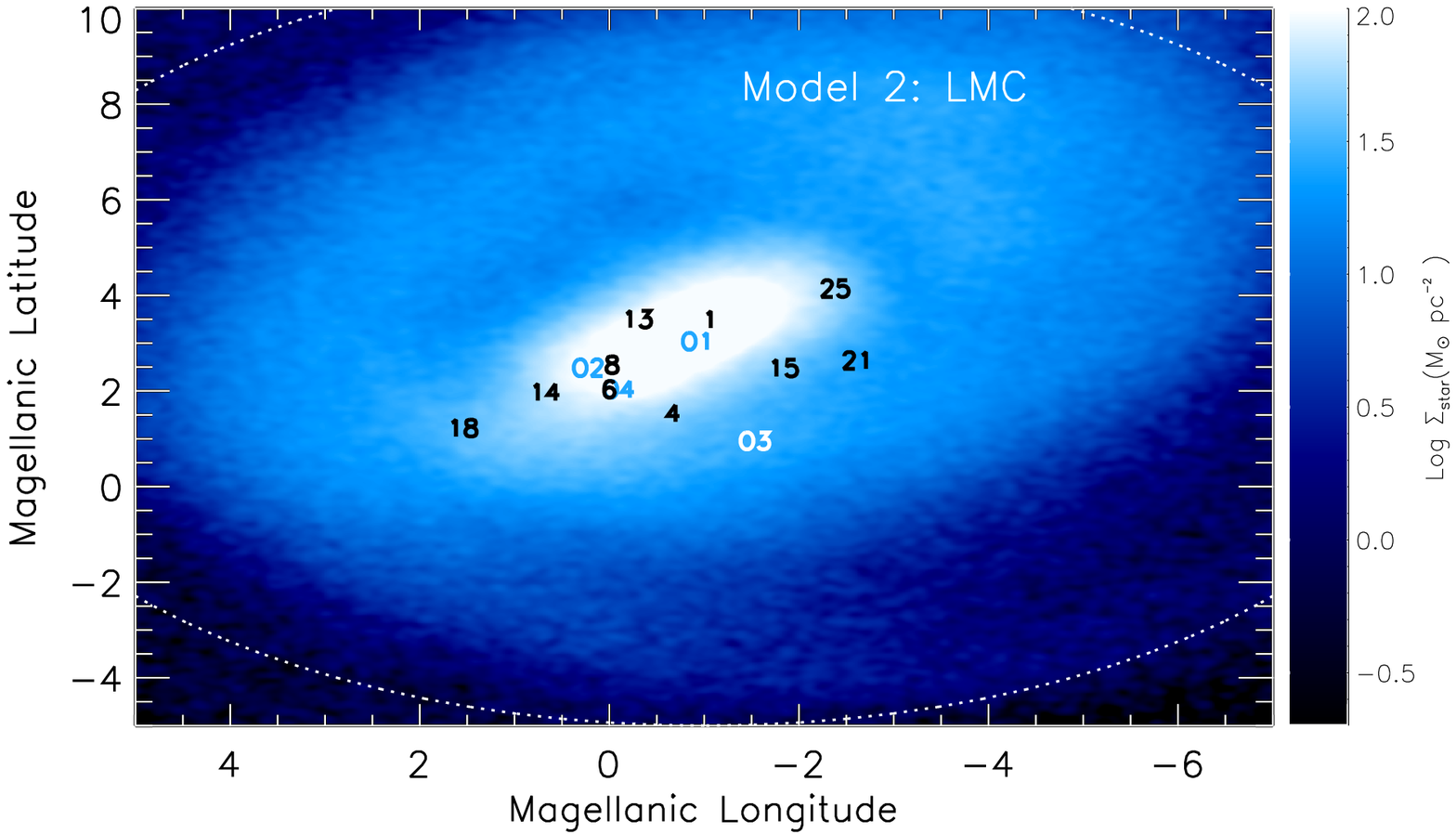}}}\\
\mbox{{\includegraphics[width=4.507in]{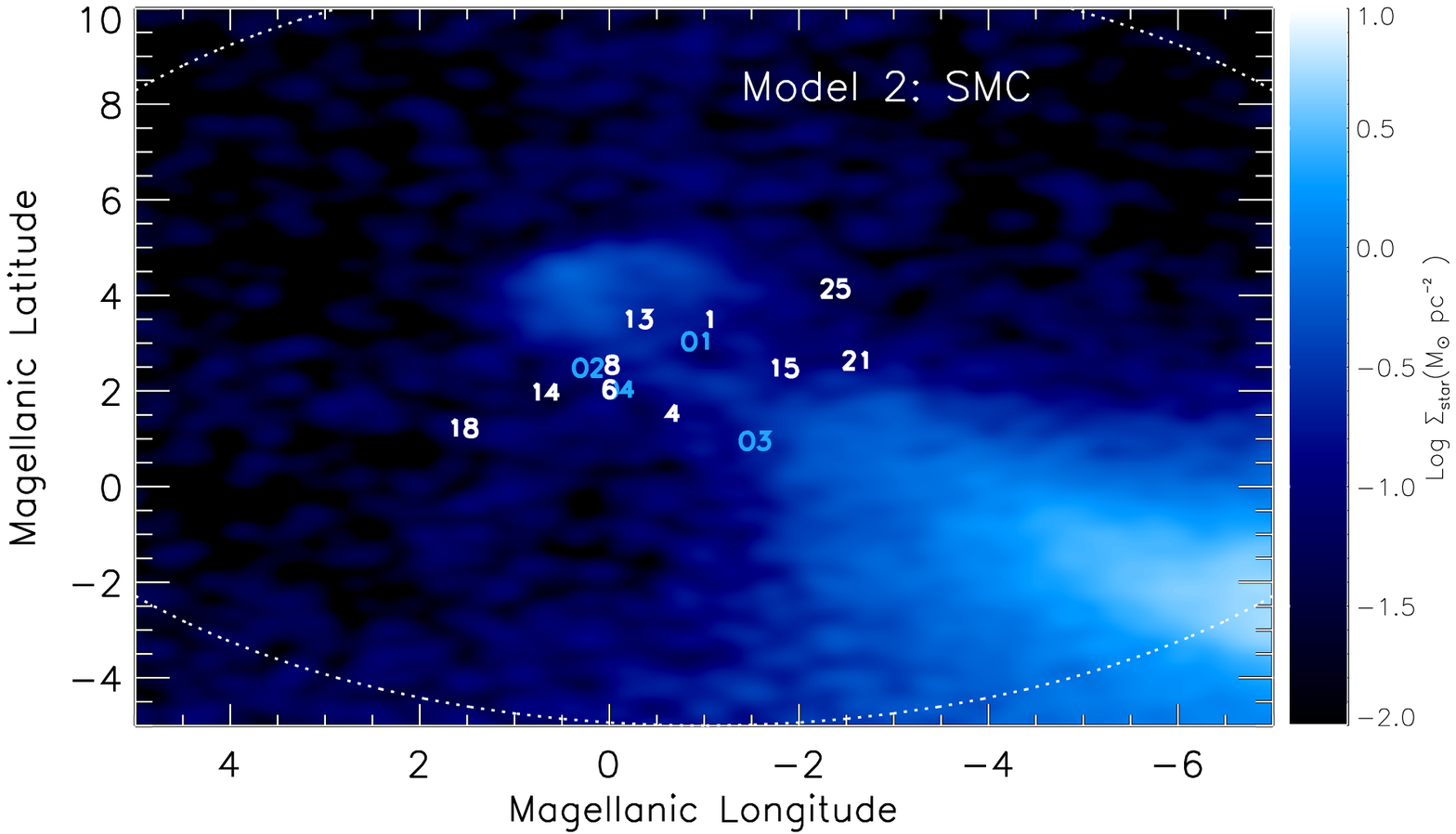}}}
 \end{center}
 \caption{\label{ch8fig:Zoom}  Same as Figure~\ref{ch8fig:ZoomModel1}, except for Model 2.  
 There are significantly more sources behind the LMC in Model 2 
 than in Model 1.  }
 \end{figure*}

In Figures~\ref{ch8fig:ZoomModel1} and~\ref{ch8fig:Zoom} we zoom in on a 10 degree ($\sim$ 10 kpc) 
field centered on the LMC for Models 1 and
2, respectively.  In the top panels we plot the stellar density of LMC particles
in that field. These are the lenses for this analysis. Overplotted are the location of the MACHO and OGLE
 microlensing candidates, labelled by their 
event number. 
 In the middle panel we show the same field but plot only the SMC stellar particles; this is the surface density of sources. 
  Again, the location of candidate microlensing events  are marked and the white dotted ellipse indicates 
  the extent of the LMC disk (this is not a circle in these images since the $x$ and $y$ axes are scaled 
  differently than in the previous figures).   There are significantly more sources in Model 2 than in Model 1: we correspondingly 
  expect a higher microlensing event frequency in Model 2. 
  
\begin{figure*}
\begin{center}
   \mbox{ {\includegraphics[width=4.1in, clip=true, trim=0 2.5in 0 2.5in]{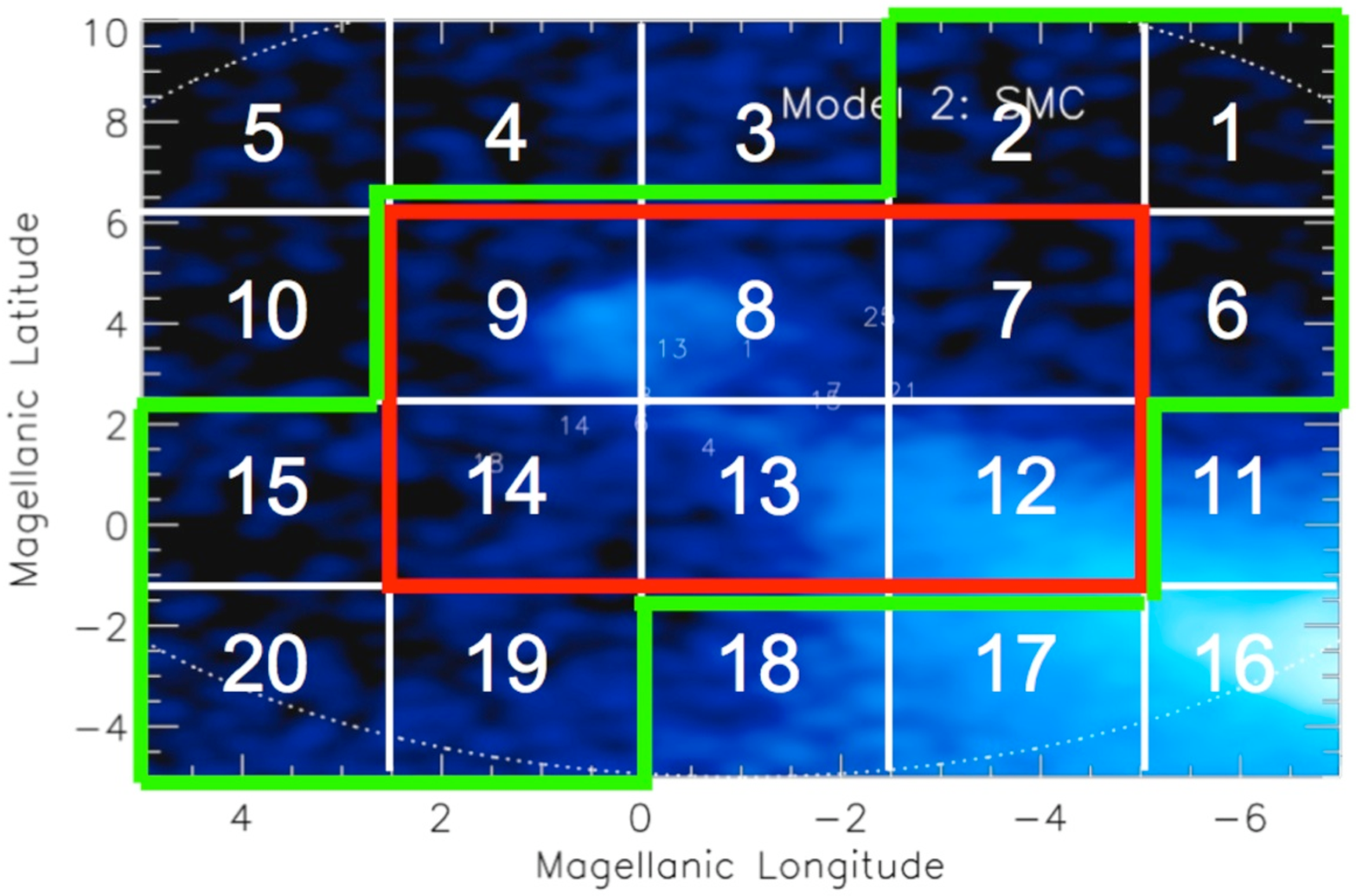}} \hspace{0.4in}}
 \end{center}
 \caption{\label{ch8fig:Grid}  The surface density 
 of sources for Model 2 is illustrated with a grid overplotted. 
 Quantities will be computed within grid cells; each grid cell is assigned a number as indicated and 
 will be referred to throughout the text. The red box indicates the cells (field of view) relevant for
 comparison with the MACHO survey. The green box highlights cells relevant for comparison with the  
 larger OGLE survey. }
 \end{figure*}

Since the density of sources and lenses varies across the field, we have gridded the field and will compute quantities, such as the event frequency 
and duration, in each grid cell.  The gridding and cell numbers are illustrated in the Figure ~\ref{ch8fig:Grid} (plotted over the surface density 
of sources for Model 2) and will be referred to as marked throughout this text.  
Each grid cell is 2.5 $\times$ 3.75 degrees in size (at the distance of the LMC, 1 degree $\sim$ 1 kpc), representing the field of view over 
which we are computing the relevant microlensing quantities listed in $\S$~\ref{sec:Definitions}.  
The grid cells for comparison with the MACHO survey are indicated by the red box.  The highlighted green region indicates 
cells used for comparison with the larger OGLE survey.

\section{The Predicted Microlensing Events in the B12 LMC-SMC Tidal Model} 
\label{sec:Results}

Next we compute the relevant quantities to determine the expected microlensing event durations and event frequencies for both
Model 1 and 2 of B12, using the properties of the simulated source and lens populations. 

\subsection{Effective Distance}
\label{sec:Dist}

We begin our analysis by first computing the effective distance between the identified lenses and sources in each grid cell, following
equation (\ref{eq:Deff}).  The mean values of $D_S$, $D_L$ and $D_{SL}$ per grid cell are listed in Table~\ref{ch8table1}.  
The weighted average of $D$ across the entire face 
of the LMC disk (weighted by the number of source stars in each grid cell)
 is  $D$ = 9.5 (4.5) kpc for Model 1 (2).  Sources are on average farther away from the lenses in Model 1 than in Model 2. 
Taking these average distances and $\Sigma_{\rm lens} = 100 \Msun/{\rm pc}^2$ as the average surface density in the 
central regions of the LMC, and following equation (\ref{ch8:eqTAU}), yields $\tau \sim 7 \times 10^{-7} $  for Model 1 
and $\tau \sim 3 \times 10^{-7}$ for Model 2. 
 This suggests that the lensing probability is higher for Model 1, since on average 
the sources are located at a larger $D$. 
However, the observable of microlensing surveys is the
number of events they detect per year. Obviously this number depends on
the product of the number of sources and the optical depth. Since Model 1
and Model 2 have different numbers of sources, the optical depth alone
does not tell the full story of which Model gives a larger number of
microlensing events. Instead we will compute the event frequency explicitly.

\begin{figure*}
\begin{center}   
\mbox{ {\includegraphics[width=2.5in]{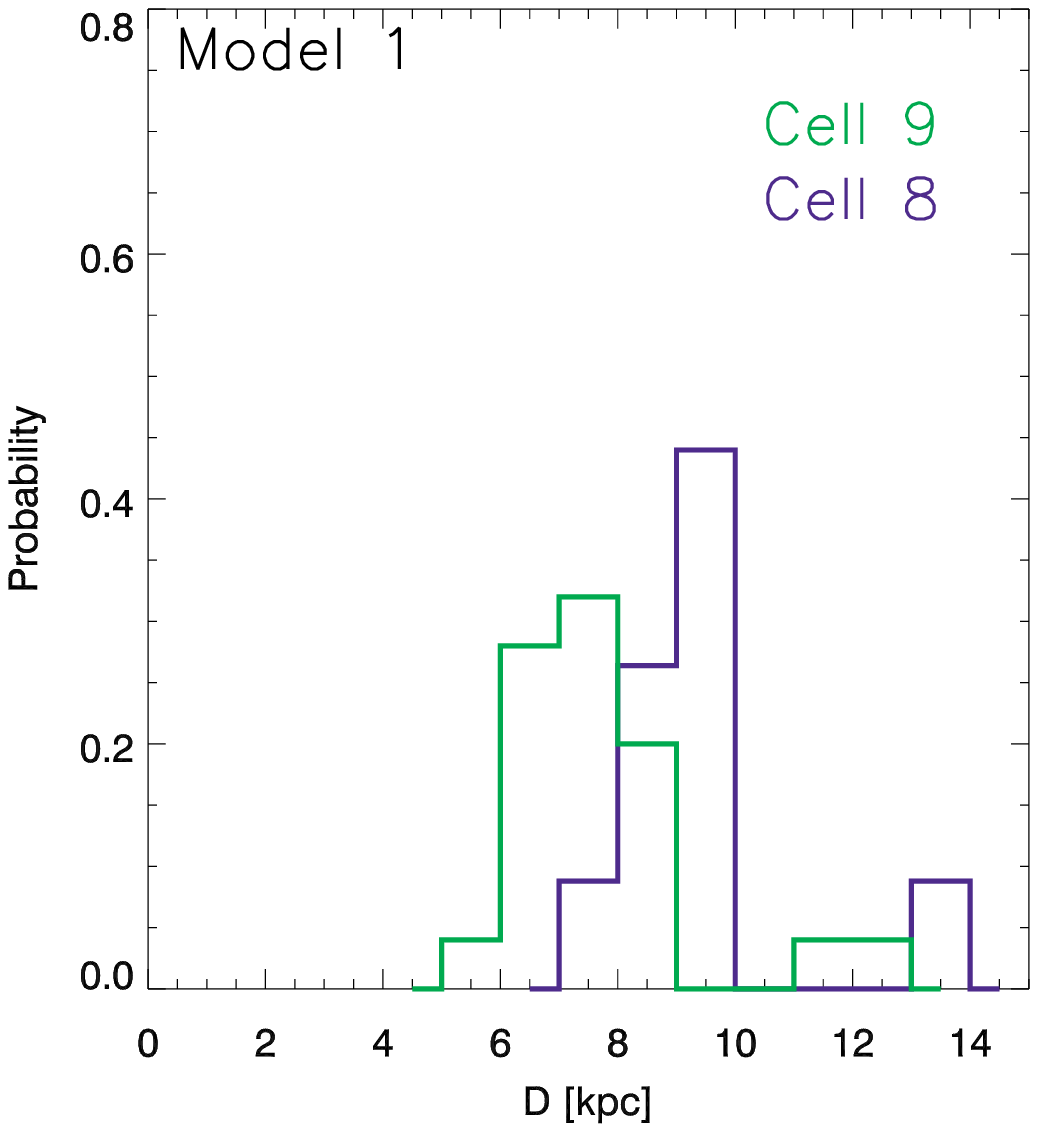}}  
\includegraphics[width=2.5in]{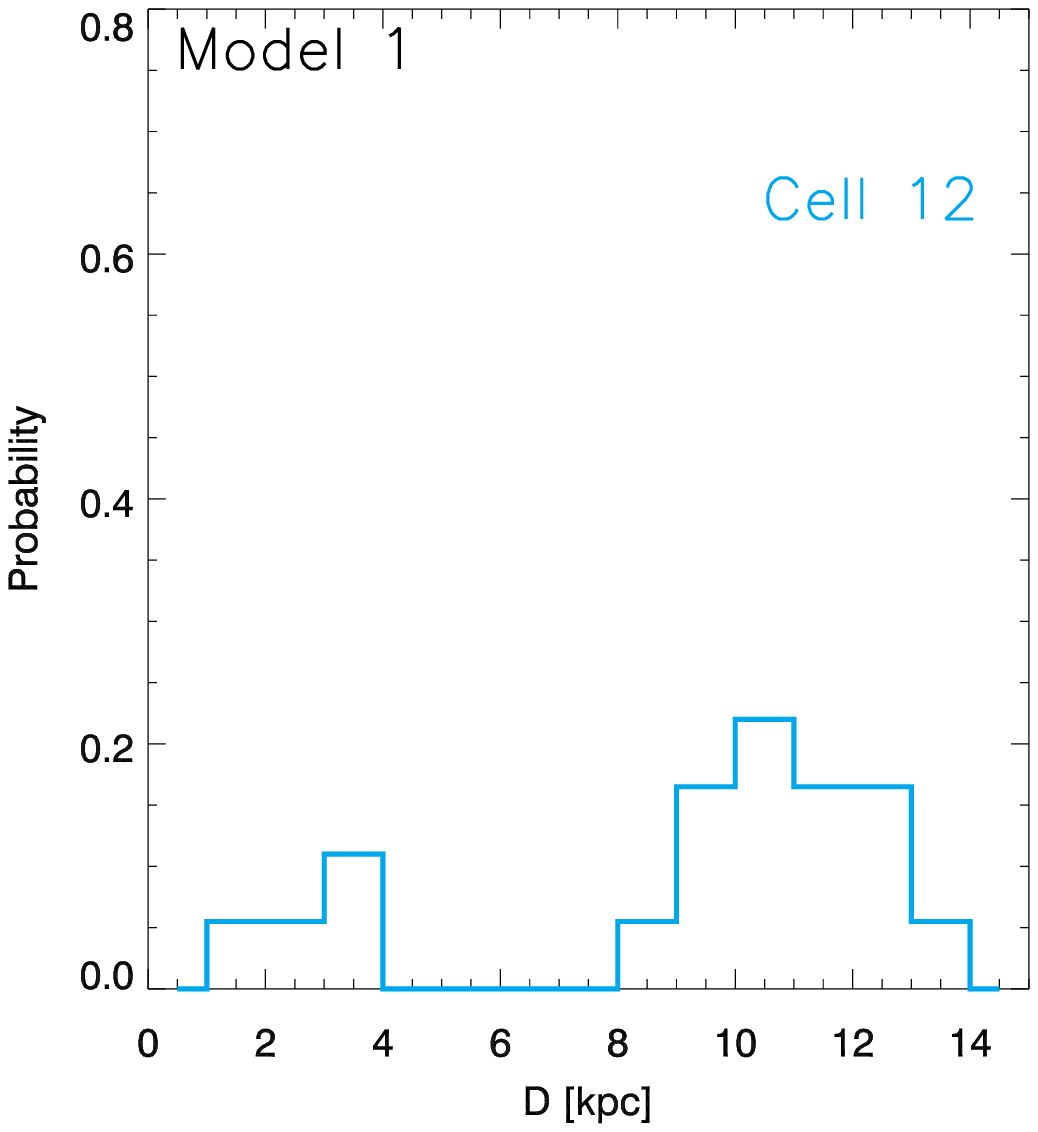} 
 \includegraphics[width=2.5in]{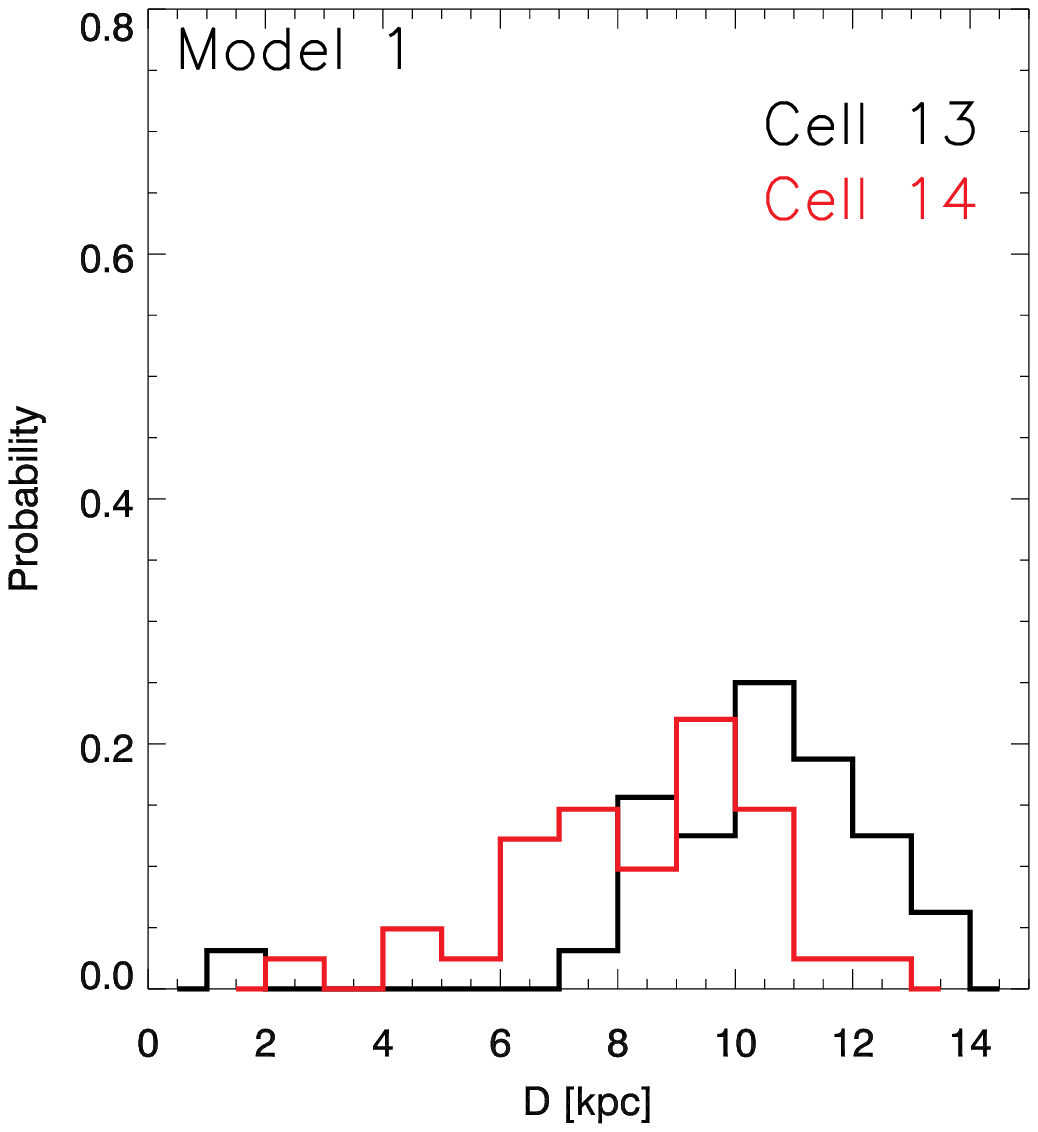}}
 \caption{\label{ch8fig:DGridModel1}    The probability distribution of effective distances between individual stellar particles
stripped from the SMC (sources) and the average lens position (LMC stellar disk particles)
  for individual grid cells, as indicated, for Model 1. Histograms have been normalized by the total number of 
  stripped SMC stellar particles in each grid cell (listed in Table~\ref{ch8table1}); integrating over the histograms
   yields a value of 1. For Model 1, Cell 7 does not have enough sources to compute a distribution. }
  \end{center}
 \end{figure*}

\begin{figure*}
\begin{center}   
\mbox{ {\includegraphics[width=2.5in]{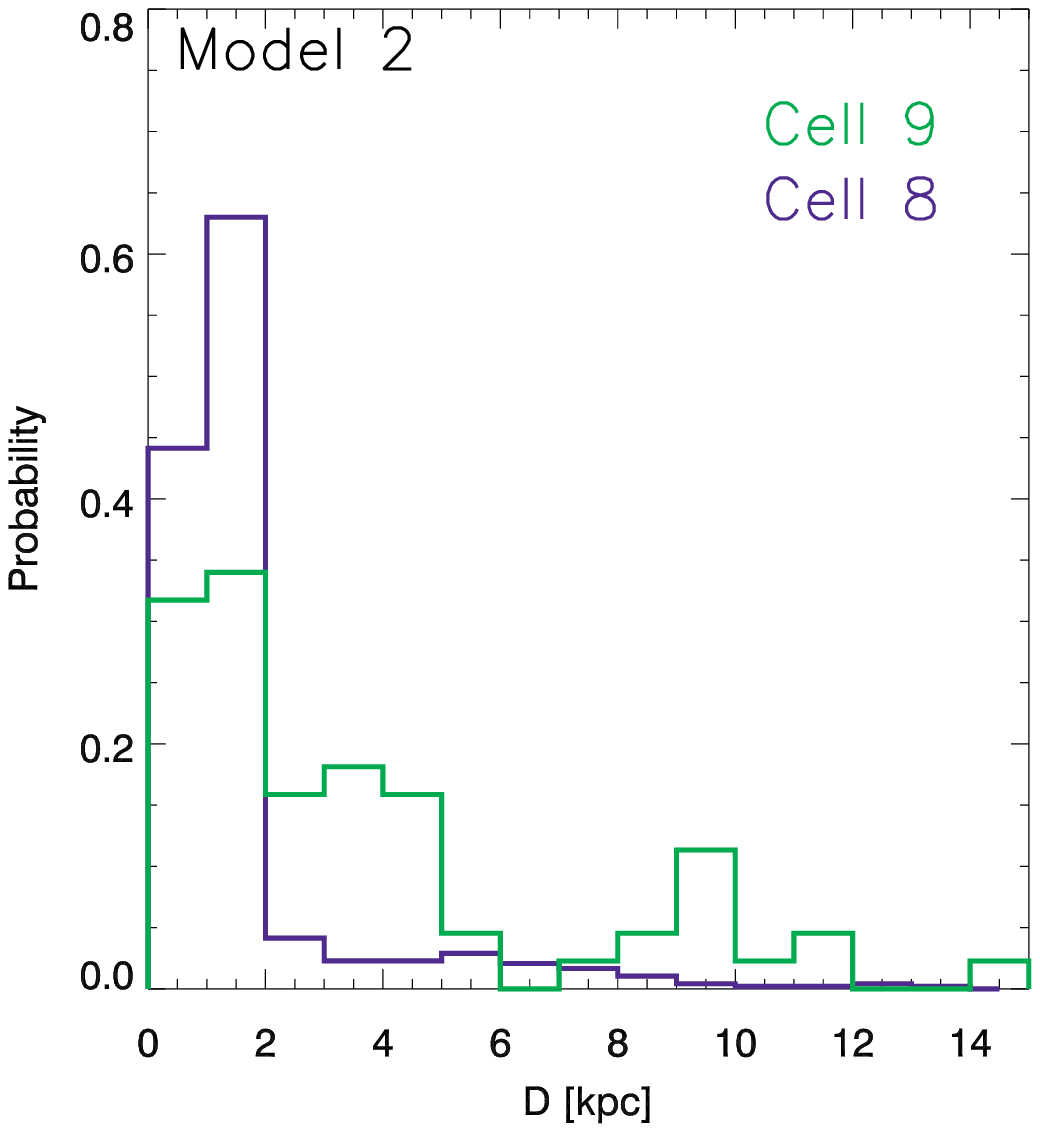}}  
\includegraphics[width=2.5in]{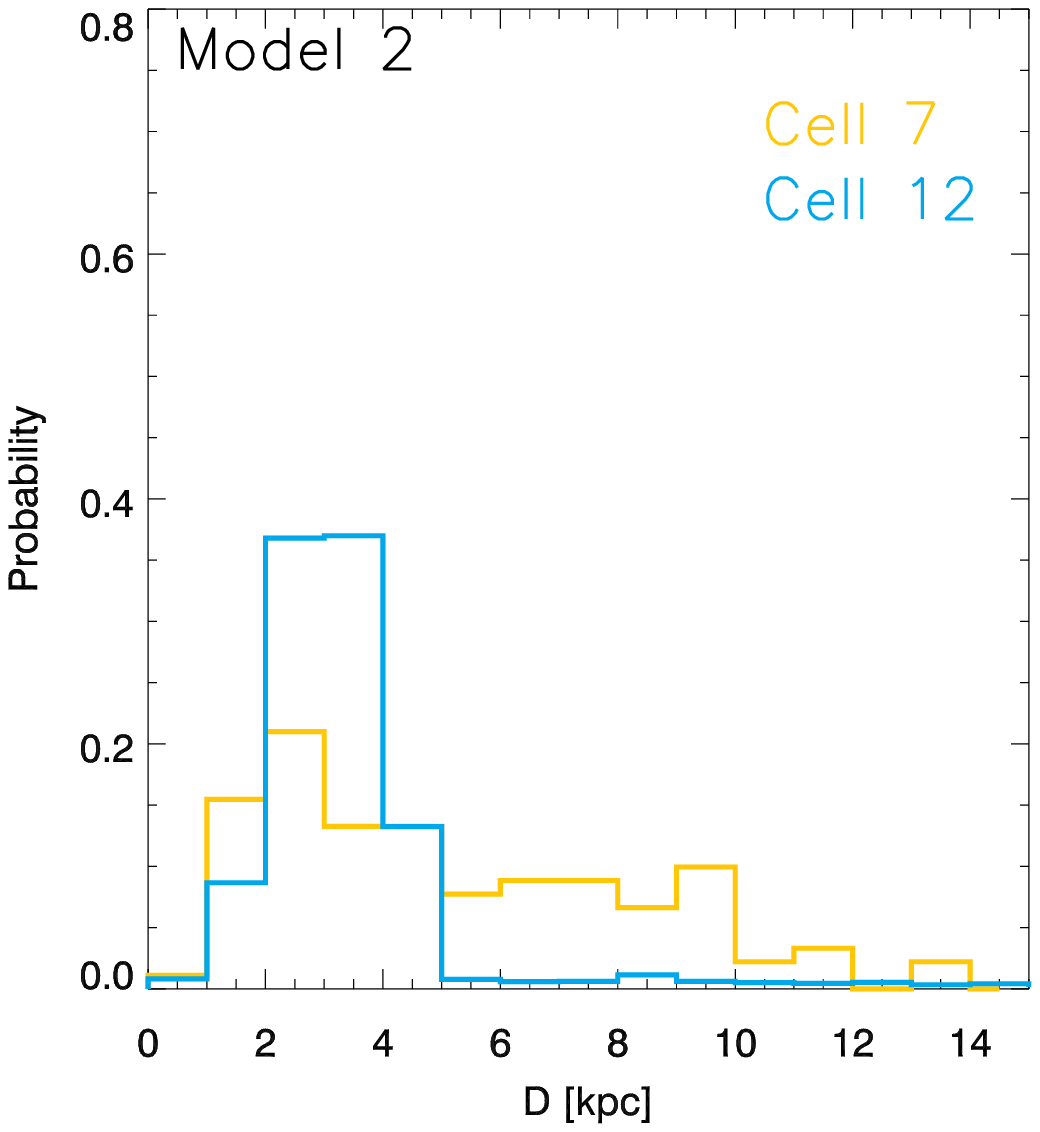}
 \includegraphics[width=2.5in]{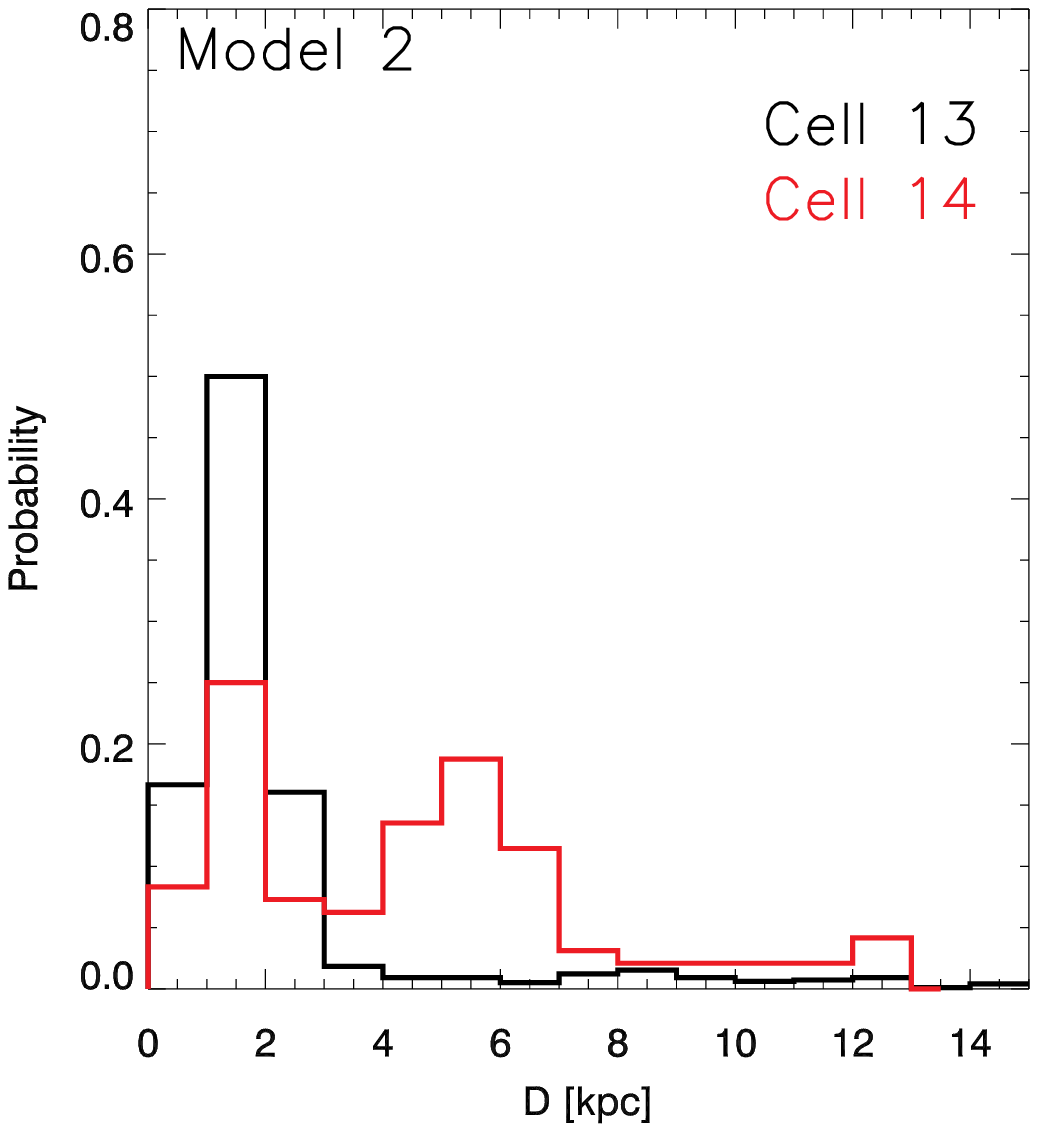}}
 \caption{\label{ch8fig:DGrid}  Same as Figure~\ref{ch8fig:DGridModel1} except for Model 2.}
  \end{center}
 \end{figure*}

The SMC debris is not at a uniform distance from the LMC; consequently, there is a range of values for $D$ per grid cell.  
The main grid cells where microlensing events have been observed are cells
  7, 8, 9, 12, 13 and 14 (as highlighted by the red box in Figure~\ref{ch8fig:Grid}). 
In Figures~\ref{ch8fig:DGridModel1} and~\ref{ch8fig:DGrid} we illustrate the probability distribution of $D$ for SMC 
stellar debris (sources) relative to the mean distance of LMC lenses in each of these main cells.

\subsection{Relative Transverse Velocity of the Sources}
\label{sec:Vperp}

The event duration is dependent on the relative velocity of the source and lens perpendicular
 to our line-of-sight and projected in the plane of the lens ($V_\perp$).   
We compute the relative velocity for all sources using equation (\ref{eq:V}).

\begin{figure*}
\begin{center}   
\mbox{ {\includegraphics[width=2.5in]{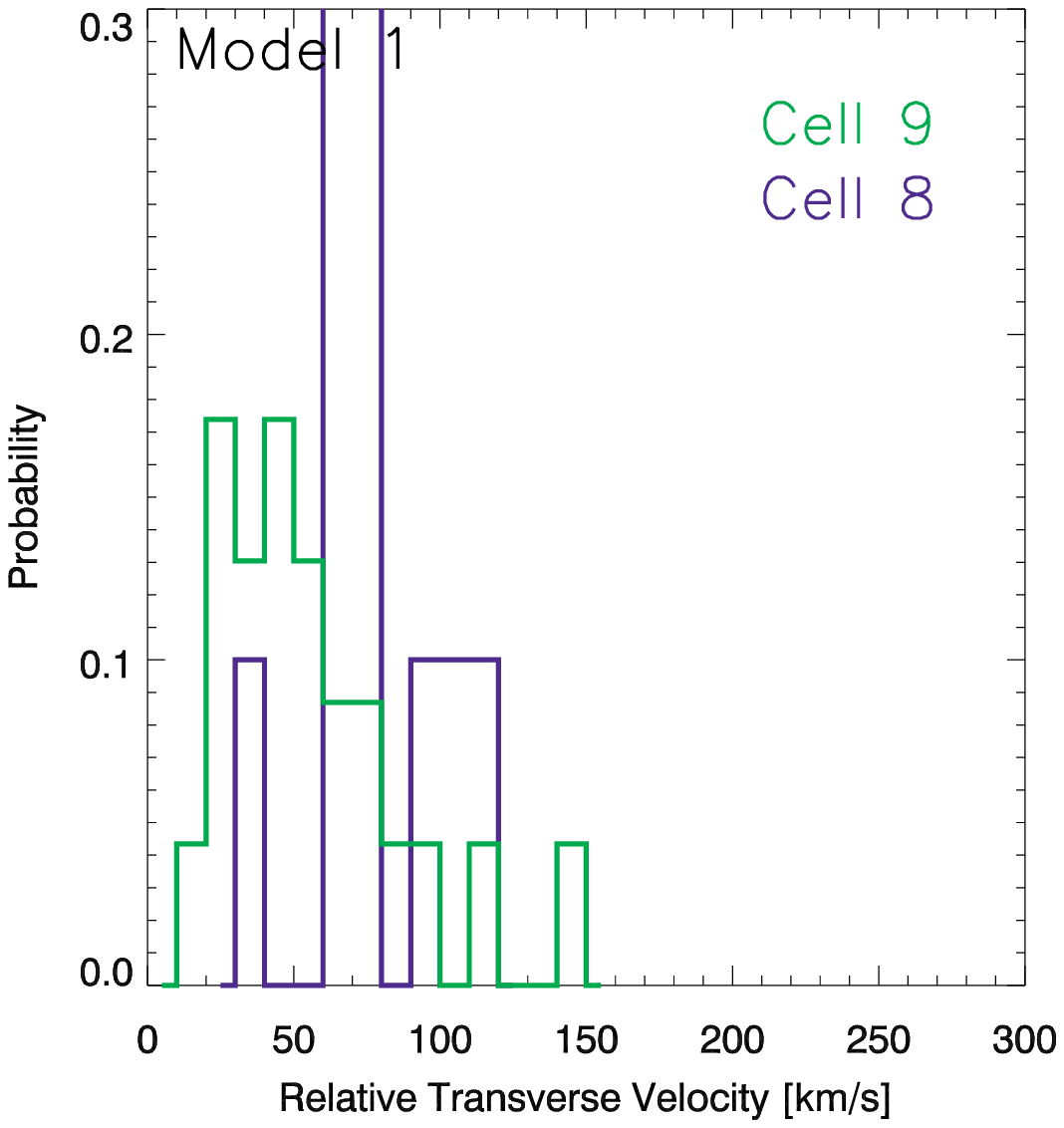}}  
\includegraphics[width=2.5in]{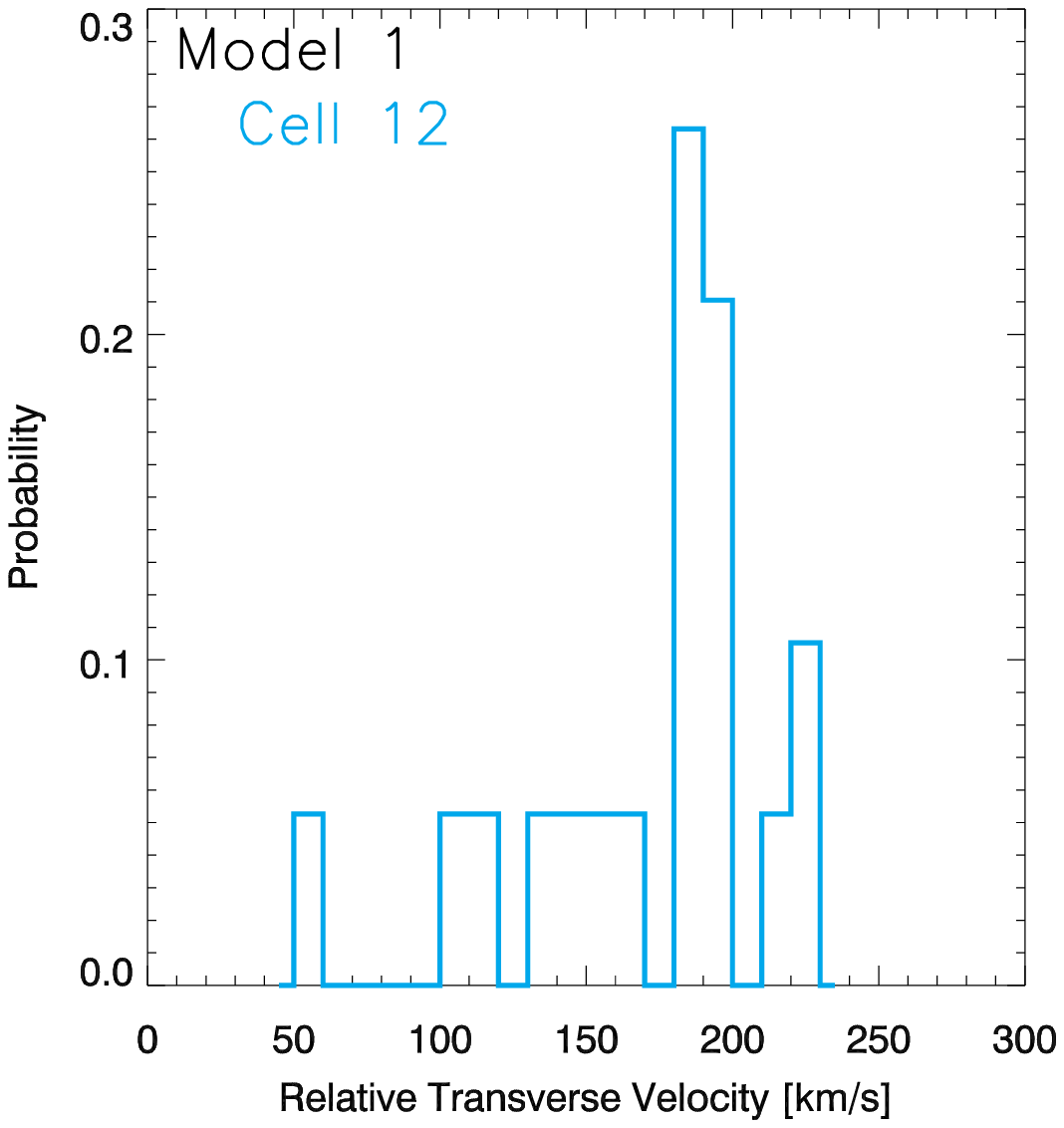}
\includegraphics[width=2.5in]{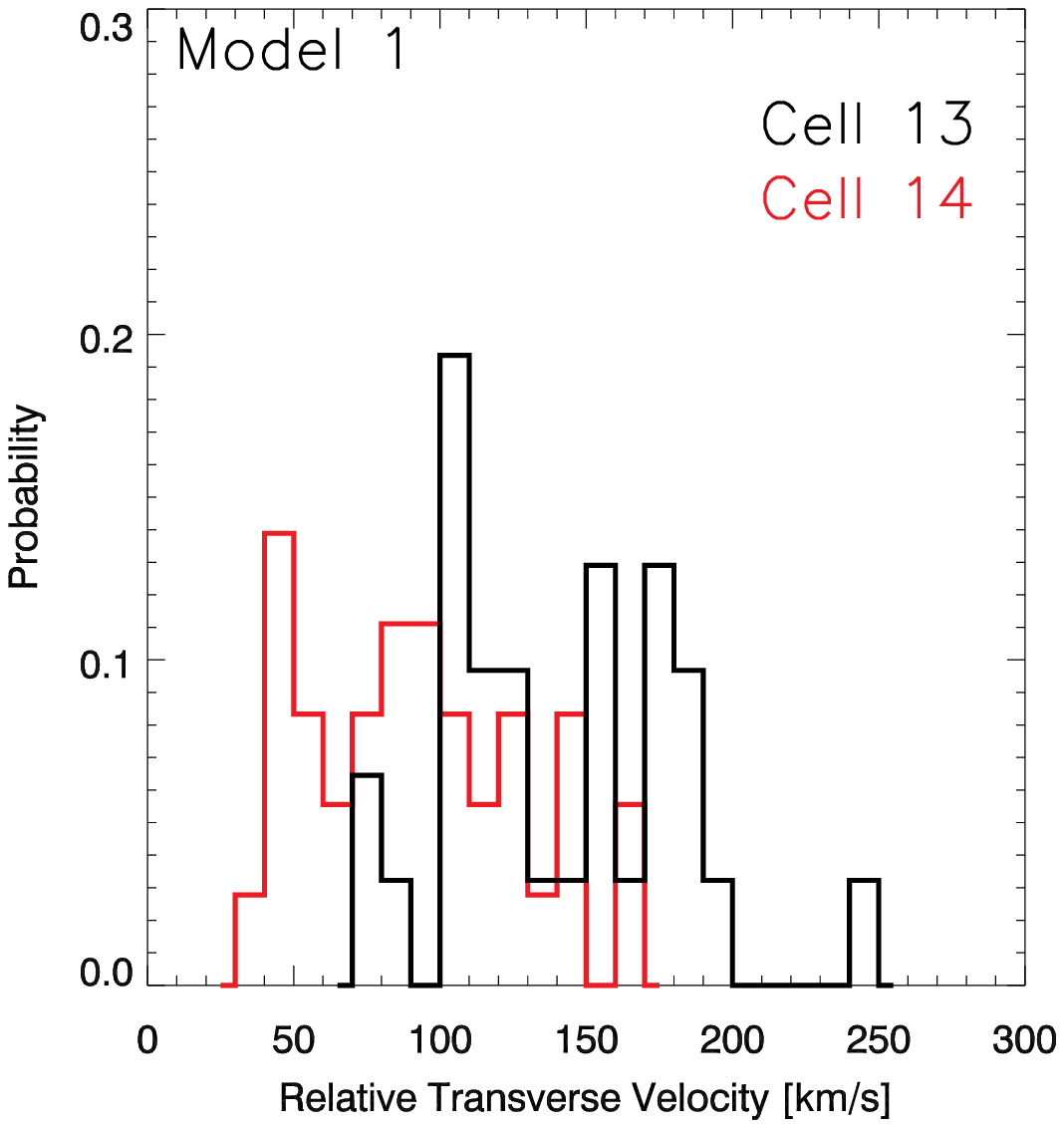}}
 \end{center}
 \caption{\label{ch8fig:VGridModel1}  Probability distribution of the relative transverse velocity ($V_\perp$) between individual 
 stellar particles stripped from the SMC (sources) and the average velocity of LMC disk stars (lenses) 
  for individual grid cells in Model 1. Histograms are normalized as described in Figure ~\ref{ch8fig:DGridModel1}. }
 \end{figure*}

\begin{figure*}
\begin{center}   
\mbox{ {\includegraphics[width=2.5in]{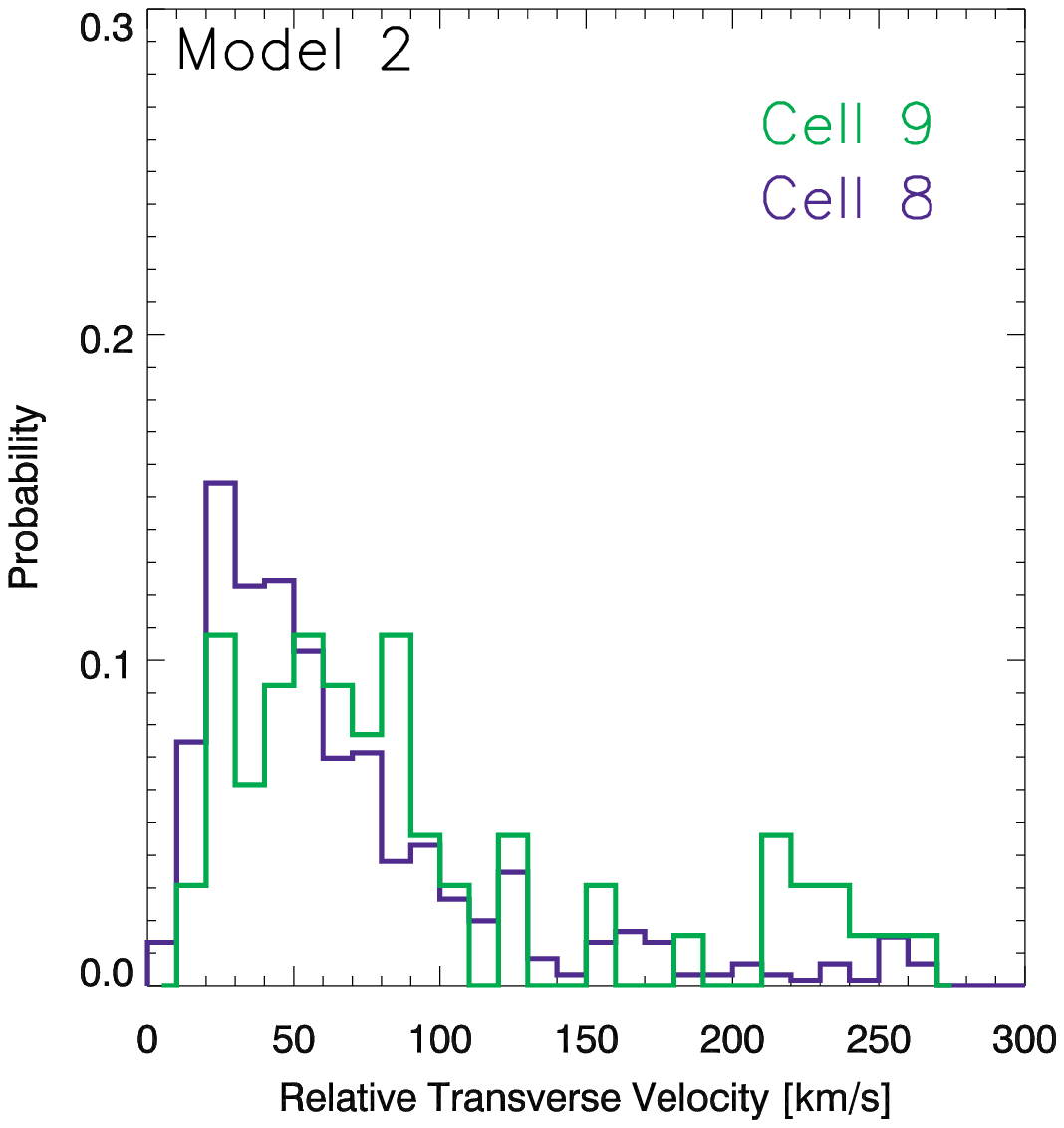}}  
\includegraphics[width=2.5in]{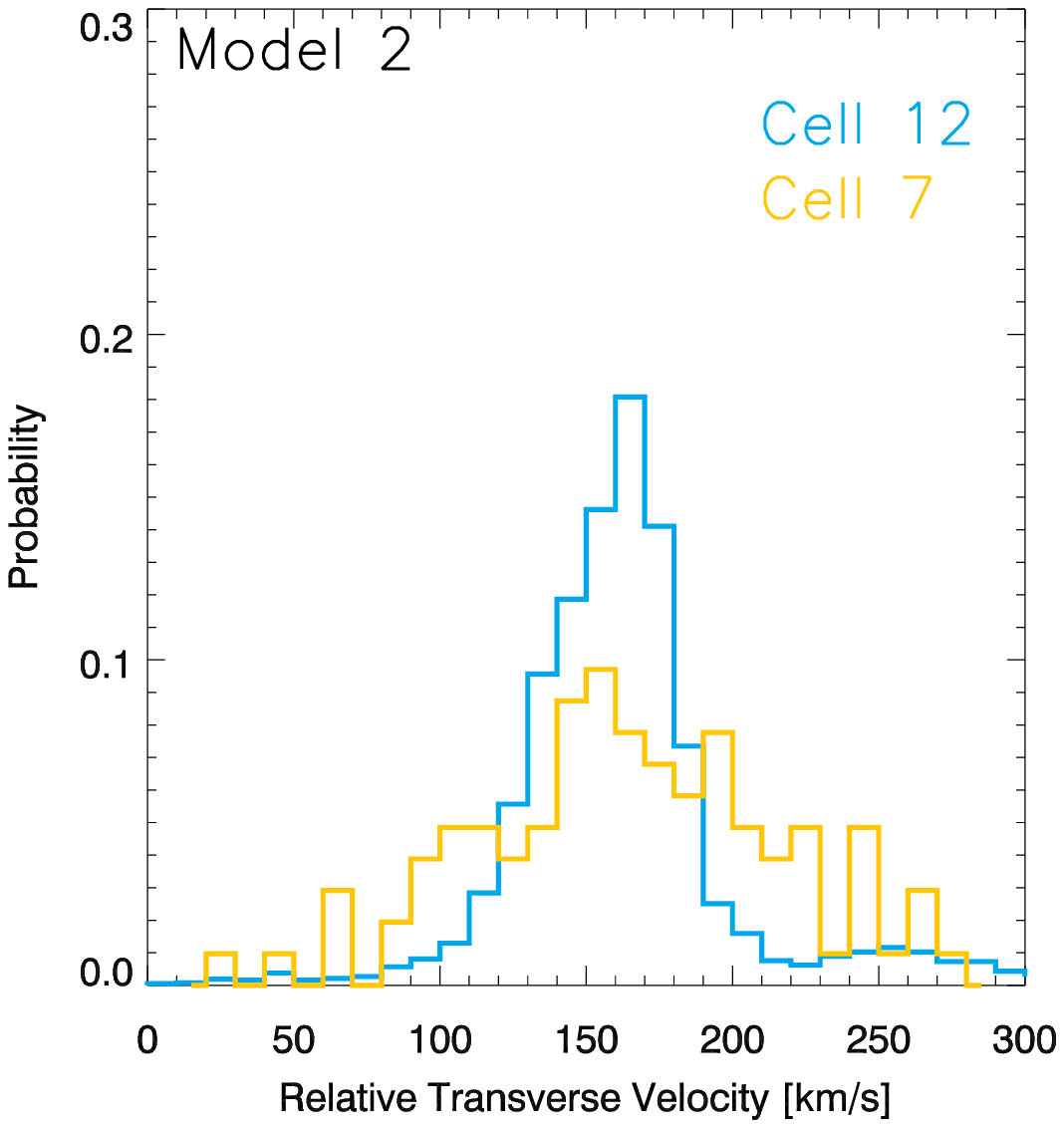}
\includegraphics[width=2.5in]{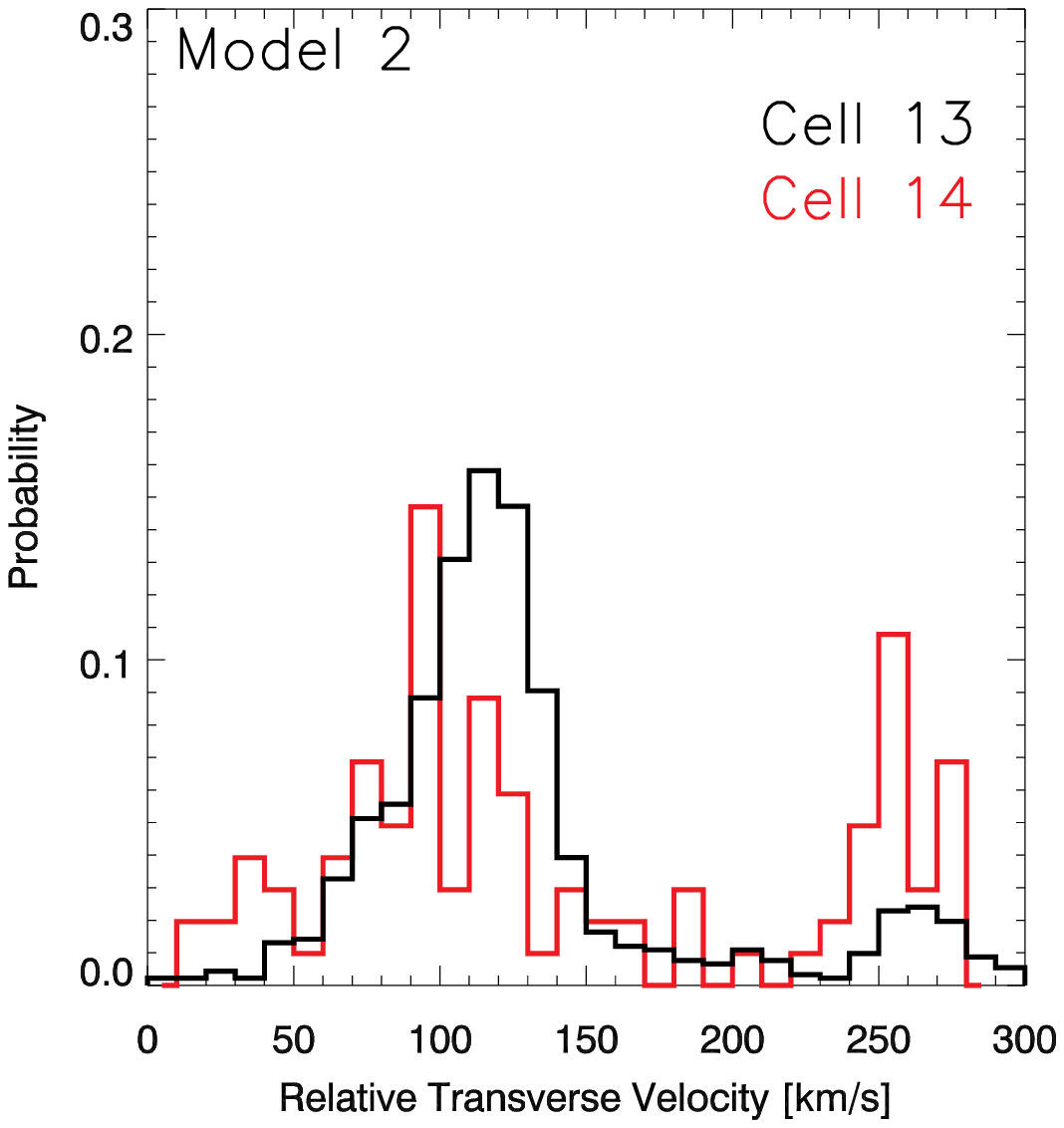}}
 \end{center}
 \caption{\label{ch8fig:VGrid}  Same as in Figure ~\ref{ch8fig:VGridModel1} but for Model 2.}
 \end{figure*}

 There is a large spread in the possible relative velocities within each grid cell: 
 the value for the lenses ($V_L$) is averaged per grid cell, but we 
   keep track of the $V_S$ for each source particle individually in
  order to determine quantify this spread.
   In Figures~\ref{ch8fig:VGridModel1}
  and~\ref{ch8fig:VGrid} we plot the normalized probability distribution of $V_\perp$ for Models 1 and 2 in the central 6
   grid cells where microlensing events have been detected.
A wide range of relative velocities are possible, although velocities in excess of 300 km/s are unlikely. 

We integrate over the probability distribution of $V_\perp$ for each grid cell to determine the average 
$\langle V_{\perp} \rangle$ per grid cell.  These values are listed in Tables~\ref{ch8table1} 
and~\ref{ch8table2} for Models 1 and 2, respectively. 
 The average weighted velocity across all grid cells for Model 1(2) is $\langle V_{\perp} \rangle \sim 120 (150)$ km/s;  the kinematics are similar in both models. 
 Given our massive LMC model (M$_{\rm LMC} = 1.8 \times 10^{11} \Msun$) and their relative spatial proximity to the 
 LMC's center of mass ($D < 10$ kpc), such relative velocities imply that these debris stars are currently bound to the LMC
 \footnote{The escape speed at 20 (10) kpc from the LMC in the B12 model is 220 (276) km/s}. 
      
 Fairly high relative velocities ($60-200$ km/s) are expected in both models, but on average the relative 
 velocities in Model 2 are higher than in Model 1 (150 km/s vs 120 km/s).
 In Model 1, source stars represent tidally stripped stars that are in orbit 
 about the LMC, forming a coherent, thin arc of stars behind the LMC. 
  In Model 2, most of the source stars were removed during a violent collision between the two galaxies.
  This low impact parameter, high speed encounter allowed the LMC to strip stars from deep within the SMC's 
  potential, naturally explaining the higher relative velocities of the stellar debris in Model 2.
  

\subsection{Event Duration}
\label{sec:Dur}

\begin{figure*}
\begin{center}   
\mbox{ {\includegraphics[width=2.5in]{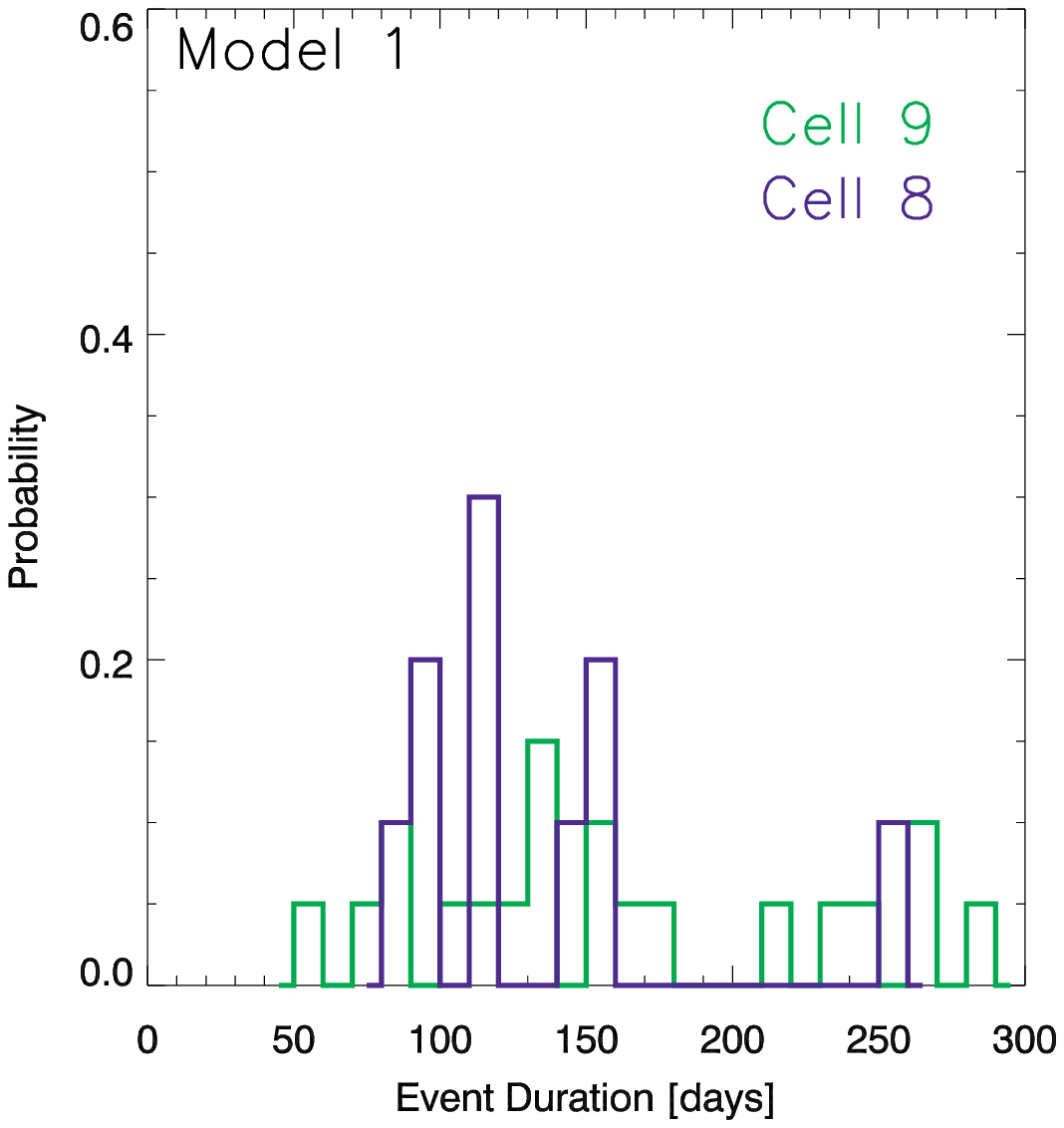}}  
\includegraphics[width=2.5in]{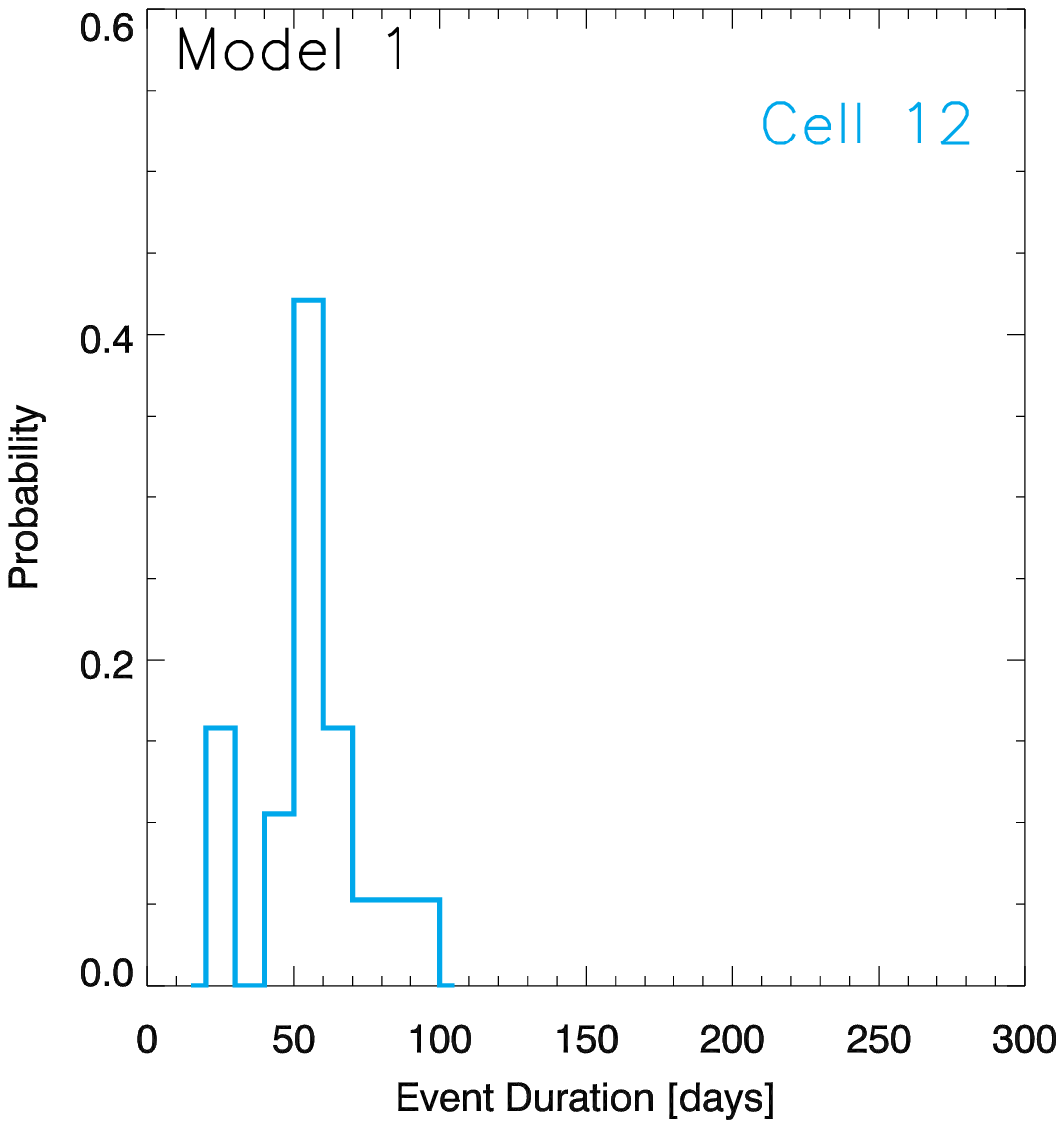}
\includegraphics[width=2.5in]{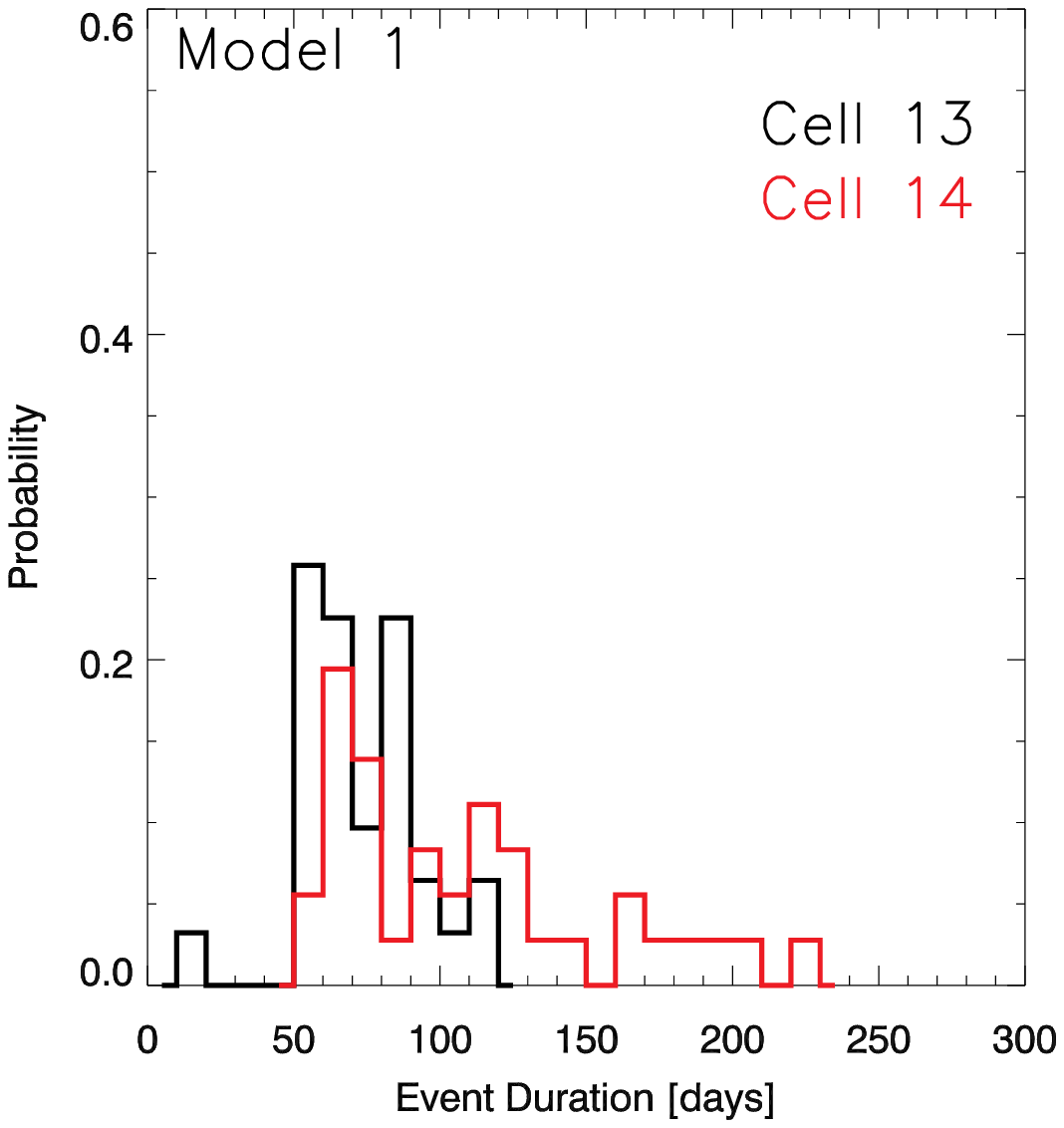}}
 \caption{\label{ch8fig:TGridModel1} Normalized probability distribution of the event durations ($t_e$) for selected grid cells in Model 1.  }
\end{center}
 \end{figure*}

\begin{figure*}
\begin{center}   
\mbox{ {\includegraphics[width=2.5in]{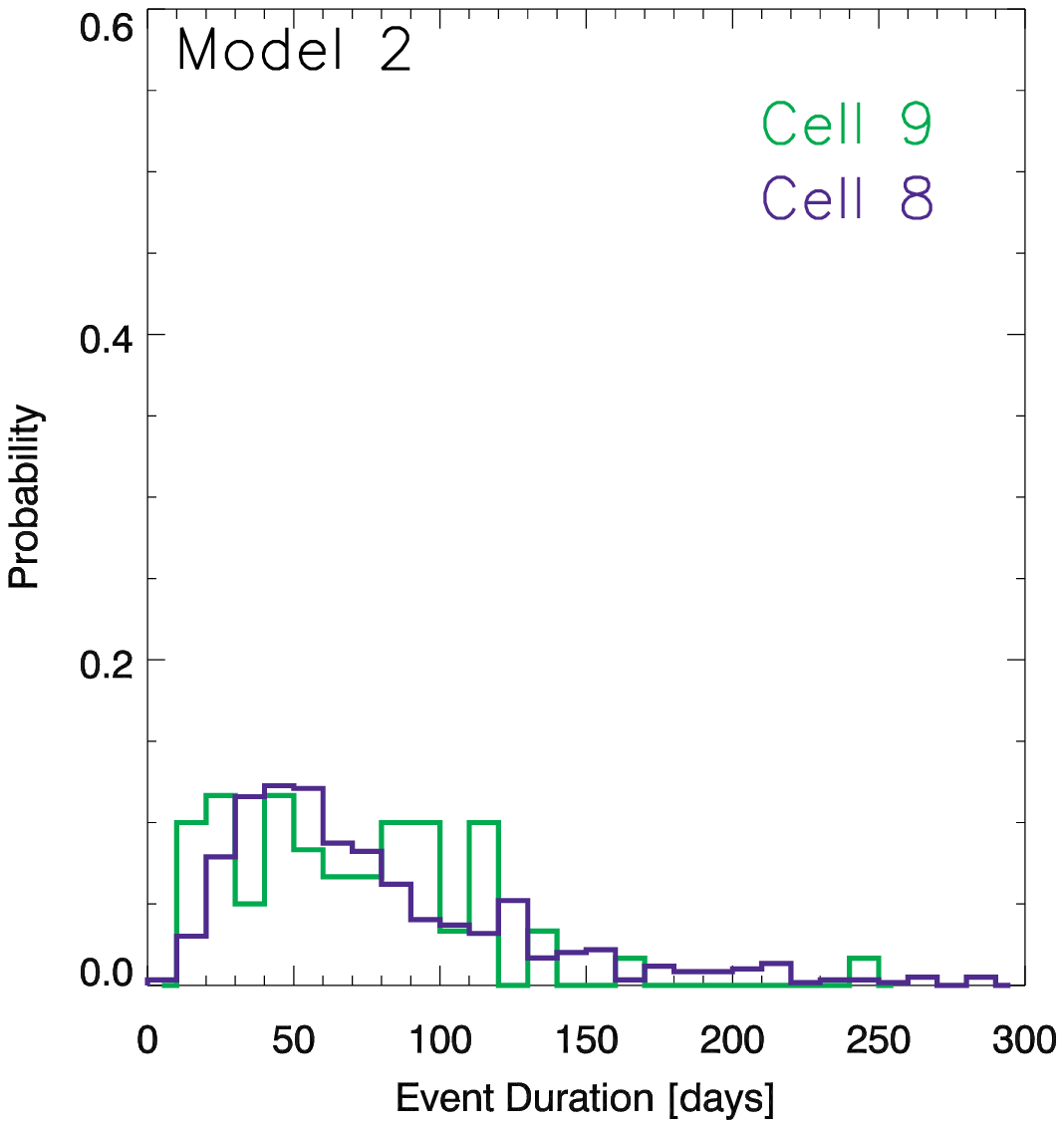}}  
\includegraphics[width=2.5in]{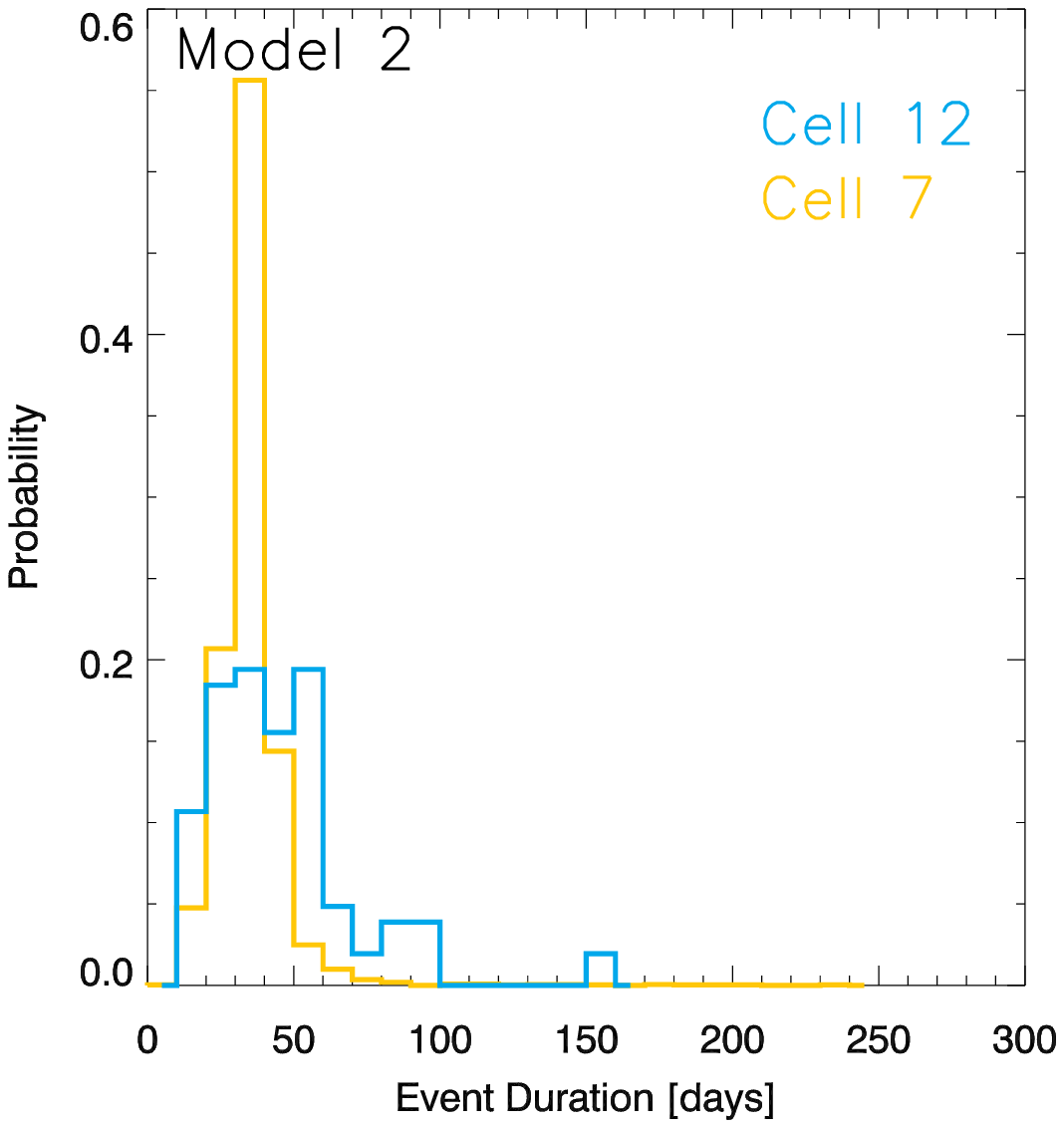}
 \includegraphics[width=2.5in]{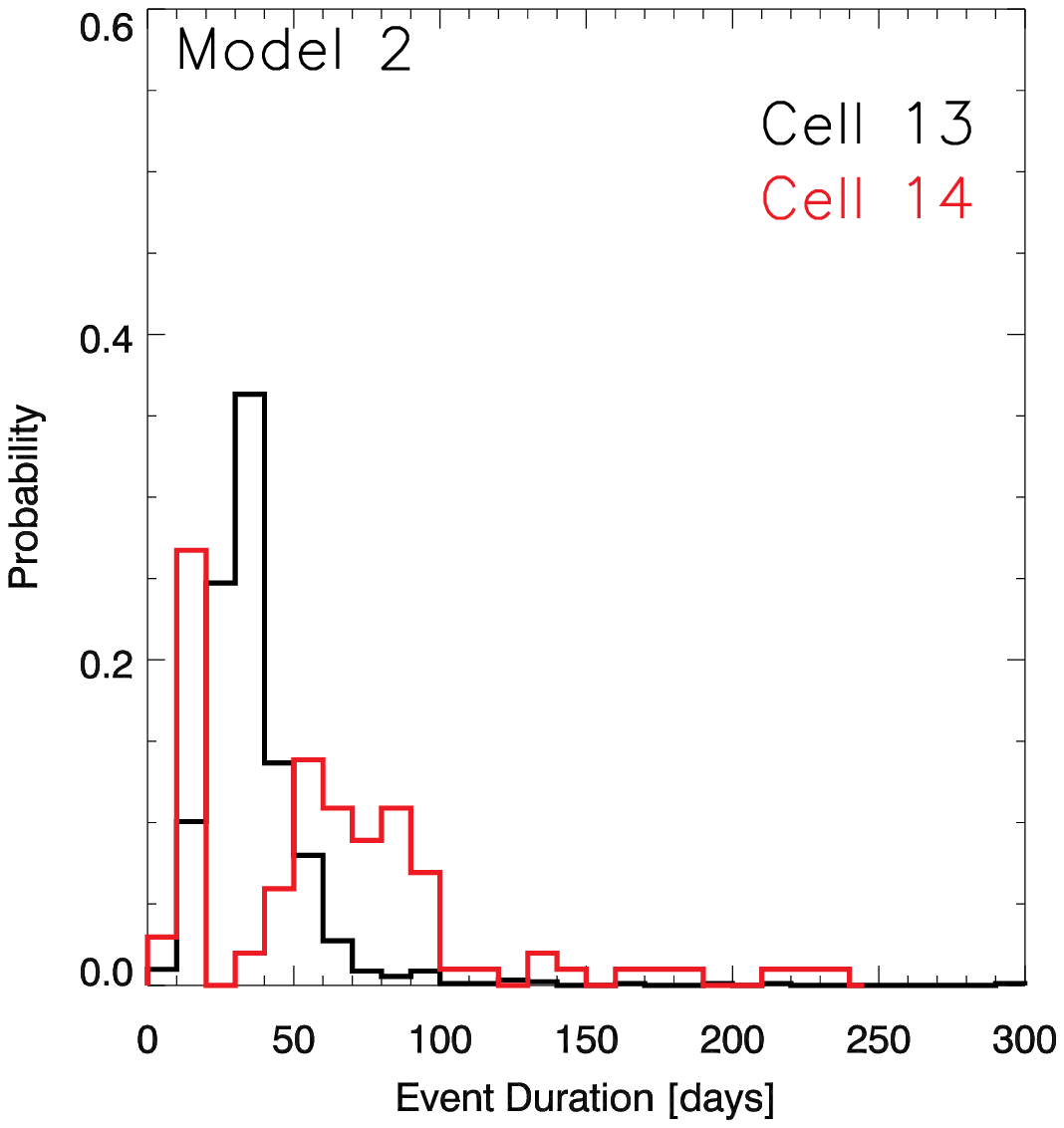}}
 \caption{\label{ch8fig:TGridModel2}  Same as Figure~\ref{ch8fig:TGridModel1}, except for Model 2. }
 \end{center}
 \end{figure*}

Following equation (\ref{eq:TeAvg}), we compute here the average event duration ($\langle t_e \rangle$) per grid cell. 
As discussed in the previous sections, the stripped SMC stellar particles (sources) exhibit a wide range of positions and velocities. 
Consequently, they will also exhibit a range of event durations.  We thus compute a normalized probability distribution for $t_e$, 
based on the probability distribution of the quantity $D^{0.5}/V_\perp$ for the stripped SMC stellar particles in each grid cell. 
We plot the distributions for the central 6 cells of interest in Figure~\ref{ch8fig:TGridModel1} for Model 1, and 
Figure~\ref{ch8fig:TGridModel2} for Model 2. 

Typical event durations range between 20-150 days for Model 1 and 20-100 days for Model 2.
Short duration events ($<$ 15 days) are not expected in these models, consistent with observations.
Both models predict event durations in excess of 70 days, the longest event duration observed; however 
event durations in excess of 100-150 days are rare. 

The observed spread in durations has often been attributed to the lensing mass function. 
Here we have assumed a single average value of $\langle M_{\rm lens}^{0.5} \rangle$
 for the lens and still obtain a large
 variation in the event duration in a given field of view. This is because of the large spread 
 in relative speeds and effective distances of the tidal debris
 that represents the source population.   
The spread in the modeled durations per grid cell also explains why events such as 
MACHO Event 13 and Event 1, which represent the longest and shortest
events, respectively, can be closely located spatially.

Integrating over these probability distributions yields an estimate for the average event duration, $\langle t_e \rangle$ 
(equation~\ref{eq:TeAvg}), per grid cell, as listed in  
Tables ~\ref{ch8table1} and ~\ref{ch8table2} for Models 1 and 2, respectively.  
Overall the event durations predicted by Model 1 are longer than those of 
Model 2.

The weighted average of the simulated event durations over the entire face of the disk yields:  80 days for Model 1 and
 42 days for Model 2.  The shorter average event durations of Model 2 are more consistent with both the OGLE and MACHO 
 results and are a direct consequence of the higher relative speeds determined in Model 2 ($\S$~\ref{sec:Vperp}).

\subsection{Event Frequency}
\label{sec:Freq}

\begin{figure*}
\begin{center}   
\mbox{ {\includegraphics[width=2.5in]{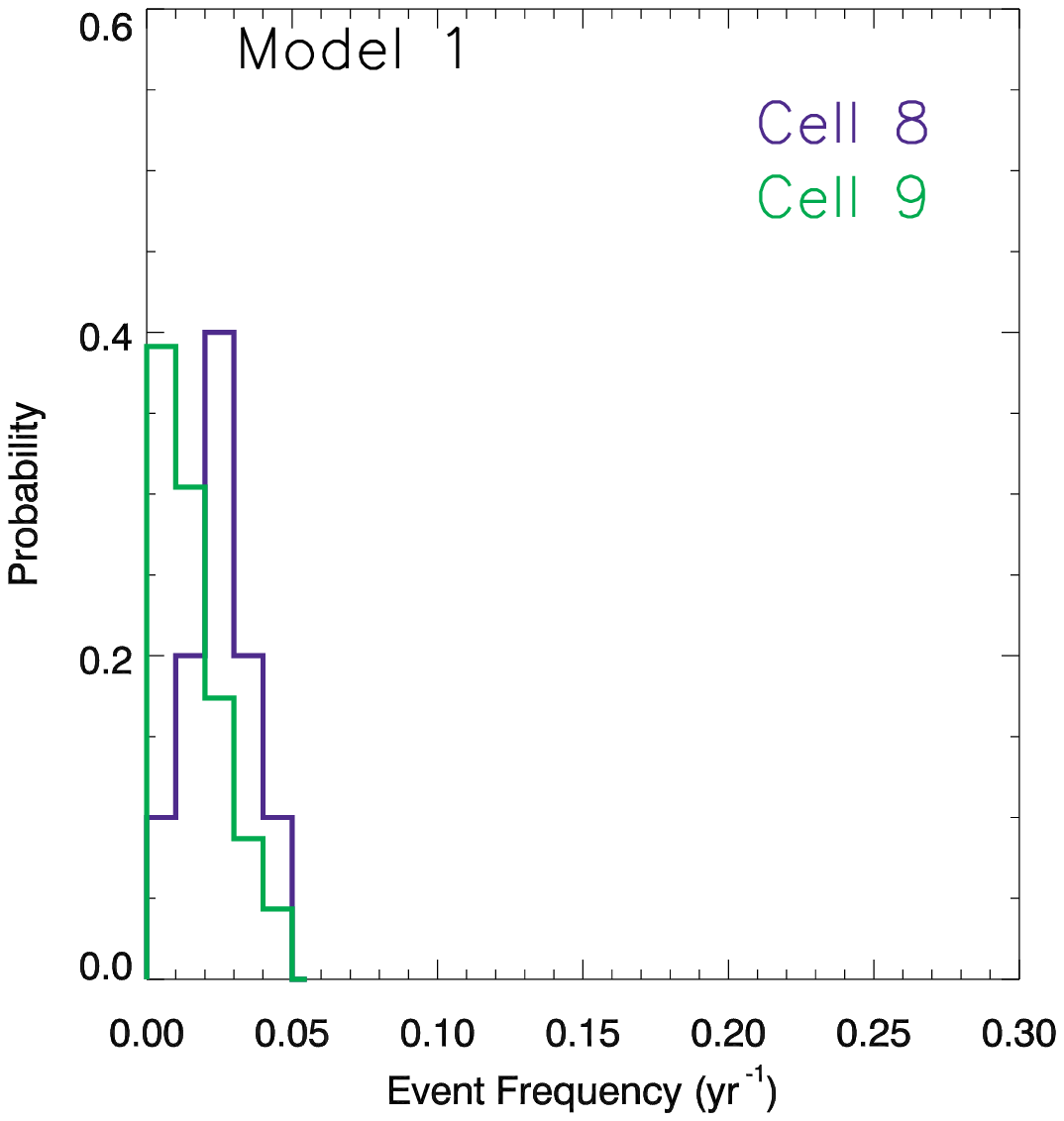}}  
\includegraphics[width=2.5in]{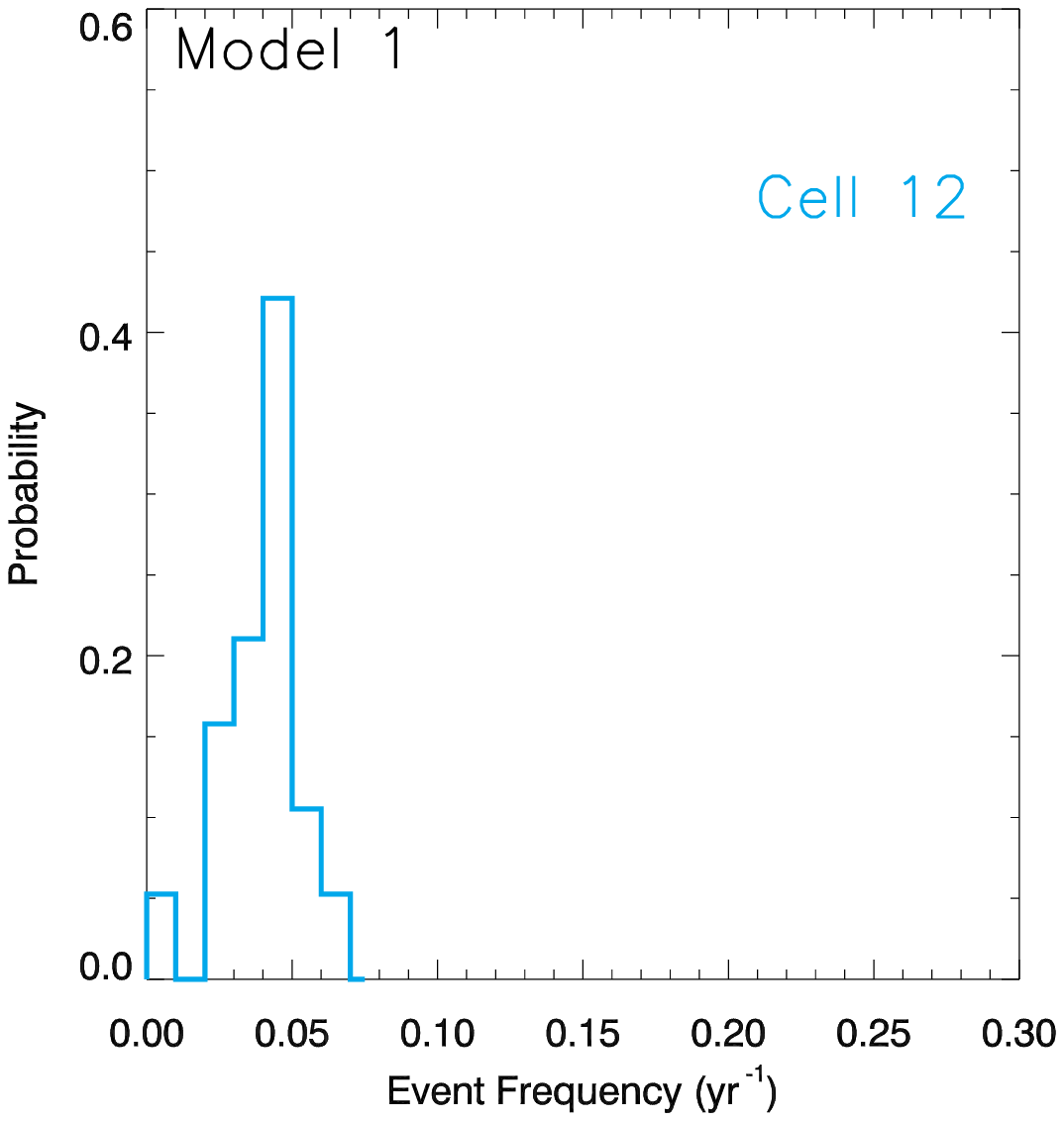}
\includegraphics[width=2.5in]{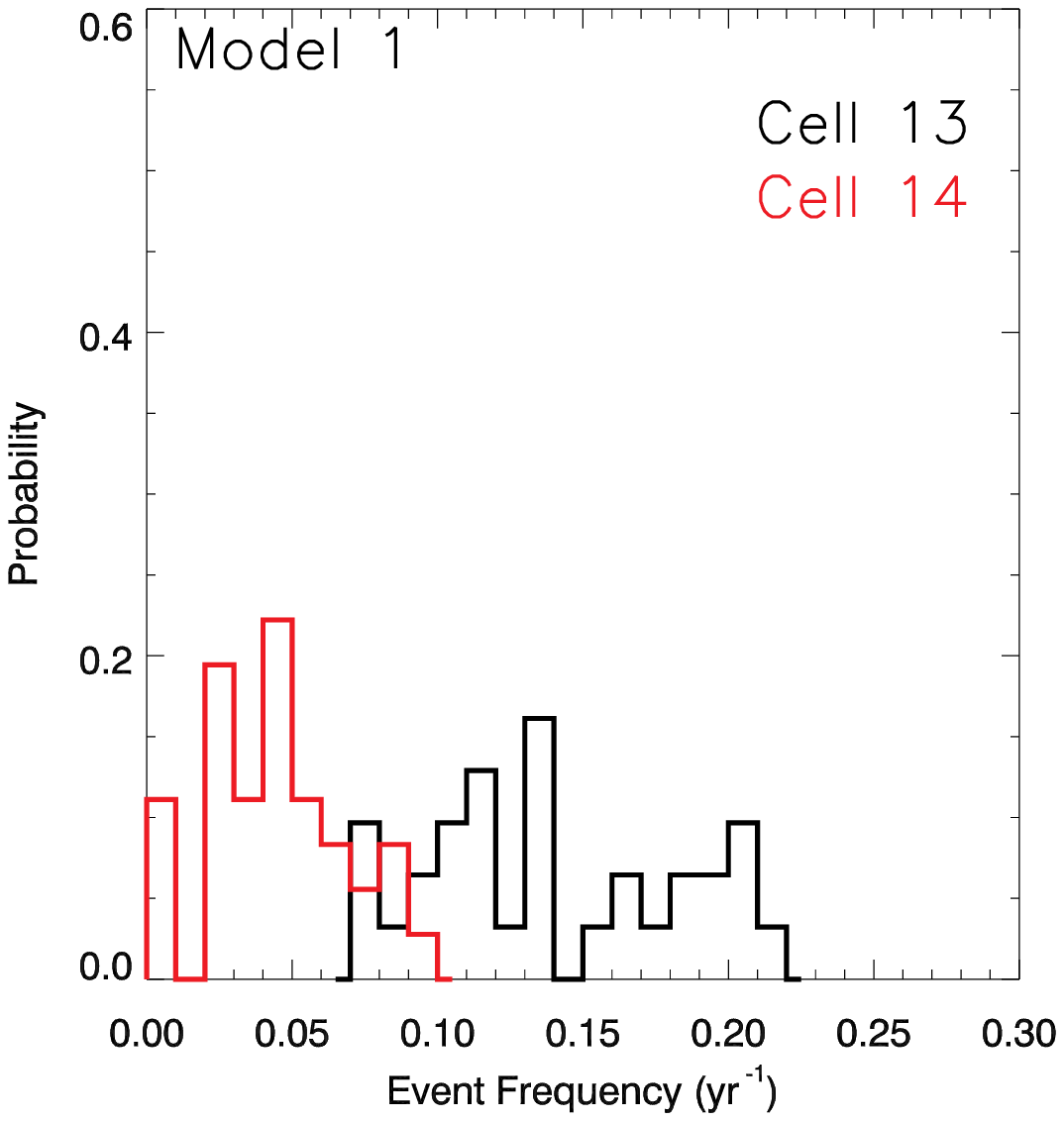}}
 \caption{\label{ch8fig:FGridModel1} Normalized probability distributions of the event frequencies for selected grid cells in Model 1.  }
\end{center}
 \end{figure*}

\begin{figure*}
\begin{center}   
\mbox{ {\includegraphics[width=2.5in]{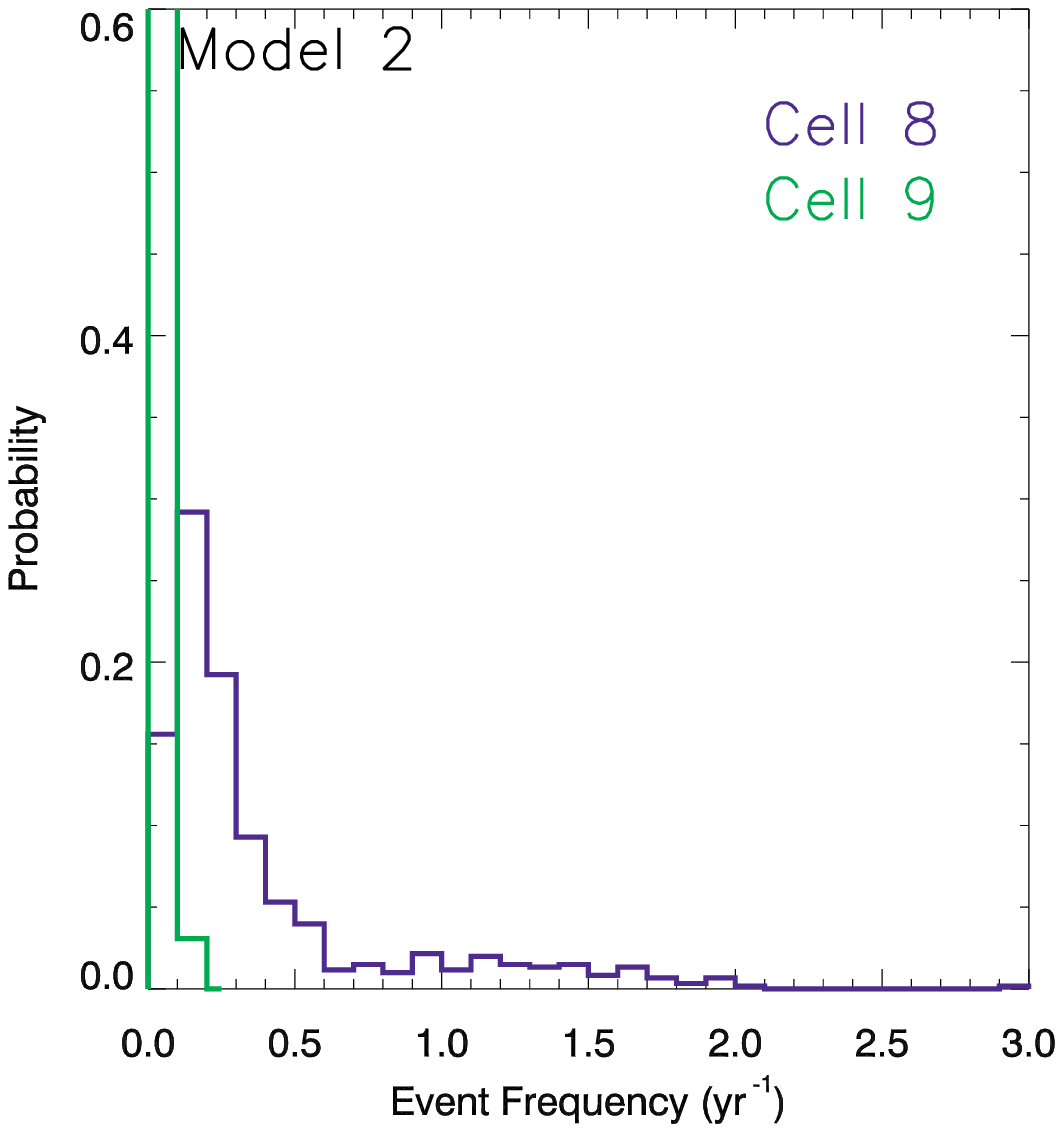}}  
\includegraphics[width=2.5in]{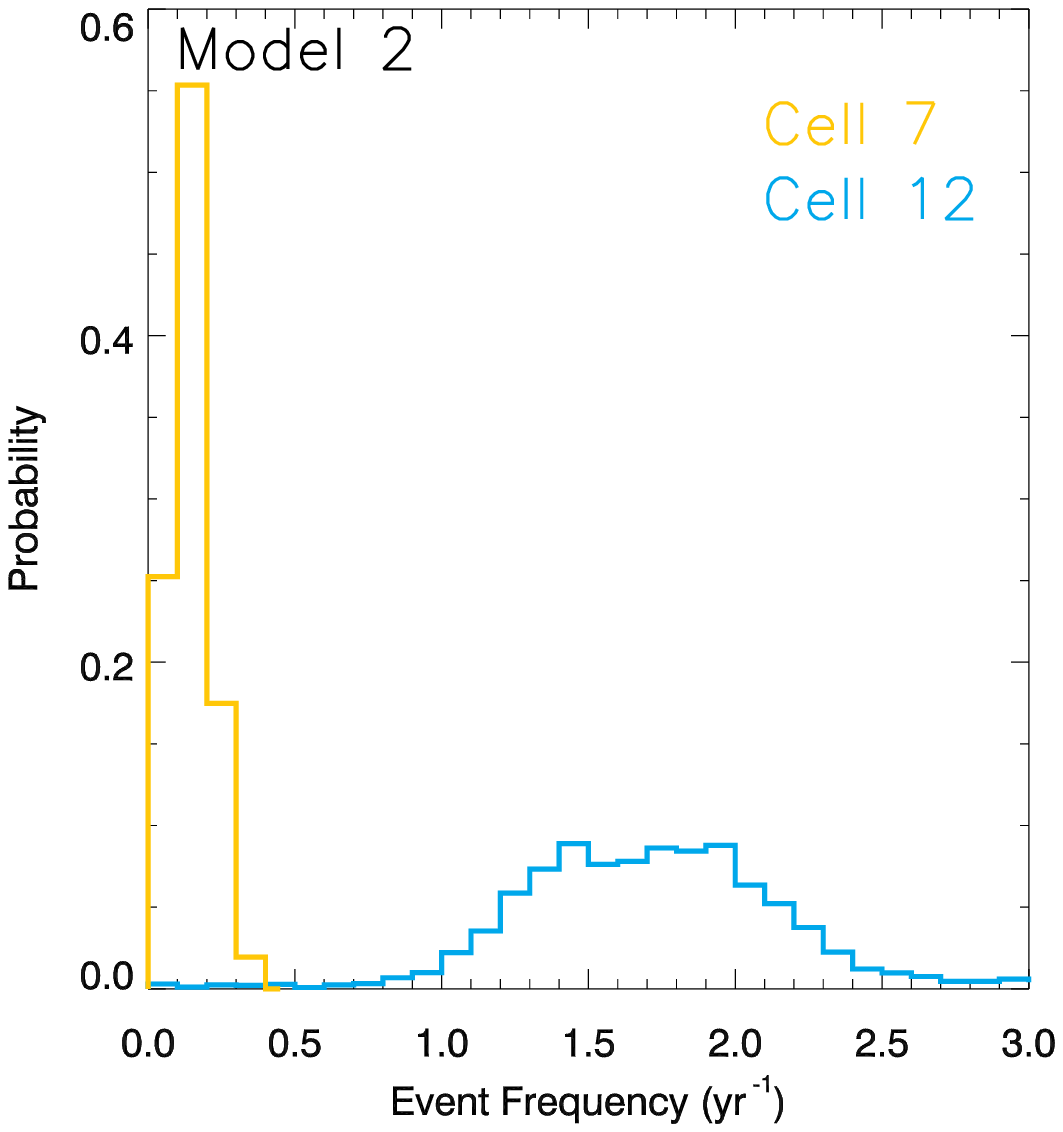}
 \includegraphics[width=2.5in]{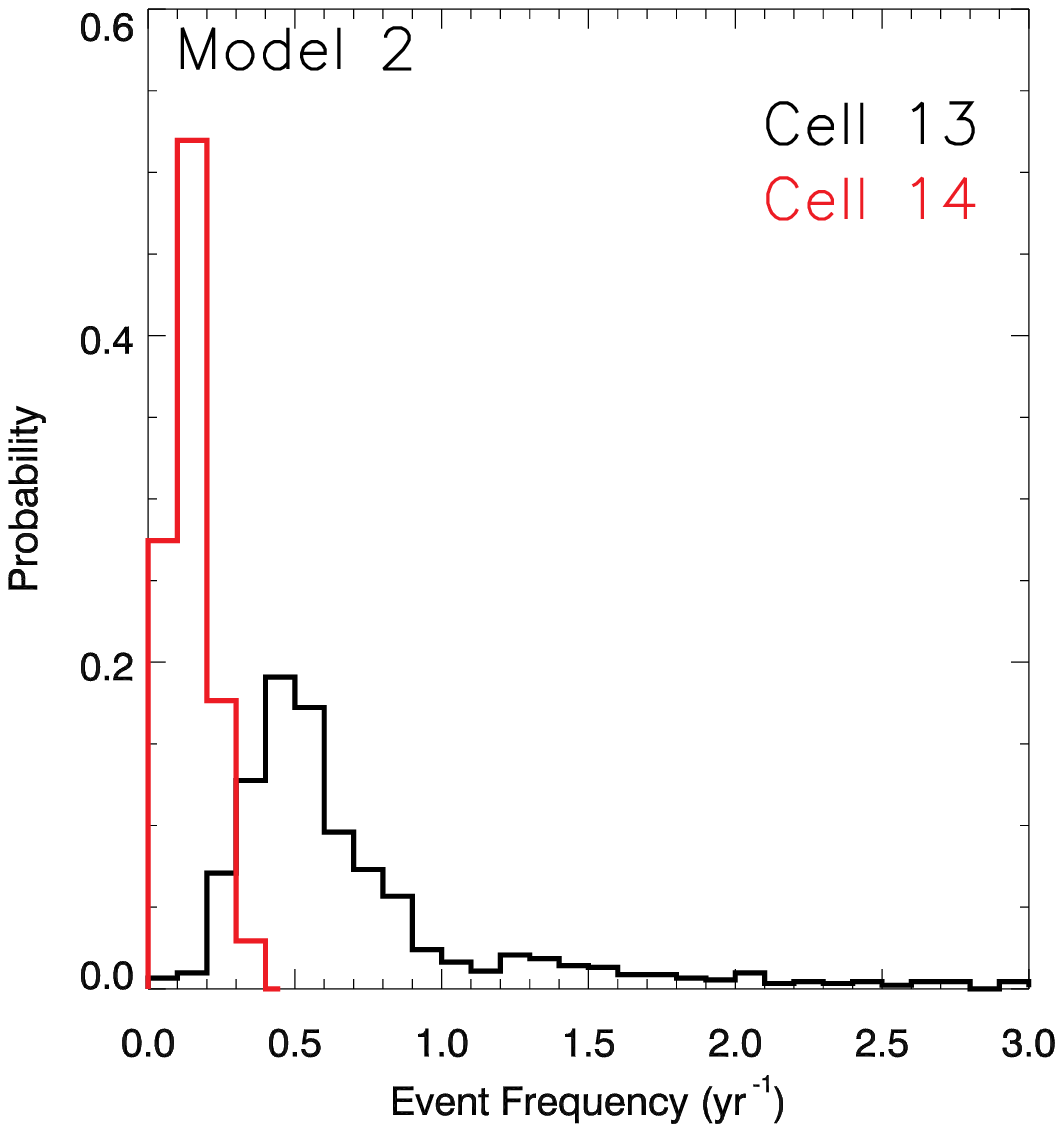}}
 \caption{\label{ch8fig:FGridModel2}  Same as Figure~\ref{ch8fig:FGridModel1}, except for Model 2. Note, however, that the x axis
 is significantly larger than for Model 1. }
 \end{center}
 \end{figure*}

Following equation (\ref{eq:FreqAvg}), we compute the average event frequency, $\langle \Gamma \rangle$, per grid cell.  
We first compute normalized probability distributions for $\Gamma$ by accounting for the variation in the quantity $D^{0.5}V_\perp$ 
exhibited by the stripped SMC stellar particles in each grid cell. These are plotted for the central 6 grid cells in Figures~\ref{ch8fig:FGridModel1}
 and~\ref{ch8fig:FGridModel2}. 

Integrating over these probability distributions yields the average event frequency in each grid cell; these
are listed in Tables ~\ref{ch8table1} and ~\ref{ch8table2} for Models 1 and 2, respectively.

Assuming 100\% detection efficiency, the total predicted event frequency in Model 1 is 
$\Gamma \sim 0.37$ yr$^{-1}$ for the entire LMC disk and 0.24 yr$^{-1}$ for 
the central 6 grid cells representing the area covered by the MACHO survey (red region in Figure~\ref{ch8fig:Grid}).  
These values are much lower than expected for the MACHO events (1.75 yr$^{-1}$).
To compare to the OGLE survey, we examine the cells in the larger green region highlighted 
in Figure~\ref{ch8fig:Grid}. We find an event frequency of 0.29 yr$^{-1}$, consistent with the results from the OGLE survey, 
which found 4 events over 12 years, implying an event frequency of 0.33 yr$^{-1}$. 

For Model 2, the total event frequency across the entire face of the disk is $\sim$5.17 yr$^{-1}$, 3.10 yr$^{-1}$ for 
the central 6 cells covered by the MACHO survey and 3.83 yr$^{-1}$ for the OGLE survey region. 
These values are consistent
with the MACHO survey result, but much higher than that of the OGLE survey.

\subsection{Detection Efficiency and Discrepancies between the OGLE and MACHO surveys} 

So far we have not accounted explicitly for the detection efficiency of the MACHO and OGLE surveys in our determination of the 
expected event frequencies.   

For the average simulated durations of $\sim$20-150 days for Models 1 and 2, the detection efficiency of the MACHO survey is roughly 50\%.  As such, the 
observed event frequency should be roughly half that determined in the previous section for the central 6 grid cells (which well represent
 the MACHO survey area). 
That is, $\Gamma \sim$1.54 yr$^{-1}$ for Model 2, which is in fact roughly the observed MACHO event frequency.  
Even taking a lower efficiency of 30\% yields a consistent result of $\Gamma \sim$1 yr$^{-1}$. 
 Since some events may turn out to be LMC self-lensing
events \citep[such as, MACHO event 14;][]{alcock2001}, this lower event frequency may be more accurate. 
However, the Model 1 event frequency is still an order of magnitude too low to match the MACHO observations (0.07-0.12 yr$^{-1}$). 

For the OGLE survey, the maximal detection efficiency for the simulated average range of durations is 10-35\%. The survey area for OGLE is
also larger, so in the previous section we considered the event frequency expected across the region highlighted
 in green in Figure~\ref{ch8fig:Grid}, finding a value of 0.37 yr$^{-1}$ for Model 1 and 3.83 yr$^{-1}$ for Model 2.  
Accounting for the OGLE survey detection efficiency yields a simulated event frequency of (0.03-0.10) yr$^{-1}$ for Model 1 
and (0.38-1.34) yr$^{-1}$ for Model 2.

Generally speaking, microlensing should occur with the same probability on every type of source star.  However, selection criteria 
adopted by the MACHO and OGLE imply a bias against detecting microlensing events on faint stars. In particular, the OGLE
survey is less sensitive than the MACHO survey.
Since our sources are expected to be faint SMC debris, a detection efficiency of 10\% for the OGLE survey is likely appropriate.  
 The observed event frequency by the OGLE survey of 0.33 yr$^{-1}$ is 
consistent with the Model 2 results of 0.38 yr$^{-1}$ at such an efficiency.  
 
 Discrepancies between the OGLE and MACHO surveys thus appear to be attributable to differences in the
 detection efficiencies of these surveys: specifically, their sensitivities to faint sources. As discussed in $\S$~\ref{sec:Surveys}, 
 the EROS team non-detections towards the LMC is likely a result of their approach of limiting their sample to only bright 
 stars to avoid uncertainties owing to blended sources.  
   
These above results favor Model 2, wherein the debris of SMC stars are non-uniformly dispersed behind the 
LMC disk, having been pulled out of the SMC
or formed in-situ in gas stripped from the SMC after a collision between these two galaxies $100-300$ Myr ago. 

\clearpage
\vfill\begin{deluxetable}{ccccccccc}
\tabletypesize{\scriptsize}
\tablewidth{0 pt}
\rotate
\centering
\tablecaption{ Microlensing Properties: Model 1 }
\tablehead{
\colhead{Grid Cell}   	&  \colhead{$\langle D_S \rangle$}    	& \colhead{$\langle D_L \rangle$}   &   \colhead{$\langle D \rangle$}        & \colhead{$\langle V_\perp \rangle$}    & \colhead{$\langle t_e \rangle$}     &   \colhead{$N_{\rm source}$}    & \colhead{$\langle \Gamma \rangle$} \\
		&      \colhead{(kpc)} 	&  \colhead{(kpc)}      &  \colhead{(kpc)}     &  \colhead{(km/s)}   & \colhead{(days)}  &     \colhead{($10^4$)}      & \colhead{(yr$^{-1}$)}   }
\startdata
		1$^\dagger$			&   -     				&  45.1		                &       -                        &        -                        &    -                 	&  -			&    -   	   \\
		2$^\dagger$			&    -  				&  46.2 		                 &      -                           &      -                        &    -                  	& -			&    - \\
		3					&    - 				         &  47.4		                &     -                              &      -                        &    -                	&   -			&  -  \\
		4					&    56.6  				&  48.7	                       &         6.9                          &     435                    &   18                	&   0.25		& 0.001    \\ 
		5					&   53.6				&  49.8		               &     3.7                             &      115                 &    80               	&   6.0		&   0.001  \\ 
		6$^\dagger$			&   -					&  44.0 	                        &       -                            	&            -                  &    -              		&  0			&   0.0 \\
		7*$^\dagger$			&   61.0				&  44.9		               &     11.9                             &      136                 &     76               & 0.13		&   0.001\\
		8*$^\dagger$			&   57.9  				&  45.9	                        &     9.4                             &         66                   &  108		&  2.52		&  0.020 	  \\ 
		9*$^\dagger$			&  56.2				&  47.1	                        &      7.6                             &  	    50                    &  146               & 5.66		&  0.012  \\
		10					&  55.3				&  48.7	                        &       5.9                            &  	   64                  &   98                 & 8.68		&  0.005  \\
		11$^\dagger$			&  59.4				&  43.1	                        &        11.7                           &      112               &     40              	  & 2.26	 	&    0.007 \\
		12*$^\dagger$			& 56.0				&  44.0	                        &        9.1                           &         147                &51                   	&  4.53		&    0.036\\
		13*$^\dagger$			&  58.6				&  45.1 	                        &       10.3                           &         132                &  65                 & 7.81		&     0.133\\
		14*$^\dagger$			& 57.0				& 46.2	                        &        8.4                          &           83                    &  103               & 9.06		&   0.041	 \\ 
		15$^\dagger$			& 57.2				&  47.6	                        &       7.9                             &         61                   &   128                 & 8.06		&    0.010  \\
		16					& 59.9				&  42.2 	                        &        12.4                           &       204                      &   47           & 7.05		&   0.008 \\
		17					& 59.	7				&  43.2		                &         11.9                        &        183                     &    51             & 11.70		&    0.035\\
		18					& 59.6				&  44.1	                       &        11.4                            &        169                     &   52       	& 5.79		&     0.030\\
		19$^\dagger$			& 58.7				&  45.2	                       &       10.3                            &         134                     &   53        	& 7.05		&   0.025  \\
		20$^\dagger$			& 57.7	      			&  46.8	                       &       8.8                            &        95                      &    80               & 5.66		& 0.007   \\
 \hline
\bf{WEIGHTED AVERAGE}         &    57.9                      & 45.5              &     9.5    			 &             	120	  		 & 	79	    	& total 92.24             	&  total 0.37   \\ 
\bf{* MACHO Cells}			&   51.1                      & 40.9               &    8.0                        &             88				&    84              	 &   total  33.22	&   total  0.24   \\ 
\bf{ $^\dagger$ OGLE Cells} 			& 57.6			& 45.9		& 9.1				& 98				& 90			&  total 52.73	&  total  0.29 \\
\enddata
\tablecomments{\label{ch8table1} Asterisks indicate the 6 grid cells that cover regions where microlensing events have been detected. Values for these six central cells
 (relevant area for the MACHO survey) are listed in the row marked `* MACHO Cells'. $^\dagger$ mark grid cells that cover
regions relevant to the OGLE survey. The final row marked `$^\dagger$ OGLE Cells' computes the total/average values relevant for the OGLE survey. 
 Averages are weighted by the number of source stars in each cell. Dashes indicate cells with no source stars. Event frequencies ($\Gamma$) quoted here assume detection 
efficiencies of 100\%.}
\end{deluxetable}

\clearpage
\vfill\begin{deluxetable}{ccccccccc}
\tabletypesize{\scriptsize}
\tablewidth{0 pt}
\rotate
\centering
\tablecaption{ Microlensing Properties: Model 2 }
\tablehead{
\colhead{Grid Cell}   	&  \colhead{$\langle D_S \rangle$}    	& \colhead{$\langle D_L \rangle$}   &   \colhead{$\langle D \rangle$}        & \colhead{$\langle V_\perp \rangle$}    & \colhead{$\langle t_e \rangle$}     &   \colhead{$N_{\rm source}$}    & \colhead{$\langle \Gamma \rangle$} \\
		&      \colhead{(kpc)} 	&  \colhead{(kpc)}      &  \colhead{(kpc)}     &  \colhead{(km/s)}   & \colhead{ (days)}  &     \colhead{($10^4$)}      & \colhead{(yr$^{-1}$)}   }
\startdata
		1$^\dagger$			&      57.8				& 	46.3	                         &    9.2                            &   38                              &    28                  & 0.50			&  0.001     	   \\
		2$^\dagger$			&      50.5				& 	47.4	                         &    2.9                              &   147                            &  23                     & 3.15		&  0.010  \\
		3					&      51.0				&	48.4	                        &     2.6                               &   130                            & 47                  &  7.81 			& 0.018  \\
		4					&      51.2				& 	49.3                         &     2.2                               &   69                            & 66                 &  4.28			& 0.003  \\ 
		5					&      52.7				&	50.1	                        &     2.6                              &    144                           & 33                   &  2.89			& 0.001   \\ 
		6$^\dagger$			& 	51.9				& 	45.3                         &     5.7                               &    144                           & 50                  &  22.01		& 0.061   \\
		7*$^\dagger$			& 	52.1				&	46.2	                        &     5.2                              &   157                             &  43                   & 23.40		& 0.136   \\
		8*$^\dagger$			& 	48.6				& 	47.1                         &     1.7                               &   66                              &   77			& 88.82		& 0.366	  \\ 
		9*$^\dagger$			& 	52.3				&	48.5                         &     3.6                               &   88	                        &   66                   & 13.46 		& 0.032    \\
		10					&  	54.2				& 	49.2                         &     4.6                               &   53	                        &  96                   &  3.40		& 0.003    \\
		11$^\dagger$			&  	50.2				& 	44.7                         &    5.0                                &   177                          &   38                  & 758.10		& 0.623     \\
		12*$^\dagger$			& 	49.0				& 	45.4                         &     3.4                               &   157                           &  35                   & 669.59		& 1.741    \\
		13*$^\dagger$			& 	48.7				& 	46.6                         &      2.3                             &  123                              &  36                   & 159.40		& 0.691     \\
		14*$^\dagger$			& 	52.9				& 	47.6                         &     4.6                             &  130                              & 60                    & 23.28		& 0.133	 \\ 
		15$^\dagger$			& 	51.5				& 	48.3                         &   3.1                                  &  129                           &  64                    & 7.43			& 0.012      \\
		16					& 	50.6				& 	44.8                         &   5.1                                 &  158                            &   43                   & 26211.20		& 0.776    \\
		17					& 	49.8				& 	45.5	                        &  4.0                                &  131                            &  47                      & 808.20		& 0.372    \\
		18					& 	50.4				& 	46.1                        &   3.9                                  &  146                            &  41                    & 163.68		& 0.157     \\
		19$^\dagger$			& 	51.7				& 	46.9                        &   4.4                                 &  181                            & 31                     & 28.32		& 0.035    \\
		20$^\dagger$			& 	49.7       			& 	47.5                        &  2.3                                  &  157                             & 24                        & 7.68		& 0.002    \\
 \hline
\bf{WEIGHTED AVERAGE}         	&    50.2                               &      45.2         		 &    4.5      			 &   153             		&  42                        &    total 5416.52          & total 5.17     \\ 
\bf{* MACHO Cells}				&  49.1                               &      45.9            		&   3.1                           	   &   142          		   & 40                      &   total 977.95   &   total  3.10  	\\
\bf {$^\dagger$ OGLE Cells}		& 49.9				& 45.6			& 4.0					& 158			& 39			& total 1797.72 	& total 3.83 \\	
\enddata
\tablecomments{\label{ch8table2}   Asterisks indicate the 6 grid cells that cover regions where microlensing events have been detected. Values for these six central 
cells are listed in the row marked as `* MACHO Cells'.   The final row marked `$^\dagger$ OGLE Cells' computes the total/average values relevant for the OGLE survey. 
Averages are weighted by the number of source stars in each cell. Event frequencies ($\Gamma$) quoted here assume detection 
efficiencies of 100\%.}
\end{deluxetable}
\clearpage

\section{Observational Consequences}
\label{sec:Discuss}

 We do not claim in this work that all observed microlensing events have sources in an SMC debris population, but that
such a population would be expected to have event frequencies and durations that are consistent with those observed.
It is likely that the lenses do not all belong to the same population \citep{jetzer2002}.
In particular, there is probably some contribution to the microlensing events from self-lensing: the \citet{gyuk2000} disk + bar self-lensing models
predict a contribution of 20\% and \citet{calchi2011} were able to explain the OGLE events in a self-lensing model. 
Furthermore, the total mass of the LMC's stellar halo is not well constrained; it may also contribute to the observed microlensing 
event frequency. In particular, on a first infall scenario the LMC is not expected to be truncated to the tidal radius, meaning that
 any putative stellar halo may be much more extended than originally considered. This is in line with the possible detection of LMC stars
  as far as 20 degrees away from the LMC center of mass \citep{munoz2006}. 
We further note that many of the initially reported microlensing events were subsequently determined to be binary lensing events; i.e. 
to have origins within the LMC disk. Thus, it is not unreasonable to consider a fraction of the lens populations as being within the LMC disk 
itself \citep{sahu2003}.  

Here we outline observationally testable predictions from our model. 

\subsection{Detecting the Stellar Counterpart to the Stream and Bridge}

 The \citet{besla2010} and B12 models for the Magellanic System predicts the existence of a
  stellar counterpart to the Magellanic Stream and Bridge. We have demonstrated here that the surface brightness of the
  predicted stellar stream is too low 
to have been detected by surveys for Stream stars conducted to date ($\S$~\ref{sec:Stellar}). 

  There are a number of on-going optical searches for stars in the Stream, including the NOAO Outer Limits Survey 
 \citep{saha2010} and the VMC survey \citep{cioni2011}, that stand a good chance
  of detecting the predicted stellar stream. 
 The NOAO Outer Limits Survey fields include non-contiguous pointings aligned with the gaseous Magellanic Stream
 and also sight lines in several directions from the MCs.  
For the published fields due north of the LMC, \citep{saha2010} report the detection of LMC stars 
out past 10 disk scale lengths. The sensitivity of these data, 
 corresponding to a surface brightness level in the Vband of $\sim$32.5 mag/arcsec$^2$ (Abi Saha, private communication, 2012), 
   may be sufficiently sensitive to detect a spatially aligned
 stellar counterpart in their Stream fields.  
  However, this survey will be unable to detect an {\it offset} stellar stream
  if one exists, as the search area is limited to regions of the highest gas 
 density.  The VMC survey has only one field on the Stream and one off the Stream.   
 Future searches for stars would be more robust if designed to search contiguously {\it across} the Stream to detect changes
 in the magnitude of diffuse emission. 
  
 The stellar Leading Arm in Model 2 (which is the model that best matches the kinematics and structure of the MCs)
  is predicted to be brighter than the trailing stream and is therefore likely the 
 best location to look for tidal debris. Furthermore, a detection of stars in the Leading Arm may serve as a way to distinguish 
 between Models 1 and 2.  
 The stellar Leading Arm is expected to be offset 
 from the gaseous Leading Arm owing to the effect of ram pressure. 
 Thus, searches for stars in the Leading Arm must survey a wide area of the 
 sky surrounding the gaseous components.

Overall, we find that the simulated stellar stream roughly follows the location of the simulated 
gaseous stream: no pronounced offsets exist
 in projection on the plane of the sky in simulations that do not include ram pressure/gas drag. 
 The tidal models of \citet{diaz2012} also suggest that a stellar counterpart must exist; however, 
 they find that the tip of the predicted stellar stream is in fact 
offset from the observed gaseous Stream, which could also explain the non-detection reported by \citet{NoStars}.
  Furthermore, if ram pressure effects were accounted for, the stellar stream would 
still appear as modeled here, while the gaseous stream may change location. 
As such, it appears reasonable to search for the stellar counterpart of the Magellanic Stream by looking 
in regions displaced from the gaseous Stream by a few degrees, rather than just where the gas column density is highest. 
For example, \citet{casetti2012} find a population of stars that is offset by 1-2 degrees from the 
high-density HI ridge in the Bridge. 

From the toy model treatment of ram pressure stripping presented in the appendix of B12, 
we find that gas drag from the motion of the system through the MW's ambient gaseous halo does strongly 
modify the appearance of the gaseous {\it Leading Arm}.  This will likely 
 result in a severe discrepancy between the location of the gaseous and stellar Leading Arm features.  
It is therefore conceivable that stellar debris from the SMC may reach the Carina dSph fields 
even though the gaseous Leading Arm does not, as suggested by \citet{munoz2006}. 

While ram pressure effects may also cause differences in the spatial distribution of stars in the Bridge relative to 
the gas, perhaps more significantly, in Model 2 the stellar debris in the simulated bridge
 is dispersed over an area $\sim$ twice as large as that in Model 1.  This is a consequence of the small impact 
 parameter collision between the Clouds, which allows for the removal of stars from deep within the SMC potential
 well. In contrast, the gas is not similarly distributed, owing to hydrodynamic effects as the two gas disks pass 
 through each other (e.g. gas drag). A spatially extended, low surface density 
 stellar bridge is a generic prediction of an LMC-SMC collision scenario.

\subsection{Nature of the Source Population: SMC Debris}

We make a number of specific predictions for the properties of the source stars of the observed microlensing events
towards the LMC. 
First, these source stars originated from the SMC and should therefore be metal poor. 
This could be testable if spectra are obtained for the source stars.  

Second, the majority of these source stars should be old (age $>$300 Myr, roughly the age of the Bridge). 
Ages as high as 1 Gyr are not implausible based on isochrone comparisons to the 
 source population from the MACHO events (Tim Axelrod, private communication 2012).  

Finally, we also predict that the sources should be located behind the LMC disk. 
A background source populations has the advantage of being difficult to detect observationally. 
\citet{zhao2000b, zhao1999b} suggest that this population can be identified by their expected reddening.
\citet{zhao2000c}  estimate this additional reddening should be $\sim$0.13 mag.
\citet{nelson2009} tested this expectation by comparing the color-magnitude diagrams (CMDs) 
of the observed sources to those of the LMC and detected no significant reddening. 
They consequently ruled out the possibility that these source stars may be behind the LMC disk.
We note two caveats to this analysis. 
First, we consider here a source population originating from the SMC, which will have different 
CMDs than LMC stars.  
Second, new reddening maps of the LMC based on red clump stars and RR Lyrae from the OGLE III data set by 
\citet{haschke2011} reveal very little reddening in the bar region of the LMC. 
Since most events are concentrated in this region, we do not expect 
the source population to exhibit significant reddening. 
Moreover, the \citet{nelson2009} analysis only excludes a toy model in which the sources for all
lensing events are in the background of the LMC and in which the extinction is uniform. 
For a more realistic model with a mixture of lens and source populations and patchy 
extinction, the constraint is even weaker. 
We conclude that the existence of a source population of SMC debris located behind the LMC cannot be ruled out by 
the currently available data.

\subsection{Kinematics of the Source Population}
\label{sec:PM}

If the sources stars are indeed a population of tidally stripped stars from the SMC, we expect the majority of these
 stars to have distinct kinematics from the mean velocities in the LMC disk. As discussed in B12, the line-of-sight velocities
of these debris stars are, in some cases, discrepant from the local line-of-sight velocities of the LMC disk. 
The observations of \citet{olsen2011} provide suggestive evidence
that such a metal poor debris population has already been observed. 

The determination of the proper motions of the sources may be a direct way to test our proposed model, as they should be discrepant from the mean proper motion 
of local LMC disk stars. In B12 we have shown that the line-of-sight velocities of this debris field are consistent with the \citet{olsen2011} results
for both Models 1 and 2.  \citet{olsen2011} suggested two possible configurations for this debris that could explain the line-of-sight 
velocities: 1) a nearly coplanar counter-rotating disk population; and 2) a disk inclined from the plane of the sky at $\sim$-19 degrees. 
Given that we expect the debris to exist {\it behind} the LMC disk, our models are more consistent with the second
proposed configuration. 

\citet{casetti2012} recently measured the proper motions of the \citet{olsen2011} data sample and concur that this population 
likely represents stripped SMC stars that are kinematically distinct from the LMC disk stars.  Their analysis disfavors 
the coplanar counter-rotating disk scenario, but are consistent with an inclined orbital plane. 
They further find relative proper motions (i.e. tangential motions of the debris population relative to the LMC disk stars) 
of on average  $\Delta\mu_\alpha = -0.51 \pm 0.30$
mas/yr and $\Delta\mu_\delta = -0.03 \pm 0.29$ mas/yr.   
   In this section we will test the predicted proper motions from our models 
against these results. 

Proper motions of source stars have been measured in a few cases, 
but most of those events are identified as having MW lenses \citep[e.g. Event 5;][]{drake2004}.
An event of relevance is Event 9, a binary caustic crossing event \citep{bennett1996, kerins1999}. 
The lens for this event is most likely a star near the LMC. 
This event was not included in our event frequency and duration analysis as it is not included in the 
\citet{bennett2005} and \citet{alcock2000} conservative samples.
The $V_\perp$ for the identified source (an A7-8 main sequence star) is low ($\sim$20 km/s). 
While this disagrees with the average 
$V_\perp$ velocities, we note that the velocities in cell 8 of Model 2 are lower on average; the 
distribution in velocities in this cells peaks at $V_\perp \sim66$ km/s.  

\begin{figure*}
\begin{center}   
\mbox{ {\includegraphics[width=4.5in, clip=true, trim=0 0.5in 0 0]{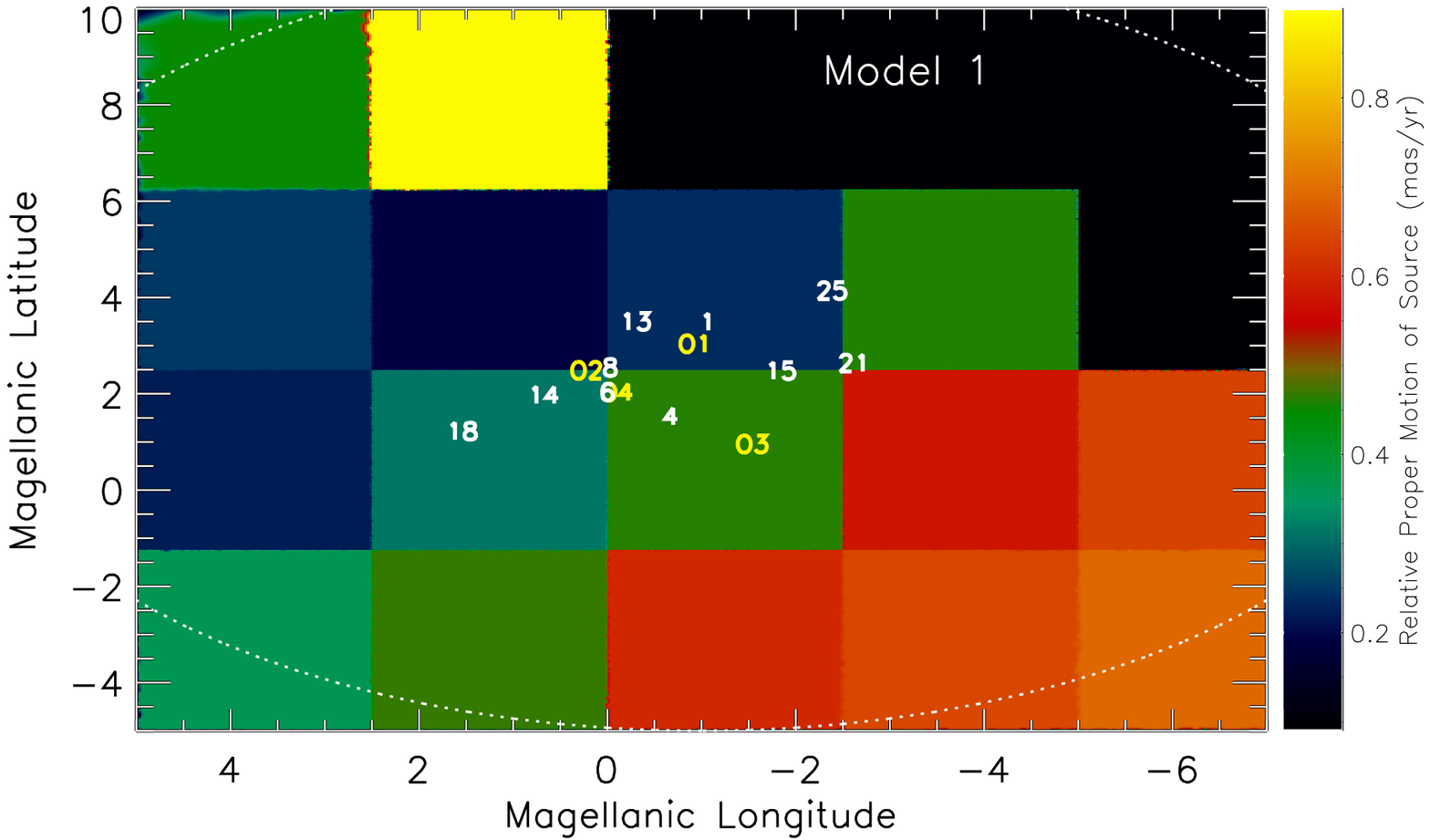}}}\\
\mbox{ {\includegraphics[width=4.5in]{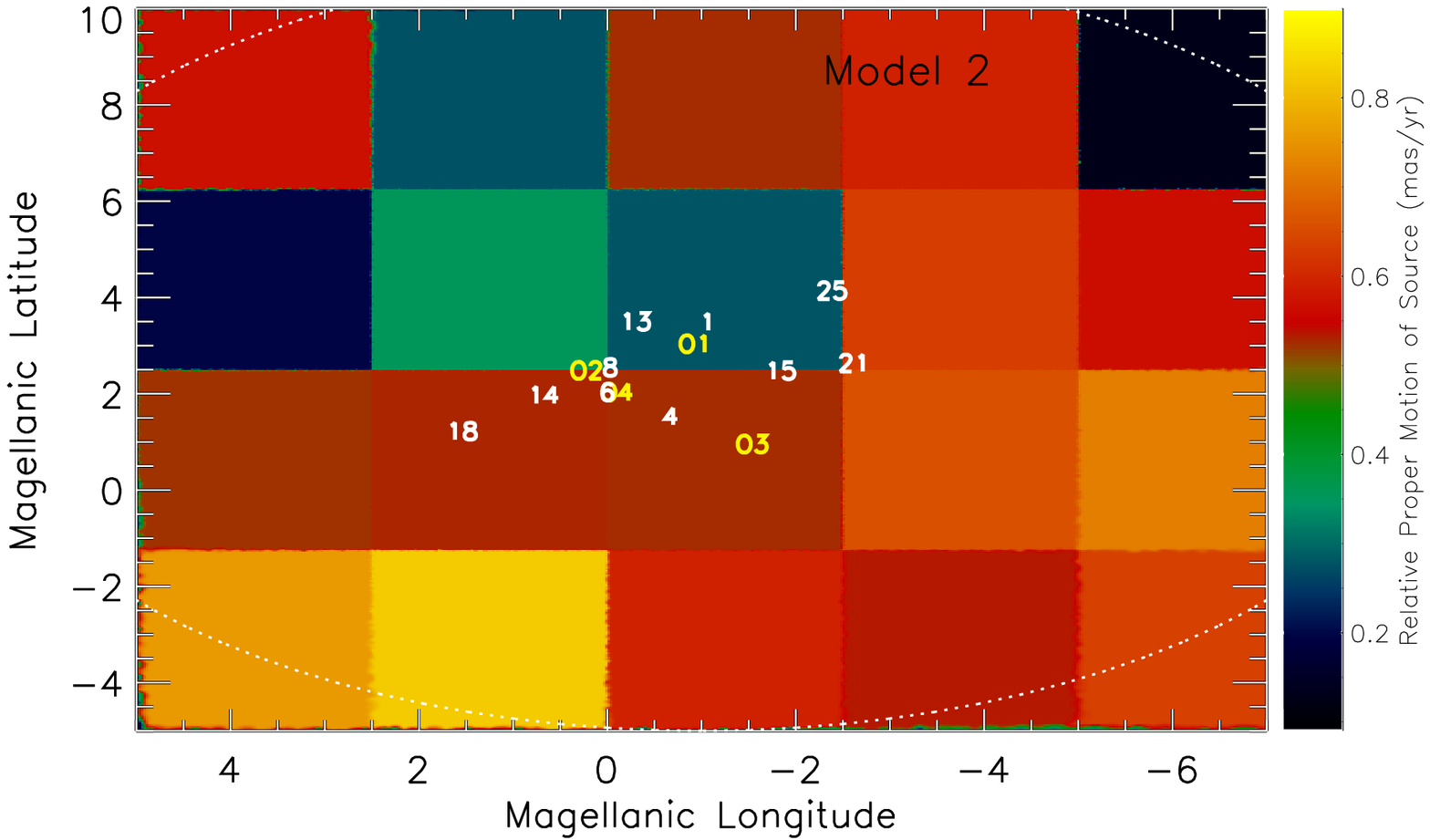}}}
 \end{center}
 \caption{\label{ch8fig:PM}  The average relative proper motion between the sources and lenses is color coded in each grid cell across the face of the LMC disk for 
 Model 1 (top) and Model 2 (bottom).  The cells are defined in Figure~\ref{ch8fig:Grid}.  }
 \end{figure*}

We translated the distribution of relative transverse velocities ($V_\perp$) 
between the sources and lenses in each grid cell ($\S$~\ref{sec:Vperp}) to a relative proper motion 
using the distribution of distances to the sources ($\S$~\ref{sec:Dist}).  
We integrate over the resulting probability distributions to determine the average 
proper motions per grid cell and plot the spatial distribution of these averages across the
 face of the LMC disk in Figure~\ref{ch8fig:PM}. 
Note that we plot only the {\it magnitude} of the proper motion rather than the vector, 
as the vector is strongly model dependent.  
The weighted average of the simulated proper motions across all grid cells covering the 
face of the LMC disk is
0.44 mas/yr for Model 1 and 0.63 mas/yr for Model 2.  
These values are consistent with the average magnitude of the proper motion 
vector reported by \citet{casetti2012} ($\sim$0.51 mas/yr).

Follow up studies of the proper motions of the observed microlensed sources would be important tests to see if the 
kinematic properties of the sources are in fact similar to the detected debris population 
of \citet{olsen2011} and \citet{casetti2012} and
ultimately with our models.  The first epoch HST observations of all the MACHO sources
have been taken in the 1990's, providing a significant time baseline.


\subsection{Spatial Distribution of Events and Event Durations}
\label{sec:Spatial}

From the results listed in Tables~\ref{ch8table1} and~\ref{ch8table2} it is clear that our models predict higher 
event frequencies in the central 6 grid cells that cover the LMC bar region.  However, it is also
 clear that events are predicted to 
occur outside the bar region. 
The event frequency could be higher in these outer regions if the underlying source population is clumpy; i.e.,  
such that there are high densities of source stars in 
fields not spatially coincident with the central bar (e.g. grid cell 16 in Model 2).  
Such off-bar events are not expected in self-lensing models \citep[e.g., ][]{calchi2009, calchi2011}, which 
rely on the high surface density of lenses in the bar region and the warped nature of the bar itself. 

\begin{figure*}
\begin{center}   
\mbox{ {\includegraphics[width=4.5in, clip=true, trim=0 0.5in 0 0]{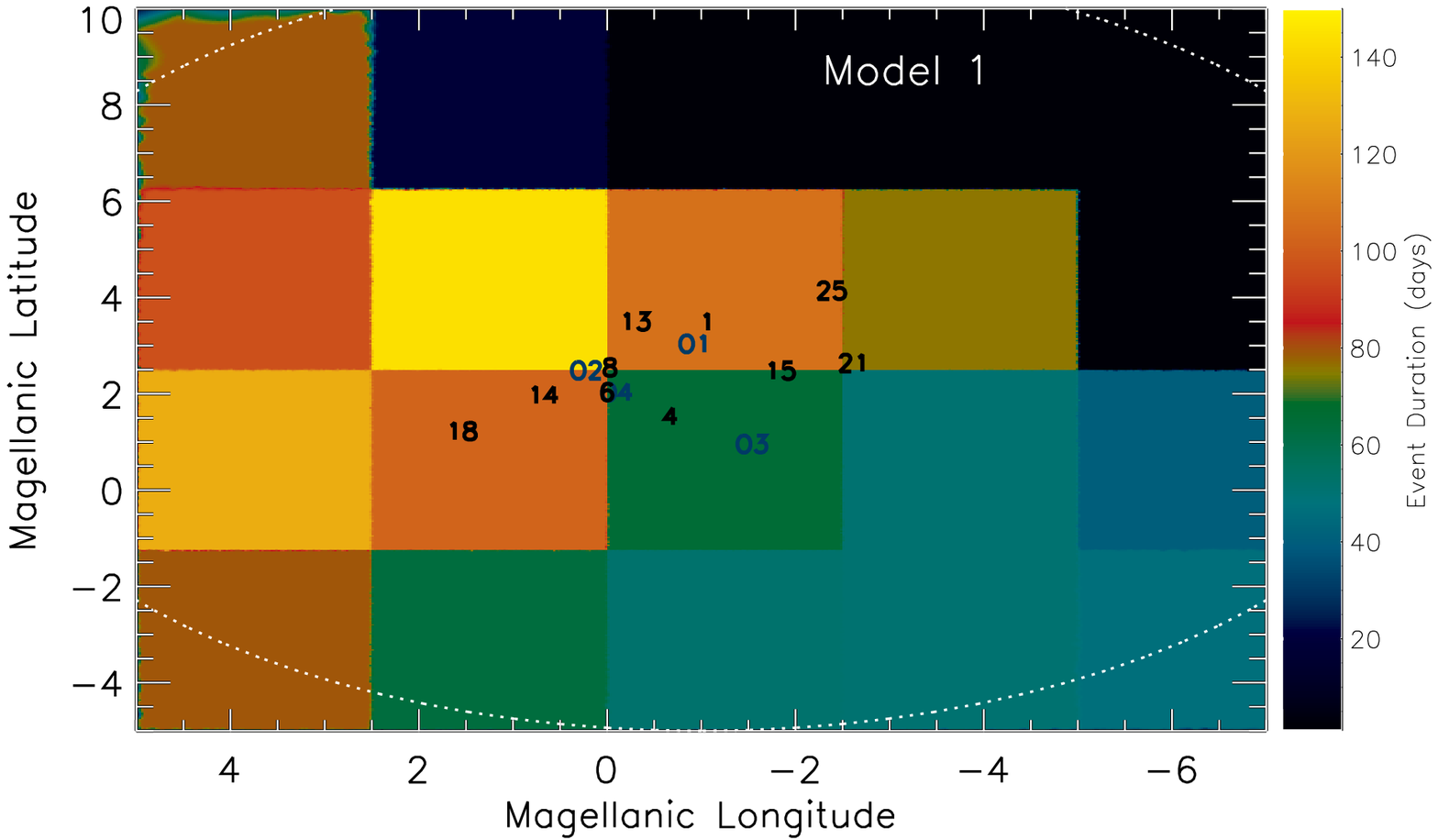}}}\\
\mbox{ { \includegraphics[width=4.5in]{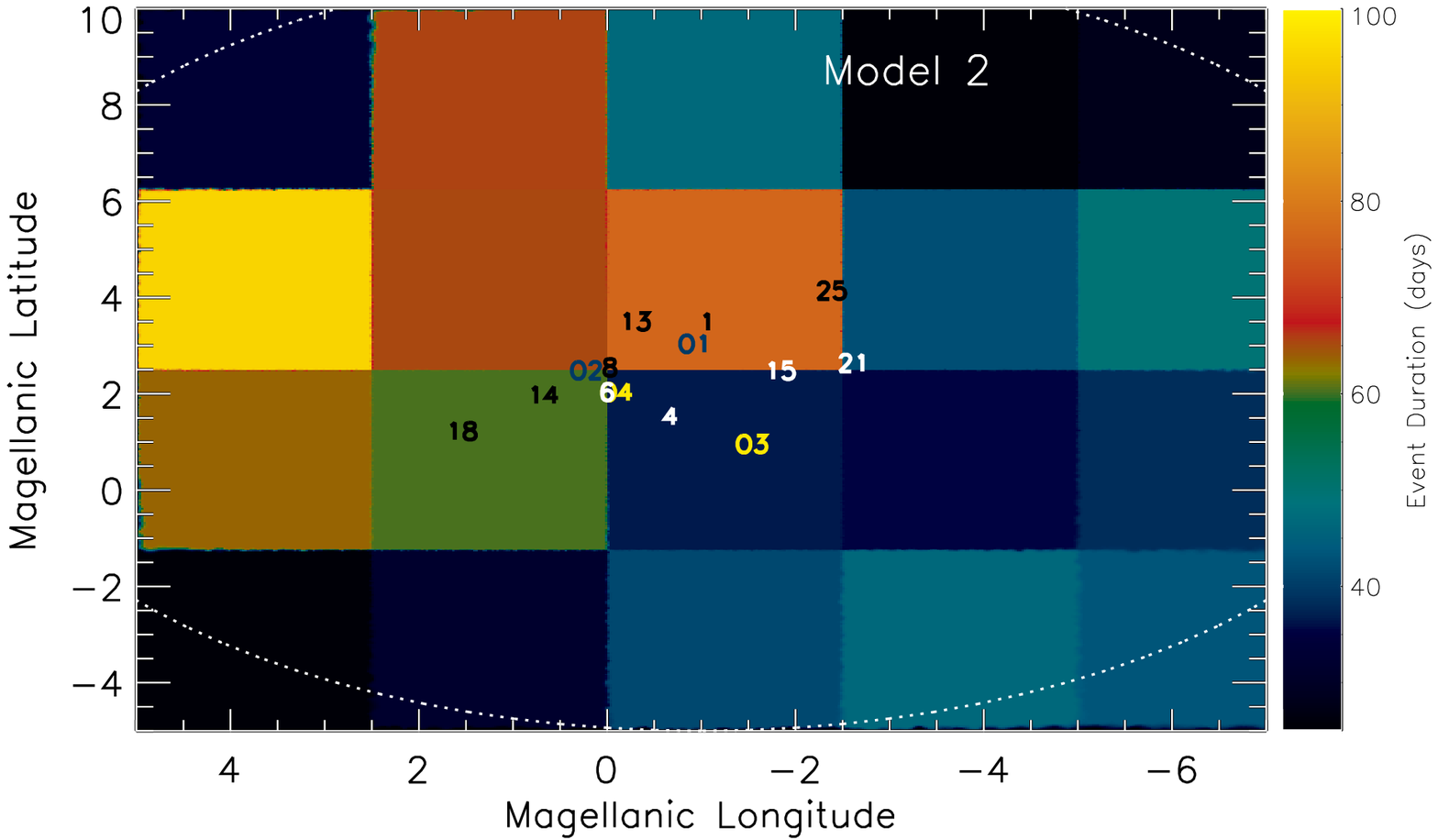}}}
 \caption{\label{ch8fig:T}  The average event duration, $\langle t_e \rangle$, is plotted in each grid cell across the LMC disk for Model 1 (top) and Model 2 (bottom) using data 
 from Tables~\ref{ch8table1} and~\ref{ch8table2}.  These figures are anti-correlated with Figure~\ref{ch8fig:PM}, since $t_e \propto V_\perp^{-1}$.
  The cells are defined in Figure~\ref{ch8fig:Grid}.}
   \end{center}
 \end{figure*}

In $\S$~\ref{sec:Dur} we computed the average event durations in each of our denoted grid cells 
(listed in Tables~\ref{ch8table1} and~\ref{ch8table2}). In Figure~\ref{ch8fig:T} we plot the spatial distribution of $\langle t_e \rangle$
 across the face of the LMC disk.  We find that both Models 1 and 2 exhibit a trend such that, on average, 
event durations are longer at larger Magellanic Longitudes and Latitudes (i.e. North-West portion of the LMC disk). 
This is because $t_e$ is inversely proportional to the relative transverse velocity between the source and lens, which 
 is on average smaller in that section of the disk (see the distributions of proper motions in Figure~\ref{ch8fig:PM}), whereas
 $\langle D \rangle$ is roughly constant across the disk.  
 
 $\langle V_\perp \rangle$ varies across the face of the LMC disk largely because the LMC disk is rotating in a known clock-wise fashion and the disk is 
 viewed roughly face-on. As such, the disk rotation is a dominant contribution to the tangential relative motion between the sources and lenses. 
 Since the kinematics of the source population is largely determined by the relative motion between the LMC and SMC (which is similar
 in both Models since it is dependent on the HST proper motions), both Models 1 and 2 exhibit similar behavior. 
 
 If more events are observed at larger distances from the bar it may be possible to observe this average difference in durations. 
 The currently observed events are too close to the central disk regions for this effect to be obviously discernible.

\subsection{Surface Density and Mass of the Source Population}
 
Model 1 fails to reproduce the observed event frequency, underestimating the rate
 by an order of magnitude.  In the previous section we have illustrated that the 
 kinematics in both models are roughly similar. Thus, the discrepancies between Models 
1 and 2  likely reflect the significantly different surface densities and mass of 
stripped SMC stellar debris behind the LMC disk in each model. Specifically, the number 
of sources are too small and too smoothly distributed in Model 1. 

In Model 1 only a modest amount of stars are stripped from the SMC to the region behind 
the LMC disk; the stellar debris comprises only 0.2 per cent of the current LMC disk mass.
In contrast, the mass of stellar debris in Model 2 is 1.5 per cent of the 
LMC disk mass. 


The failure of Model 1 provides a lower limit of $7 \times 10^{6} \Msun$
for the mass of SMC debris behind the LMC disk. 
The distribution of this material could be clumpy, but the average 
surface density of this debris must be larger than that of Model 1, i.e. 
$\Sigma_{\rm debris} > 0.03 \Msun$ pc$^{-2}$.

Uncertainties in the star formation prescription adopted can change the amount of stellar
debris predicted in the simulations.  Regardless of such uncertainties, the limits discussed 
here are still realistic constraints on the stellar densities {\it required} to reproduce the
 observed microlensing properties.
 
\citet{olsen2011} find a kinematically distinct population of stars in the LMC 
disk field that comprise 5 per cent of their sample. Such studies
can provide important tests of the 
limits placed here and better constraints on the density and distribution of the stellar debris
from the SMC.

\subsection{RR Lyrae in the Magellanic Stream, Bridge and Leading Arm}

The existence of a stellar counterpart to the extended gaseous 
components of the Magellanic System may be inferred via the detection of RR
 Lyrae stars through e.g. the ongoing OGLE surveys. 
 The variability of these stars and their reliability as distance indicators
   make them readily identifiable tracers of a faint stellar population that can be
   associated with Magellanic System. 

Given the stellar mass in the Bridge, Stream and Leading Arm from the simulations
presented in this study, we can make a simple estimate for the expected number of RR Lyrae.

Using the OGLE III Catalog for Variable Stars, \citet{soszynski2010} find 2445 RR Lyrae in the SMC. 1933 of these
pulsate in the fundamental mode (RRab).  If we assume that the RR Lyrae are uniformly 
distributed over the face of the SMC, we can estimate the number of RR Lyrae per unit 
stellar mass in the SMC.  

The total stellar mass of the SMC is $3\times10^{8} L_{\odot} \times \frac{\Msun}{L_\odot} = 3 \times 10^8 \Msun$
(adopting a mass to light ratio of $1\Msun/L_\odot$).  This means that there are 

\begin{equation}
\frac{1933 \, \rm {RR Lyrae}}{ 3\times 10^8 \Msun} = 6.4 \times 10^{-6} \, \rm{RR Lyrae/}\Msun 
\label{eq:RRL}
\end{equation}

In the B12 model the Stream, Bridge and Leading Arm all originate from the SMC.  Thus, we can use 
this conversion factor to estimate the corresponding number of RR Lyrae in each of these components. 
The results are listed in Table~\ref{Table:RRL}

\begin{table*}
\centering
\begin{minipage}{200mm}
\caption{RR Lyrae Number Counts in the Magellanic System \label{Table:RRL}}
\begin{tabular}{@{}ccc@{}}
\hline

Component				& Stellar Mass	&  Number of RR Lyrae\tablenotemark{a}   \\
						&  ($10^6 \Msun$)	& 						\\
\hline
\hline
SMC	\tablenotemark{b}		&  300	&  1933	\\
Model 1 Stream			&  6	& 38 \\ 
Model 2 Stream			& 4	& 26 \\
Model 1 Leading Arm		& 5	& 32 \\ 
Model 2 Leading Arm		&  25 &  160 \\
Model 1 Bridge				& 2    &   13  \\
Model 2 Bridge				& 40 	 &   250 \\ 
\hline
\tablenotetext{a}{Number of expected RR Lyrae pulsating in the fundamental mode (RRab).\\
Values are computed from the stellar mass using the conversion factor of  \\
$6.4\times 10^{-6}$ RRLyrae/$\Msun$.  }
\tablenotetext{b}{Stellar mass from \citet{Stan2004}, RR Lyrae number counts from \citet{soszynski2010}. }
\end{tabular}
\end{minipage}
\end{table*}

Number counts of RR Lyrae in the Stream are expected to be low in both models ($< 40$). These stars would 
also be scattered across the $\sim$100 degree extent of the Stream, making the number density of RR Lyrae
undetectably low.

RR Lyrae in the Leading Arm are too low in Model 1. However, the stellar counterpart to the Arm in Model 2 is 
more massive (see $\S$~\ref{sec:StellarStream}). As such, the RR Lyrae counts predicted by Model 2
may be detectable: 160 RR Lyrae are expected over the 70 degree extent of the Arm. 

Detecting RR Lyrae looks most promising for the Magellanic Bridge in Model 2, where we expect  $\sim$250 
RR Lyrae.  The quoted values are estimates that invoke a number of assumptions, the corresponding error bars are expected 
to be large. However, they do suggest that searches for RR Lyrae should focus on the Magellanic Bridge connecting 
the MCs. 

If RR Lyrae are detected, reliable distances can be obtained to map out the 3D structure of the Magellanic Bridge. 
Such observations can significantly constrain models for the recent interaction history of the Magellanic Clouds and 
validate our proposed theoretical model for the origin of the microlensing events.

\subsection{Metal Poor Globular Clusters in the MW Halo}

The SMC is known to harbor an anomalously low number of old globular clusters. 
 Although most SMC clusters may have 
 formed in two main events, 2 and 8 Gyr ago \citep{rich2000}, there are also clusters at intermediate 
 ages \citep{piatti2011}. This is not true of the LMC, where a pronounced age gap exists; 
  clusters in the LMC are either as old as the universe or 1-3 Gyr old
  \citep{DaCosta, vandenbergh1991, girardi1995, olszewski1996, piatti2002}. However, 
 the oldest SMC cluster is at the young and metal-rich extreme of the LMC globular cluster distribution \citep{piatti2012};
  there are no analogs of galactic halo globular clusters in the SMC\footnote{The oldest cluster in the SMC is NGC 121,
  which is 2 Gyr younger than the oldest clusters in the LMC \citep{shara1998}}, but they do exist in the LMC \citep{DaCosta}.

We suggest that the fundamental difference in the cluster age distribution in the LMC and SMC
 might be a result of their interaction history.  
 
 The recent dramatic collision between the two galaxies predicted by 
 Model 2 could have resulted in the destruction of globular clusters in the SMC, explaining its overall paucity.\footnote{\citet{glatt2010}
 find no evidence of dissolution of clusters younger than 1 Gyr, but have no constraints for older clusters.}  
Furthermore, older clusters might have been stripped by the action of LMC tides over time. Some could be 
accreted by the LMC, but stellar material is not transferred as efficiently between the two galaxies as 
gaseous material.  The stripped clusters would instead most likely reside in the galactic halo, tracing the extended 
Magellanic Stream and Leading Arm on the plane of the sky. 

It is thus possible that old globular clusters in the galactic halo that exist at Galactocentric distances of $\sim$50-120 kpc 
and are spatially coincident with the Magellanic Stream or Leading Arm might have once resided in the SMC. 
If such clusters exist, their kinematics could confirm an association with the Magellanic System. 


\section{Conclusions} 

We have demonstrated that a population of tidally stripped stars from the SMC located $\sim$4 -10 kpc behind a lensing population of LMC disk stars 
can naturally explain the observed event durations, event frequencies and spatial distribution of microlensing events from the MACHO and OGLE surveys.  
 These results favor the B12 Model 2 scenario for the interaction history of the MCs, wherein the MCs are on their first infall towards the MW and the 
SMC has recently collided with the LMC, leading to a large number of sources distributed non-uniformly behind the LMC disk. 
We note that the models presented in this study were not engineered to address the microlensing puzzle, but were 
rather constructed to explain the structure and kinematics of the Magellanic System more generally. Despite this fact, Model 2 
seems to provide a natural explanation for the observed properties of the reported microlensing events.

The models presented in this study are consistent with the upcoming 3rd epoch HST proper motion data 
(Kallivayalil et al. 2012, in prep).  Note that, since the microlensing observables are dependent on the 
relative motions of the sources and lenses, the exact 3D velocity vector adopted for the system is not a 
major source of error. 
Both Models 1 and 2 use cosmologically motivated models for the dark matter content of the SMC and LMC
 and can explain the origin of the Magellanic Stream \citep{besla2010}.
 However, in B12 we showed that only the direct collision scenario of Model 2 
 reproduces the internal kinematics and structure of both the LMC and SMC.   Support for such an off-center, moderate to high-inclination 
  collision has recently been proposed by \citet{casetti2012} based on the configuration of OB candidates in the LMC. 
In this study we find that Model 2 also provides a better match to the observed event durations 
and event frequencies reported by both the OGLE and MACHO microlensing surveys. 
Model 1, on the other hand, fails to reproduce the event frequency.

The lower event frequencies reported by the OGLE survey appears to be naturally 
accounted for by the survey's lower detection efficiencies and 
sensitivity to faint sources, compared to the MACHO survey.  
Assuming a detection efficiency of 30-50\% we find event frequencies of $\Gamma 
\sim$1-1.54 events/yr in the central regions of the LMC disk in Model 2; this is consistent with the
  observed rate for the MACHO survey of $\Gamma \sim$1.75 yr$^{-1}$. 
If instead we assume a much lower detection efficiency of 10\%, we find event an 
event frequency of $\Gamma \sim$0.38 yr$^{-1}$ across a larger region 
comparable to that covered by the OGLE survey; 
this is consistent with the OGLE survey result of $\Gamma \sim$0.33 yr$^{-1}$. 
Model 1 underpredicts the MACHO observed event frequency by a factor of $\sim$10, although this
predicted rate may still be consistent with the OGLE results for a high detection
 efficiency ($\sim$0.29 yr$^{-1}$ across the entire face of the disk, 100\% efficiency). 

Models 1 and 2 both reproduce the observed range of event durations ($t_e \sim$ 17-70 days) 
without appealing to variations in lensing mass.  Instead, 
we find that the simulated sources exhibit a range of effective distances and relative transverse 
velocities within a given field of view owing to their tidal origin. 
On average, event durations are shorter for Model 2, reflecting the higher relative velocities of the 
stellar debris removed during a high speed, low impact parameter collision between the Clouds.

The simulated lensing events are expected to occur predominantly, but 
not exclusively, in fields of view in the central regions of the LMC disk. 
This is in sharp contrast to self-lensing models. Self-lensing models may be able to 
explain the lower event frequency determined by the OGLE survey \citep{calchi2011, calchi2009}, 
but they fail to explain the existence of events in fields off the LMC's stellar bar (such as MACHO 
events 21, 18); there is a strong decrease of the LMC self-lensing rate with increasing distance
from the LMC stellar bar. 
The exact event frequency expected in off-bar fields in our proposed model 
strongly depends on the spatial distribution/surface density of the underlying source debris field.
Studies like that of \citet{olsen2011} may be able to constrain this unknown. 
 
Model 2 can thus reproduce the observed microlensing quantities reported by both the MACHO and OGLE surveys.
The EROS team chose to limit their sample such that they monitored only the brightest stars in the LMC; we do not 
expect that this approach could detect the lensing signal from a population of faint SMC debris, potentially explaining
the EROS survey non-detection towards the LMC.  We do not compare directly to the microlensing
optical depth reported by these three surveys, as this is a derived quantity that depends 
sensitively on what is assumed about the source population.  
These surveys have assumed that the sources represent a population of LMC stars, 
whereas we assume here that the sources represent SMC stellar debris. 
As such, our derived optical depth cannot be compared directly to the values quoted by the MACHO, EROS and OGLE teams. 
Instead we have computed the observable quantities (event duration and event frequency) directly. 

The presented scenario makes a number of testable predictions. In particular, 
invoking a stellar debris field as the source population implies that there 
must exist a stellar counterpart to the gaseous Magellanic Stream \citep[see also,][]{GN, diaz2012}.
 Searches for stars in the Stream have thus far yielded null results. In $\S$~\ref{sec:Stellar}
we have shown that the 
predicted stellar stream has a surface brightness in the Vband that is $> 34$ mag/arcsec$^2$.  This is
 well beyond the sensitivities of these surveys and so our model does not violate known observational constraints. 
  We further predict number counts of RR Lyrae in the simulated Magellanic Bridge, Stream and Leading 
 Arm;  number counts in the Bridge and Leading Arm in Model 2 may be detectable. 
The variability of RR Lyrae make them easily identifiable and their reliability as distance indicators 
provides an accurate method for confirming association with the Magellanic System.  The detection 
of RR Lyrae may thus provide the most direct method of confirming the tidal nature of the extended 
gaseous components of the Magellanic System and validating the theoretical model presented in this study.

Model 1 and Model 2 also differ in the predicted spatial distribution of debris in the Bridge. In the Model 2 
SMC-LMC collision scenario, LMC tides are able to remove stars from the inner regions of the SMC 
potential.  This debris will consequently be more energetic and spread out over a larger area than that 
predicted by Model 1 or that traced by the gas. 
 
This model predicts that the sources should originate from the SMC and that 
they should make up more than $\sim$0.2 per cent of the LMC's disk mass and be distributed such that 
the surface density is larger than 0.03$\Msun$ pc$^{-2}$. This lower limit is derived from 
the failure of Model 1 to reproduce the observed event frequencies. The Model 2 sources make up 
1.5 per cent of the LMC's current disk mass and naturally explain the observed microlensing events. 
Such a population of low-metallicity stars may have already been detected; 
 \citet{olsen2011} have identified a population of low metallicity stars with distinct kinematics from the LMC disk stars in the same field of view.  
 These stars comprise 5 per cent of their sample and we have already shown in B12 that their line-of-sight kinematics 
 are consistent with expectations of our Models 1 and 2.  
 In this study we have made further predictions for the expected proper motions of these source
  stars with respect to local LMC disk stars, which could be 
 observationally testable ($\S$~\ref{sec:PM}). 

In addition, owing to the rotation of the LMC disk, we find that there is a gradient in the relative 
tangential velocities of the source population across the face of the LMC disk. 
Consequently, we also expect to see, on average, a gradient in the observed event 
durations of the corresponding microlensing events (since $t_e \propto V_\perp^{-1}$, 
$\S$~\ref{sec:Spatial}). 
 It may be possible to observe this gradient if a number of 
microlensing events are detected at locations a few degrees away from either side of the LMC's stellar bar.  


This study provides a natural explanation for the long-standing puzzle concerning the 
origin of microlensing events towards the LMC, which, if unexplained by normal stellar populations, would imply the existence of
 exotic MACHO objects populating the MW halo.  The success of Model 2 in reproducing the 
 observed lensing event properties supports a 
 scenario in which the LMC and SMC have recently ($<$300 Myr ago) collided directly, causing the  
removal of stars from deep within the SMC potential and the formation of the Magellanic Bridge. 
Moreover, the agreement between our simulations and the MACHO and OGLE survey data also suggests that 
there exists a stellar counterpart to the Magellanic Stream and Bridge that has yet to be discovered.

\section*{Acknowledgments}

We thank Dave Bennett, Charles Alcock, Roeland P. van der Marel, Nitya Kallivayalil,
 Knut Olsen, Ed Olszewski, Tim Axelrod, Abi Saha, Kailash Sahu, Raja Guhathakurta, 
David Nidever, Du\v{s}an Kere\v{s}, Kathryn Johnston, Ruben Sanchez-Janssen, 
Renyue Cen, Jenny Green and Martin Weinberg
 for useful discussions that have contributed to this paper.
 The simulations in this paper were run on the Odyssey cluster
supported by the FAS Science Division Research Computing
Group at Harvard University.
GB acknowledges support from NASA through Hubble Fellowship grant
 HST-HF-51284.01-A. This work was also supported in part by NSF grant 
 AST-0907890 and NASA grants NNX08AL43G and NNA09DB30A.

\bibliography{ms}

\begin{thebibliography}{135}
\expandafter\ifx\csname natexlab\endcsname\relax\def\natexlab#1{#1}\fi

\bibitem[{{Afonso} {et~al}\mbox{.}(2000){Afonso}, {Alard}, {Albert},
  {Andersen}, {Ansari}, {Aubourg}, {Bareyre}, {Bauer}, {Beaulieu}, {Bouquet},
  {Char}, {Charlot}, {Couchot}, {Coutures}, {Derue}, {Ferlet}, {Glicenstein},
  {Goldman}, {Gould}, {Graff}, {Gros}, {Haissinski}, {Hamilton}, {Hardin}, {de
  Kat}, {Kim}, {Lasserre}, {Lesquoy}, {Loup}, {Magneville}, {Marquette},
  {Maurice}, {Milsztajn}, {Moniez}, {Palanque-Delabrouille}, {Perdereau},
  {Pr{\'e}vot}, {Regnault}, {Rich}, {Spiro}, {Vidal-Madjar}, {Vigroux},
  {Zylberajch}, {Alcock}, {Allsman}, {Alves}, {Axelrod}, {Becker}, {Cook},
  {Drake}, {Freeman}, {Griest}, {King}, {Lehner}, {Marshall}, {Minniti},
  {Peterson}, {Pratt}, {Quinn}, {Rodgers}, {Stetson}, {Stubbs}, {Sutherland},
  {Tomaney}, {Vandehei}, {Rhie}, {Bennett}, {Fragile}, {Johnson}, {Quinn},
  {Udalski}, {Kubiak}, {Szyma{\'n}ski}, {Pietrzy{\'n}ski}, {Wo{\'z}niak},
  {Zebru{\'n}}, {Albrow}, {Caldwell}, {DePoy}, {Dominik}, {Gaudi}, {Greenhill},
  {Hill}, {Kane}, {Martin}, {Menzies}, {Naber}, {Pogge}, {Pollard}, {Sackett},
  {Sahu}, {Vermaak}, {Watson}, \& {Williams}}]{afonso2000}
{Afonso} C. {et~al.}, 2000, \apj, 532, 340

\bibitem[{{Alard} {et~al}\mbox{.}(1995){Alard}, {Mao}, \&
  {Guibert}}]{alard1995}
{Alard} C., {Mao} S., {Guibert} J., 1995, \aap, 300, L17

\bibitem[{{Alcock}(2009)}]{Alcock2009}
{Alcock} C., 2009, in Astronomical Society of the Pacific Conference Series,
  Vol. 403, The Variable Universe: A Celebration of Bohdan Paczynski,
  {K.~Z.~Stanek}, ed., p.~71

\bibitem[{{Alcock} {et~al}\mbox{.}(2000){Alcock}, {Allsman}, {Alves},
  {Axelrod}, {Becker}, {Bennett}, {Cook}, {Dalal}, {Drake}, {Freeman}, {Geha},
  {Griest}, {Lehner}, {Marshall}, {Minniti}, {Nelson}, {Peterson}, {Popowski},
  {Pratt}, {Quinn}, {Stubbs}, {Sutherland}, {Tomaney}, {Vandehei}, \&
  {Welch}}]{alcock2000}
{Alcock} C. {et~al.}, 2000, \apj, 542, 281

\bibitem[{{Alcock} {et~al}\mbox{.}(2001){Alcock}, {Allsman}, {Alves},
  {Axelrod}, {Becker}, {Bennett}, {Cook}, {Dalal}, {Drake}, {Freeman}, {Geha},
  {Griest}, {Lehner}, {Marshall}, {Minniti}, {Nelson}, {Peterson}, {Popowski},
  {Pratt}, {Quinn}, {Stubbs}, {Sutherland}, {Tomaney}, \&
  {Vandehei}}]{alcock2001}
{Alcock} C. {et~al.}, 2001, \apj, 552, 582

\bibitem[{{Alcock} {et~al}\mbox{.}(1996){Alcock}, {Allsman}, {Axelrod},
  {Bennett}, {Cook}, {Freeman}, {Griest}, {Guern}, {Lehner}, {Marshall},
  {Park}, {Perlmutter}, {Peterson}, {Pratt}, {Quinn}, {Rodgers}, {Stubbs}, \&
  {Sutherland}}]{alcock1996}
{Alcock} C. {et~al.}, 1996, \apj, 461, 84

\bibitem[{{Alves} \& {Nelson}(2000)}]{alves2000}
{Alves} D.~R., {Nelson} C.~A., 2000, \apj, 542, 789

\bibitem[{{Ansari} {et~al}\mbox{.}(1995){Ansari}, {Cavalier}, {Couchot},
  {Moniez}, {Perdereau}, {Aubourg}, {Bareyre}, {Brehin}, {Delabrouille},
  {Gros}, {de Kat}, {Lachieze-Rey}, {Laurent}, {Lesquoy}, {Magneville},
  {Milsztajn}, {Moscoso}, {Queinnec}, {Renault}, {Rich}, {Spiro}, {Vigroux},
  {Zylberajch}, {Beaulieu}, {Ferlet}, {Grison}, {Vidal-Madjar}, {Alard},
  {Guibert}, {Moreau}, {Tajahmady}, {Maurice}, {Prevot}, {Gry}, {Blecha}, \&
  {Burki}}]{ansari1995}
{Ansari} R. {et~al.}, 1995, \aap, 299, L21

\bibitem[{{Aubourg} {et~al}\mbox{.}(1995){Aubourg}, {Bareyre}, {Brehin},
  {Gros}, {de Kat}, {Lachieze-Rey}, {Laurent}, {Lesquoy}, {Magneville},
  {Milsztajn}, {Moscoso}, {Queinnec}, {Renault}, {Rich}, {Spiro}, {Vigroux},
  {Zylberajch}, {Ansari}, {Cavalier}, {Moniez}, {Beaulieu}, {Ferlet}, {Grison},
  {Vidal-Madjar}, {Guibert}, {Moreau}, {Tajahmady}, {Maurice}, {Prevot}, \&
  {Gry}}]{aubourg1995}
{Aubourg} E. {et~al.}, 1995, \aap, 301, 1

\bibitem[{{Aubourg} {et~al}\mbox{.}(1993){Aubourg}, {Bareyre}, {Br{\'e}hin},
  {Gros}, {Lachi{\`e}ze-Rey}, {Laurent}, {Lesquoy}, {Magneville}, {Milsztajn},
  {Moscoso}, {Queinnec}, {Rich}, {Spiro}, {Vigroux}, {Zylberajch}, {Ansari},
  {Cavalier}, {Moniez}, {Beaulieu}, {Ferlet}, {Grison}, {Vidal-Madjar},
  {Guibert}, {Moreau}, {Tajahmady}, {Maurice}, {Pr{\'e}v{\^o}t}, \&
  {Gry}}]{aubourg1993}
{Aubourg} E. {et~al.}, 1993, \nat, 365, 623

\bibitem[{{Bastian} {et~al}\mbox{.}(2010){Bastian}, {Covey}, \&
  {Meyer}}]{Bastian2010}
{Bastian} N., {Covey} K.~R., {Meyer} M.~R., 2010, \araa, 48, 339

\bibitem[{{Beaulieu} \& {Sackett}(1998)}]{beaulieu1998}
{Beaulieu} J.-P., {Sackett} P.~D., 1998, \aj, 116, 209

\bibitem[{{Bekki} \& {Chiba}(2005)}]{Bekki}
{Bekki} K., {Chiba} M., 2005, \mnras, 356, 680

\bibitem[{{Belokurov} {et~al}\mbox{.}(2004){Belokurov}, {Evans}, \& {Le
  Du}}]{belokurov2004}
{Belokurov} V., {Evans} N.~W., {Le Du} Y., 2004, \mnras, 352, 233

\bibitem[{{Bennett}(1998)}]{bennett1998}
{Bennett} D.~P., 1998, \apjl, 493, L79

\bibitem[{{Bennett}(2005)}]{bennett2005}
{Bennett} D.~P., 2005, \apj, 633, 906

\bibitem[{{Bennett} {et~al}\mbox{.}(1996){Bennett}, {Alcock}, {Allsman},
  {Alves}, {Axelrod}, {Becker}, {Cook}, {Freeman}, {Griest}, {Guern}, {Lehner},
  {Marshall}, {Minniti}, {Peterson}, {Pratt}, {Quinn}, {Rhie}, {Rodgers},
  {Stubbs}, {Sutherland}, \& {Welch}}]{bennett1996}
{Bennett} D.~P. {et~al.}, 1996, Nuclear Physics B Proceedings Supplements,
  Vol.~51, 51, 152

\bibitem[{{Bennett} {et~al}\mbox{.}(2005){Bennett}, {Becker}, \&
  {Tomaney}}]{bennettbecker2005}
{Bennett} D.~P., {Becker} A.~C., {Tomaney} A., 2005, \apj, 631, 301

\bibitem[{{Besla} {et~al}\mbox{.}(2007){Besla}, {Kallivayalil}, {Hernquist},
  {Robertson}, {Cox}, {van der Marel}, \& {Alcock}}]{besla2007}
{Besla} G., {Kallivayalil} N., {Hernquist} L., {Robertson} B., {Cox} T.~J.,
  {van der Marel} R.~P., {Alcock} C., 2007, \apj, 668, 949

\bibitem[{{Besla} {et~al}\mbox{.}(2010){Besla}, {Kallivayalil}, {Hernquist},
  {van der Marel}, {Cox}, \& {Kere{\v s}}}]{besla2010}
{Besla} G., {Kallivayalil} N., {Hernquist} L., {van der Marel} R.~P., {Cox}
  T.~J., {Kere{\v s}} D., 2010, \apjl, 721, L97, [B10]

\bibitem[{{Besla} {et~al}\mbox{.}(2012){Besla}, {Kallivayalil}, {Hernquist},
  {van der Marel}, {Cox}, \& {Kere{\v s}}}]{besla2012}
{Besla} G., {Kallivayalil} N., {Hernquist} L., {van der Marel} R.~P., {Cox}
  T.~J., {Kere{\v s}} D., 2012, \mnras, 2457, [B12]

\bibitem[{{Boylan-Kolchin} {et~al}\mbox{.}(2011){Boylan-Kolchin}, {Besla}, \&
  {Hernquist}}]{BoylanBesla2011}
{Boylan-Kolchin} M., {Besla} G., {Hernquist} L., 2011, \mnras, 414, 1560

\bibitem[{{Brook} {et~al}\mbox{.}(2003){Brook}, {Kawata}, \&
  {Gibson}}]{brook2003}
{Brook} C.~B., {Kawata} D., {Gibson} B.~K., 2003, \mnras, 343, 913

\bibitem[{{Brueck} \& {Hawkins}(1983)}]{brueck1983}
{Brueck} M.~T., {Hawkins} M.~R.~S., 1983, \aap, 124, 216

\bibitem[{{Busha} {et~al}\mbox{.}(2011){Busha}, {Marshall}, {Wechsler},
  {Klypin}, \& {Primack}}]{busha2010}
{Busha} M.~T., {Marshall} P.~J., {Wechsler} R.~H., {Klypin} A., {Primack} J.,
  2011, \apj, 743, 40

\bibitem[{{Calchi Novati} \& {Mancini}(2011)}]{calchi2011}
{Calchi Novati} S., {Mancini} L., 2011, \mnras, 416, 1292

\bibitem[{{Calchi Novati} {et~al}\mbox{.}(2009){Calchi Novati}, {Mancini},
  {Scarpetta}, \& {Wyrzykowski}}]{calchi2009}
{Calchi Novati} S., {Mancini} L., {Scarpetta} G., {Wyrzykowski} {\L}., 2009,
  \mnras, 400, 1625

\bibitem[{{Casetti-Dinescu} {et~al}\mbox{.}(2012){Casetti-Dinescu}, {Vieira},
  {Girard}, \& {van Altena}}]{casetti2012}
{Casetti-Dinescu} D.~I., {Vieira} K., {Girard} T.~M., {van Altena} W.~F., 2012,
  \apj, 753, 123

\bibitem[{{Cioni} {et~al}\mbox{.}(2011){Cioni}, {Clementini}, {Girardi},
  {Guandalini}, {Gullieuszik}, {Miszalski}, {Moretti}, {Ripepi}, {Rubele},
  {Bagheri}, {Bekki}, {Cross}, {de Blok}, {de Grijs}, {Emerson}, {Evans},
  {Gibson}, {Gonzales-Solares}, {Groenewegen}, {Irwin}, {Ivanov}, {Lewis},
  {Marconi}, {Marquette}, {Mastropietro}, {Moore}, {Napiwotzki}, {Naylor},
  {Oliveira}, {Read}, {Sutorius}, {van Loon}, {Wilkinson}, \&
  {Wood}}]{cioni2011}
{Cioni} M.-R.~L. {et~al.}, 2011, \aap, 527, A116+

\bibitem[{{Connors} {et~al}\mbox{.}(2005){Connors}, {Kawata}, \&
  {Gibson}}]{Connors}
{Connors} T.~W., {Kawata} D., {Gibson} B.~K., 2005, ArXiv Astrophysics e-prints

\bibitem[{{Connors} {et~al}\mbox{.}(2006){Connors}, {Kawata}, \&
  {Gibson}}]{Connors2006}
{Connors} T.~W., {Kawata} D., {Gibson} B.~K., 2006, \mnras, 371, 108

\bibitem[{{Crowl} {et~al}\mbox{.}(2001){Crowl}, {Sarajedini}, {Piatti},
  {Geisler}, {Bica}, {Clari{\'a}}, \& {Santos}}]{crowl2001}
{Crowl} H.~H., {Sarajedini} A., {Piatti} A.~E., {Geisler} D., {Bica} E.,
  {Clari{\'a}} J.~J., {Santos}, Jr. J.~F.~C., 2001, \aj, 122, 220

\bibitem[{{Da Costa}(1991)}]{DaCosta}
{Da Costa} G.~S., 1991, in IAU Symp. 148: The Magellanic Clouds, {Haynes} R.,
  {Milne} D., eds., p. 183

\bibitem[{{Diaz} \& {Bekki}(2011)}]{diaz2011}
{Diaz} J., {Bekki} K., 2011, \mnras, 413, 2015

\bibitem[{{Diaz} \& {Bekki}(2012)}]{diaz2012}
{Diaz} J.~D., {Bekki} K., 2012, \apj, 750, 36

\bibitem[{{Dominik} \& {Hirshfeld}(1996)}]{dominik1996}
{Dominik} M., {Hirshfeld} A.~C., 1996, \aap, 313, 841

\bibitem[{{Drake} {et~al}\mbox{.}(2004){Drake}, {Cook}, \&
  {Keller}}]{drake2004}
{Drake} A.~J., {Cook} K.~H., {Keller} S.~C., 2004, \apjl, 607, L29

\bibitem[{{Evans} \& {Kerins}(2000)}]{evans2000}
{Evans} N.~W., {Kerins} E., 2000, \apj, 529, 917

\bibitem[{{Flynn} {et~al}\mbox{.}(1996){Flynn}, {Gould}, \&
  {Bahcall}}]{flynn1996}
{Flynn} C., {Gould} A., {Bahcall} J.~N., 1996, \apjl, 466, L55+

\bibitem[{{Flynn} {et~al}\mbox{.}(2003){Flynn}, {Holopainen}, \&
  {Holmberg}}]{flynn2003}
{Flynn} C., {Holopainen} J., {Holmberg} J., 2003, \mnras, 339, 817

\bibitem[{{Fox} {et~al}\mbox{.}(2010){Fox}, {Wakker}, {Smoker}, {Richter},
  {Savage}, \& {Sembach}}]{fox2010}
{Fox} A.~J., {Wakker} B.~P., {Smoker} J.~V., {Richter} P., {Savage} B.~D.,
  {Sembach} K.~R., 2010, \apj, 718, 1046

\bibitem[{{Gallart}(1998)}]{gallart1998}
{Gallart} C., 1998, \apjl, 495, L43

\bibitem[{{Gardiner} \& {Noguchi}(1996)}]{GN}
{Gardiner} L.~T., {Noguchi} M., 1996, \mnras, 278, 191

\bibitem[{{Gibson} \& {Mould}(1997)}]{gibson1997}
{Gibson} B.~K., {Mould} J.~R., 1997, \apj, 482, 98

\bibitem[{{Girardi} {et~al}\mbox{.}(1995){Girardi}, {Chiosi}, {Bertelli}, \&
  {Bressan}}]{girardi1995}
{Girardi} L., {Chiosi} C., {Bertelli} G., {Bressan} A., 1995, \aap, 298, 87

\bibitem[{{Glatt} {et~al}\mbox{.}(2010){Glatt}, {Grebel}, \&
  {Koch}}]{glatt2010}
{Glatt} K., {Grebel} E.~K., {Koch} A., 2010, \aap, 517, A50+

\bibitem[{{Gould}(1995)}]{gould1995a}
{Gould} A., 1995, \apj, 441, 77

\bibitem[{{Gould}(1998)}]{gould1998}
{Gould} A., 1998, \apj, 499, 728

\bibitem[{{Graff} {et~al}\mbox{.}(2000){Graff}, {Gould}, {Suntzeff},
  {Schommer}, \& {Hardy}}]{graff2000}
{Graff} D.~S., {Gould} A.~P., {Suntzeff} N.~B., {Schommer} R.~A., {Hardy} E.,
  2000, \apj, 540, 211

\bibitem[{{Green} \& {Jedamzik}(2002)}]{green2002}
{Green} A.~M., {Jedamzik} K., 2002, \aap, 395, 31

\bibitem[{{Guhathakurta} \& {Reitzel}(1998)}]{NoStars}
{Guhathakurta} P., {Reitzel} D.~B., 1998, in ASP Conf. Ser. 136: Galactic
  Halos, {Zaritsky} D., ed., p.~22

\bibitem[{{Gyuk} {et~al}\mbox{.}(2000){Gyuk}, {Dalal}, \& {Griest}}]{gyuk2000}
{Gyuk} G., {Dalal} N., {Griest} K., 2000, \apj, 535, 90

\bibitem[{{Han} \& {Gould}(1996)}]{han1996}
{Han} C., {Gould} A., 1996, \apj, 467, 540

\bibitem[{{Harris}(2007)}]{harris2007}
{Harris} J., 2007, \apj, 658, 345

\bibitem[{{Haschke} {et~al}\mbox{.}(2011){Haschke}, {Grebel}, \&
  {Duffau}}]{haschke2011}
{Haschke} R., {Grebel} E.~K., {Duffau} S., 2011, \aj, 141, 158

\bibitem[{{Haschke} {et~al}\mbox{.}(2012){Haschke}, {Grebel}, \&
  {Duffau}}]{haschke2012}
{Haschke} R., {Grebel} E.~K., {Duffau} S., 2012, \aj, 144, 107

\bibitem[{{Heller} \& {Rohlfs}(1994)}]{Heller}
{Heller} P., {Rohlfs} K., 1994, \aap, 291, 743

\bibitem[{{Jetzer} {et~al}\mbox{.}(2002){Jetzer}, {Mancini}, \&
  {Scarpetta}}]{jetzer2002}
{Jetzer} P., {Mancini} L., {Scarpetta} G., 2002, \aap, 393, 129

\bibitem[{{Jonsson} \& {Primack}(2010)}]{jonsson2010}
{Jonsson} P., {Primack} J.~R., 2010, New Astronomy, 15, 509

\bibitem[{{Kallivayalil} {et~al}\mbox{.}(2006){Kallivayalil}, {Patten},
  {Marengo}, {Alcock}, {Werner}, \& {Fazio}}]{kallivayalil2006}
{Kallivayalil} N., {Patten} B.~M., {Marengo} M., {Alcock} C., {Werner} M.~W.,
  {Fazio} G.~G., 2006, \apjl, 652, L97

\bibitem[{{Kallivayalil} {et~al}\mbox{.}(2006b){Kallivayalil}, {van der Marel},
  \& {Alcock}}]{Nitya2}
{Kallivayalil} N., {van der Marel} R.~P., {Alcock} C., 2006b, \apj, 652, 1213

\bibitem[{{Kallivayalil} {et~al}\mbox{.}(2006a){Kallivayalil}, {van der Marel},
  {Alcock}, {Axelrod}, {Cook}, {Drake}, \& {Geha}}]{Nitya1}
{Kallivayalil} N., {van der Marel} R.~P., {Alcock} C., {Axelrod} T., {Cook}
  K.~H., {Drake} A.~J., {Geha} M., 2006a, \apj, 638, 772

\bibitem[{{Kerins} \& {Evans}(1999)}]{kerins1999}
{Kerins} E.~J., {Evans} N.~W., 1999, \apj, 517, 734

\bibitem[{{Kerr} {et~al}\mbox{.}(1954){Kerr}, {Hindman}, \&
  {Robinson}}]{kerr1954}
{Kerr} F.~J., {Hindman} J.~F., {Robinson} B.~J., 1954, Australian Journal of
  Physics, 7, 297

\bibitem[{{Kerr} \& {Lynden-Bell}(1986)}]{kerr1986}
{Kerr} F.~J., {Lynden-Bell} D., 1986, \mnras, 221, 1023

\bibitem[{{Kinman} {et~al}\mbox{.}(1991){Kinman}, {Stryker}, {Hesser},
  {Graham}, {Walker}, {Hazen}, \& {Nemec}}]{kinman1991}
{Kinman} T.~D., {Stryker} L.~L., {Hesser} J.~E., {Graham} J.~A., {Walker}
  A.~R., {Hazen} M.~L., {Nemec} J.~M., 1991, \pasp, 103, 1279

\bibitem[{{Komatsu} {et~al}\mbox{.}(2011){Komatsu}, {Smith}, {Dunkley},
  {Bennett}, {Gold}, {Hinshaw}, {Jarosik}, {Larson}, {Nolta}, {Page},
  {Spergel}, {Halpern}, {Hill}, {Kogut}, {Limon}, {Meyer}, {Odegard}, {Tucker},
  {Weiland}, {Wollack}, \& {Wright}}]{komatsu2011}
{Komatsu} E. {et~al.}, 2011, \apjs, 192, 18

\bibitem[{{Kunkel} {et~al}\mbox{.}(1997){Kunkel}, {Demers}, {Irwin}, \&
  {Albert}}]{kunkel1997}
{Kunkel} W.~E., {Demers} S., {Irwin} M.~J., {Albert} L., 1997, \apjl, 488,
  L129+

\bibitem[{{Lasserre} {et~al}\mbox{.}(2000){Lasserre}, {Afonso}, {Albert},
  {Andersen}, {Ansari}, {Aubourg}, {Bareyre}, {Bauer}, {Beaulieu}, {Blanc},
  {Bouquet}, {Char}, {Charlot}, {Couchot}, {Coutures}, {Derue}, {Ferlet},
  {Glicenstein}, {Goldman}, {Gould}, {Graff}, {Gros}, {Haissinski}, {Hamilton},
  {Hardin}, {de Kat}, {Kim}, {Lesquoy}, {Loup}, {Magneville}, {Mansoux},
  {Marquette}, {Maurice}, {Milsztajn}, {Moniez}, {Palanque-Delabrouille},
  {Perdereau}, {Pr{\'e}vot}, {Regnault}, {Rich}, {Spiro}, {Vidal-Madjar},
  {Vigroux}, {Zylberajch}, \& {EROS Collaboration}}]{laserre2000}
{Lasserre} T. {et~al.}, 2000, \aap, 355, L39

\bibitem[{{Lin} {et~al}\mbox{.}(1995){Lin}, {Jones}, \& {Klemola}}]{Lin95}
{Lin} D.~N.~C., {Jones} B.~F., {Klemola} A.~R., 1995, \apj, 439, 652

\bibitem[{{Majewski} {et~al}\mbox{.}(2009){Majewski}, {Nidever}, {Mu{\~n}oz},
  {Patterson}, {Kunkel}, \& {Carlin}}]{majewski2009}
{Majewski} S.~R., {Nidever} D.~L., {Mu{\~n}oz} R.~R., {Patterson} R.~J.,
  {Kunkel} W.~E., {Carlin} J.~L., 2009, in IAU Symposium, Vol. 256, IAU
  Symposium, {J.~T.~van Loon \& J.~M.~Oliveira}, ed., pp. 51--56

\bibitem[{{Mancini} {et~al}\mbox{.}(2004){Mancini}, {Calchi Novati}, {Jetzer},
  \& {Scarpetta}}]{mancini2004}
{Mancini} L., {Calchi Novati} S., {Jetzer} P., {Scarpetta} G., 2004, \aap, 427,
  61

\bibitem[{{Mao} \& {Di Stefano}(1995)}]{mao1995}
{Mao} S., {Di Stefano} R., 1995, \apj, 440, 22

\bibitem[{{Maraston}(2005)}]{maraston2005}
{Maraston} C., 2005, \mnras, 362, 799

\bibitem[{{Mastropietro} {et~al}\mbox{.}(2005){Mastropietro}, {Moore}, {Mayer},
  {Wadsley}, \& {Stadel}}]{Mastro}
{Mastropietro} C., {Moore} B., {Mayer} L., {Wadsley} J., {Stadel} J., 2005,
  \mnras, 363, 509

\bibitem[{{Mathewson} {et~al}\mbox{.}(1974){Mathewson}, {Cleary}, \&
  {Murray}}]{Mathewson}
{Mathewson} D.~S., {Cleary} M.~N., {Murray} J.~D., 1974, \apj, 190, 291

\bibitem[{{McMillan}(2011)}]{McMillan2011}
{McMillan} P.~J., 2011, \mnras, 414, 2446

\bibitem[{{Monelli} {et~al}\mbox{.}(2011){Monelli}, {Carrera}, {Gallart},
  {Meschin}, {Aparicio}, {Hidalgo}, {Bono}, {Stetson}, \&
  {Walker}}]{monelli2011}
{Monelli} M. {et~al.}, 2011, in EAS Publications Series, Vol.~48, EAS
  Publications Series, {Koleva} M., {Prugniel} P., {Vauglin} I., eds., pp.
  43--49

\bibitem[{{Moniez}(2010)}]{Moniez2010}
{Moniez} M., 2010, General Relativity and Gravitation, 42, 2047

\bibitem[{{Moore} \& {Davis}(1994)}]{Moore}
{Moore} B., {Davis} M., 1994, \mnras, 270, 209

\bibitem[{{Mu{\~n}oz} {et~al}\mbox{.}(2006){Mu{\~n}oz}, {Majewski}, {Zaggia},
  {Kunkel}, {Frinchaboy}, {Nidever}, {Crnojevic}, {Patterson}, {Crane},
  {Johnston}, {Sohn}, {Bernstein}, \& {Shectman}}]{munoz2006}
{Mu{\~n}oz} R.~R. {et~al.}, 2006, \apj, 649, 201

\bibitem[{{Nelson} {et~al}\mbox{.}(2009){Nelson}, {Drake}, {Cook}, {Bennett},
  {Popowski}, {Dalal}, {Nikolaev}, {Alcock}, {Axelrod}, {Becker}, {Freeman},
  {Geha}, {Griest}, {Keller}, {Lehner}, {Marshall}, {Minniti}, {Pratt},
  {Quinn}, {Stubbs}, {Sutherland}, {Tomaney}, {Vandehei}, \&
  {Welch}}]{nelson2009}
{Nelson} C.~A. {et~al.}, 2009, ArXiv e-prints

\bibitem[{{Nguyen} {et~al}\mbox{.}(2004){Nguyen}, {Kallivayalil}, {Werner},
  {Alcock}, {Patten}, \& {Stern}}]{nguyen2004}
{Nguyen} H.~T., {Kallivayalil} N., {Werner} M.~W., {Alcock} C., {Patten} B.~M.,
  {Stern} D., 2004, \apjs, 154, 266

\bibitem[{{Nidever} {et~al}\mbox{.}(2008){Nidever}, {Majewski}, \&
  {Burton}}]{nidever2008}
{Nidever} D.~L., {Majewski} S.~R., {Burton} W.~B., 2008, \apj, 679, 432

\bibitem[{{Nidever} {et~al}\mbox{.}(2010){Nidever}, {Majewski}, {Butler
  Burton}, \& {Nigra}}]{nidever2010}
{Nidever} D.~L., {Majewski} S.~R., {Butler Burton} W., {Nigra} L., 2010, \apj,
  723, 1618

\bibitem[{{Olano}(2004)}]{olano2004}
{Olano} C.~A., 2004, \aap, 423, 895

\bibitem[{{Olsen} {et~al}\mbox{.}(2011){Olsen}, {Zaritsky}, {Blum}, {Boyer}, \&
  {Gordon}}]{olsen2011}
{Olsen} K.~A.~G., {Zaritsky} D., {Blum} R.~D., {Boyer} M.~L., {Gordon} K.~D.,
  2011, \apj, 737, 29

\bibitem[{{Olszewski} {et~al}\mbox{.}(1996){Olszewski}, {Suntzeff}, \&
  {Mateo}}]{olszewski1996}
{Olszewski} E.~W., {Suntzeff} N.~B., {Mateo} M., 1996, \araa, 34, 511

\bibitem[{{Oppenheimer} {et~al}\mbox{.}(2001){Oppenheimer}, {Saumon},
  {Hodgkin}, {Jameson}, {Hambly}, {Chabrier}, {Filippenko}, {Coil}, \&
  {Brown}}]{oppenheimer2001}
{Oppenheimer} B.~R. {et~al.}, 2001, \apj, 550, 448

\bibitem[{{Paczynski}(1986)}]{paczynski1986}
{Paczynski} B., 1986, \apj, 304, 1

\bibitem[{{Pejcha} \& {Stanek}(2009)}]{pejcha2009}
{Pejcha} O., {Stanek} K.~Z., 2009, \apj, 704, 1730

\bibitem[{{Piatti}(2011)}]{piatti2011}
{Piatti} A.~E., 2011, \mnras, 416, L89

\bibitem[{{Piatti} \& {Geisler}(2012)}]{piatti2012}
{Piatti} A.~E., {Geisler} D., 2012, ArXiv e-prints

\bibitem[{{Piatti} {et~al}\mbox{.}(2002){Piatti}, {Sarajedini}, {Geisler},
  {Bica}, \& {Clari{\'a}}}]{piatti2002}
{Piatti} A.~E., {Sarajedini} A., {Geisler} D., {Bica} E., {Clari{\'a}} J.~J.,
  2002, \mnras, 329, 556

\bibitem[{{Popowski} {et~al}\mbox{.}(2003){Popowski}, {Nelson}, {Bennett},
  {Drake}, {Vandehei}, {Griest}, {Cook}, {Alcock}, {Allsman}, {Alves},
  {Axelrod}, {Becker}, {Freeman}, {Geha}, {Lehner}, {Marshall}, {Minniti},
  {Peterson}, {Quinn}, {Stubbs}, {Sutherland}, \& {Welch}}]{popowski2003}
{Popowski} P. {et~al.}, 2003, ArXiv Astrophysics e-prints

\bibitem[{{Putman} {et~al}\mbox{.}(2003){Putman}, {Staveley-Smith}, {Freeman},
  {Gibson}, \& {Barnes}}]{Putman2003}
{Putman} M.~E., {Staveley-Smith} L., {Freeman} K.~C., {Gibson} B.~K., {Barnes}
  D.~G., 2003, \apj, 586, 170

\bibitem[{{Rahvar}(2004)}]{rahvar2004}
{Rahvar} S., 2004, \mnras, 347, 213

\bibitem[{{Reid} \& {Brunthaler}(2004)}]{reid2004}
{Reid} M.~J., {Brunthaler} A., 2004, \apj, 616, 872

\bibitem[{{Reid} {et~al}\mbox{.}(2009){Reid}, {Menten}, {Zheng}, {Brunthaler},
  {Moscadelli}, {Xu}, {Zhang}, {Sato}, {Honma}, {Hirota}, {Hachisuka}, {Choi},
  {Moellenbrock}, \& {Bartkiewicz}}]{reid2009}
{Reid} M.~J. {et~al.}, 2009, \apj, 700, 137

\bibitem[{{Renault} {et~al}\mbox{.}(1997){Renault}, {Afonso}, {Aubourg},
  {Bareyre}, {Bauer}, {Brehin}, {Coutures}, {Gaucherel}, {Glicenstein},
  {Goldman}, {Gros}, {Hardin}, {de Kat}, {Lachieze-Rey}, {Laurent}, {Lesquoy},
  {Magneville}, {Milsztajn}, {Moscoso}, {Palanque-Delabrouille}, {Queinnec},
  {Rich}, {Spiro}, {Vigroux}, {Zylberajch}, {Ansari}, {Cavalier}, {Couchot},
  {Mansoux}, {Moniez}, {Perdereau}, {Beaulieu}, {Ferlet}, {Grison},
  {Vidal-Madjar}, {Guibert}, {Moreau}, {Maurice}, {Prevot}, {Gry}, {Char}, \&
  {Fernandez}}]{renault1997}
{Renault} C. {et~al.}, 1997, \aap, 324, L69

\bibitem[{{Rich} {et~al}\mbox{.}(2000){Rich}, {Shara}, {Fall}, \&
  {Zurek}}]{rich2000}
{Rich} R.~M., {Shara} M., {Fall} S.~M., {Zurek} D., 2000, \aj, 119, 197

\bibitem[{{Ru{\v z}i{\v c}ka} {et~al}\mbox{.}(2010){Ru{\v z}i{\v c}ka},
  {Theis}, \& {Palou{\v s}}}]{ruzicka2010}
{Ru{\v z}i{\v c}ka} A., {Theis} C., {Palou{\v s}} J., 2010, \apj, 725, 369

\bibitem[{{Saha} {et~al}\mbox{.}(2010){Saha}, {Olszewski}, {Brondel}, {Olsen},
  {Knezek}, {Harris}, {Smith}, {Subramaniam}, {Claver}, {Rest}, {Seitzer},
  {Cook}, {Minniti}, \& {Suntzeff}}]{saha2010}
{Saha} A. {et~al.}, 2010, \aj, 140, 1719

\bibitem[{{Sahu}(1994)}]{sahu1994}
{Sahu} K.~C., 1994, \nat, 370, 275

\bibitem[{{Sahu}(2003)}]{sahu2003}
{Sahu} K.~C., 2003, in The Dark Universe: Matter, Energy and Gravity, {Livio}
  M., ed., pp. 14--23

\bibitem[{{Salpeter}(1955)}]{Salpeter1955}
{Salpeter} E.~E., 1955, \apj, 121, 161

\bibitem[{{Sch{\"o}nrich} {et~al}\mbox{.}(2010){Sch{\"o}nrich}, {Binney}, \&
  {Dehnen}}]{Schonrich2010}
{Sch{\"o}nrich} R., {Binney} J., {Dehnen} W., 2010, \mnras, 403, 1829

\bibitem[{{Shara} {et~al}\mbox{.}(1998){Shara}, {Fall}, {Rich}, \&
  {Zurek}}]{shara1998}
{Shara} M.~M., {Fall} S.~M., {Rich} R.~M., {Zurek} D., 1998, \apj, 508, 570

\bibitem[{{Shattow} \& {Loeb}(2009)}]{shattow2009}
{Shattow} G., {Loeb} A., 2009, \mnras, 392, L21

\bibitem[{{Soszy{\~n}ski} {et~al}\mbox{.}(2010){Soszy{\~n}ski}, {Udalski},
  {Szyma{\~n}ski}, {Kubiak}, {Pietrzy{\~n}ski}, {Wyrzykowski}, {Ulaczyk}, \&
  {Poleski}}]{soszynski2010}
{Soszy{\~n}ski} I., {Udalski} A., {Szyma{\~n}ski} M.~K., {Kubiak} J.,
  {Pietrzy{\~n}ski} G., {Wyrzykowski} {\L}., {Ulaczyk} K., {Poleski} R., 2010,
  Acta Astronomica, 60, 165

\bibitem[{{Spagna} {et~al}\mbox{.}(2004){Spagna}, {Carollo}, {Lattanzi}, \&
  {Bucciarelli}}]{spagna2004}
{Spagna} A., {Carollo} D., {Lattanzi} M.~G., {Bucciarelli} B., 2004, \aap, 428,
  451

\bibitem[{{Springel}(2005)}]{Gadget2}
{Springel} V., 2005, \mnras, 364, 1105

\bibitem[{{Stanimirovi{\'c}} {et~al}\mbox{.}(2004){Stanimirovi{\'c}},
  {Staveley-Smith}, \& {Jones}}]{Stan2004}
{Stanimirovi{\'c}} S., {Staveley-Smith} L., {Jones} P.~A., 2004, \apj, 604, 176

\bibitem[{{Subramanian} \& {Subramaniam}(2012)}]{sub2012}
{Subramanian} S., {Subramaniam} A., 2012, \apj, 744, 128

\bibitem[{{Sumi} {et~al}\mbox{.}(2011){Sumi}, {Kamiya}, {Bennett}, {Bond},
  {Abe}, {Botzler}, {Fukui}, {Furusawa}, {Hearnshaw}, {Itow}, {Kilmartin},
  {Korpela}, {Lin}, {Ling}, {Masuda}, {Matsubara}, {Miyake}, {Motomura},
  {Muraki}, {Nagaya}, {Nakamura}, {Ohnishi}, {Okumura}, {Perrott},
  {Rattenbury}, {Saito}, {Sako}, {Sullivan}, {Sweatman}, {Tristram}, {Udalski},
  {Szyma{\'n}ski}, {Kubiak}, {Pietrzy{\'n}ski}, {Poleski}, {Soszy{\'n}ski},
  {Wyrzykowski}, {Ulaczyk}, \& {Microlensing Observations in Astrophysics (MOA)
  Collaboration}}]{sumi2011}
{Sumi} T. {et~al.}, 2011, \nat, 473, 349

\bibitem[{{Tisserand}(2005)}]{tisserand2005}
{Tisserand} P., 2005, in SF2A-2005: Semaine de l'Astrophysique Francaise,
  {F.~Casoli, T.~Contini, J.~M.~Hameury, \& L.~Pagani}, ed., pp. 569--+

\bibitem[{{Tisserand} {et~al}\mbox{.}(2007){Tisserand}, {Le Guillou}, {Afonso},
  {Albert}, {Andersen}, {Ansari}, {Aubourg}, {Bareyre}, {Beaulieu}, {Charlot},
  {Coutures}, {Ferlet}, {Fouqu{\'e}}, {Glicenstein}, {Goldman}, {Gould},
  {Graff}, {Gros}, {Haissinski}, {Hamadache}, {de Kat}, {Lasserre}, {Lesquoy},
  {Loup}, {Magneville}, {Marquette}, {Maurice}, {Maury}, {Milsztajn}, {Moniez},
  {Palanque-Delabrouille}, {Perdereau}, {Rahal}, {Rich}, {Spiro},
  {Vidal-Madjar}, {Vigroux}, {Zylberajch}, \& {The EROS-2
  Collaboration}}]{tisserand2007}
{Tisserand} P. {et~al.}, 2007, \aap, 469, 387

\bibitem[{{Torres} {et~al}\mbox{.}(2002){Torres}, {Garc{\'{\i}}a-Berro},
  {Burkert}, \& {Isern}}]{torres2002}
{Torres} S., {Garc{\'{\i}}a-Berro} E., {Burkert} A., {Isern} J., 2002, \mnras,
  336, 971

\bibitem[{{Udalski} {et~al}\mbox{.}(1997){Udalski}, {Kubiak}, \&
  {Szymanski}}]{udalski1997}
{Udalski} A., {Kubiak} M., {Szymanski} M., 1997, Acta Astronomica, 47, 319

\bibitem[{{Udalski} {et~al}\mbox{.}(1994){Udalski}, {Szymanski}, {Mao}, {Di
  Stefano}, {Kaluzny}, {Kubiak}, {Mateo}, \& {Krzeminski}}]{udalski1994}
{Udalski} A., {Szymanski} M., {Mao} S., {Di Stefano} R., {Kaluzny} J., {Kubiak}
  M., {Mateo} M., {Krzeminski} W., 1994, \apjl, 436, L103

\bibitem[{{van den Bergh}(1991)}]{vandenbergh1991}
{van den Bergh} S., 1991, \apj, 369, 1

\bibitem[{{van der Marel}(2001)}]{vanderMarel2001}
{van der Marel} R.~P., 2001, \aj, 122, 1827

\bibitem[{{van der Marel} {et~al}\mbox{.}(2002){van der Marel}, {Alves},
  {Hardy}, \& {Suntzeff}}]{vanderMarel}
{van der Marel} R.~P., {Alves} D.~R., {Hardy} E., {Suntzeff} N.~B., 2002, \aj,
  124, 2639

\bibitem[{{Weinberg}(2000)}]{weinberg2000}
{Weinberg} M.~D., 2000, \apj, 532, 922

\bibitem[{{Weinberg} \& {Nikolaev}(2001)}]{weinberg2001}
{Weinberg} M.~D., {Nikolaev} S., 2001, \apj, 548, 712

\bibitem[{{Wu}(1994)}]{wu1994}
{Wu} X.-P., 1994, \apj, 435, 66

\bibitem[{{Wyrzykowski} {et~al}\mbox{.}(2009){Wyrzykowski}, {Koz{\l}owski},
  {Skowron}, {Belokurov}, {Smith}, {Udalski}, {Szyma{\'n}ski}, {Kubiak},
  {Pietrzy{\'n}ski}, {Soszy{\'n}ski}, {Szewczyk}, \&
  {{\.Z}ebru{\'n}}}]{wyrzy2009}
{Wyrzykowski} {\L}. {et~al.}, 2009, \mnras, 397, 1228

\bibitem[{{Wyrzykowski} {et~al}\mbox{.}(2011{\natexlab{a}}){Wyrzykowski},
  {Koz{\l}owski}, {Skowron}, {Udalski}, {Szyma{\'n}ski}, {Kubiak},
  {Pietrzy{\'n}ski}, {Soszy{\'n}ski}, {Szewczyk}, {Ulaczyk}, \&
  {Poleski}}]{OGLEIII}
{Wyrzykowski} {\L}. {et~al.}, 2011{\natexlab{a}}, \mnras, 413, 493

\bibitem[{{Wyrzykowski} {et~al}\mbox{.}(2011{\natexlab{b}}){Wyrzykowski},
  {Skowron}, {Koz{\l}owski}, {Udalski}, {Szyma{\'n}ski}, {Kubiak},
  {Pietrzy{\'n}ski}, {Soszy{\'n}ski}, {Szewczyk}, {Ulaczyk}, {Poleski}, \&
  {Tisserand}}]{Wyrzy2011}
{Wyrzykowski} L. {et~al.}, 2011{\natexlab{b}}, \mnras, 416, 2949

\bibitem[{{Zaritsky} \& {Lin}(1997)}]{zaritsky1997}
{Zaritsky} D., {Lin} D.~N.~C., 1997, \aj, 114, 2545

\bibitem[{{Zhao}(1998)}]{zhao1998}
{Zhao} H., 1998, \mnras, 294, 139, [Z98]

\bibitem[{{Zhao}(1999)}]{zhao1999b}
{Zhao} H., 1999, \apj, 527, 167

\bibitem[{{Zhao} \& {Evans}(2000)}]{zhao2000}
{Zhao} H., {Evans} N.~W., 2000, \apjl, 545, L35

\bibitem[{{Zhao} {et~al}\mbox{.}(2000){Zhao}, {Graff}, \&
  {Guhathakurta}}]{zhao2000c}
{Zhao} H., {Graff} D.~S., {Guhathakurta} P., 2000, \apjl, 532, L37

\bibitem[{{Zhao}(2000)}]{zhao2000b}
{Zhao} H.~S., 2000, \apj, 530, 299

\end{thebibliography}

\label{lastpage}
\end{document}